\documentclass[english,floatfix,twocolumn,showpacs,aps, reprint, superscriptaddress, longbibliography]{revtex4-2}

\usepackage{amsmath}

\usepackage{hyperref}
\hypersetup{colorlinks=true, citecolor=blue, urlcolor=blue, linkcolor=blue, breaklinks=true}
\usepackage{amssymb}
\usepackage{graphicx}
\usepackage{siunitx}
\usepackage{booktabs, footnote}
\usepackage{physics}
\usepackage{float}
\usepackage[capitalise,compress]{cleveref}
\crefname{section}{Sec.}{Secs.}
\Crefname{section}{Section}{Sections}
\crefrangelabelformat{equation}{\textup{(#3#1#4)}--\textup{(#5#2#6)}}

\begin{document}

\title{Trimer quantum spin liquid in a honeycomb array of Rydberg atoms}

\author{Milan Kornja\v ca}
\thanks{These authors contributed equally to this work.\\
Email address:
\href{mailto:rhine\textunderscore samajdar@princeton.edu}{rhine\textunderscore samajdar@princeton.edu};\,\href{mailto:swang@quera.com}{swang@quera.com};
\href{mailto:fliu@quera.com}{fliu@quera.com}
}
\affiliation{QuEra Computing Inc., 1284 Soldiers Field Road, Boston, MA, 02135, USA}
\affiliation{Department of Physics and Astronomy, Iowa State University, 12 Physics Hall, Ames, Iowa 50011, USA}

\author{Rhine Samajdar$^\dagger$}
\thanks{These authors contributed equally to this work.\\
Email address:
\href{mailto:rhine\textunderscore samajdar@princeton.edu}{rhine\textunderscore samajdar@princeton.edu};\,\href{mailto:swang@quera.com}{swang@quera.com};
\href{mailto:fliu@quera.com}{fliu@quera.com}
}
\affiliation{Department of Physics, Princeton University, Princeton, NJ, 08544, USA}
\affiliation{Princeton Center for Theoretical Science, Princeton University, Princeton, NJ, 08544, USA}

\author{Tommaso Macr\` i}
\affiliation{QuEra Computing Inc., 1284 Soldiers Field Road, Boston, MA, 02135, USA}
\affiliation{ITAMP, Harvard-Smithsonian Center for Astrophysics, Cambridge, Massachusetts 02138, USA}
\affiliation{Departamento de F\'isica Te\'orica e Experimental, Federal University of Rio Grande do Norte  59078-950 Natal-RN, Brazil}

\author{Nathan Gemelke}
\affiliation{QuEra Computing Inc., 1284 Soldiers Field Road, Boston, MA, 02135, USA}

\author{Sheng-Tao Wang$^\dagger$}
\affiliation{QuEra Computing Inc., 1284 Soldiers Field Road, Boston, MA, 02135, USA}

\author{Fangli Liu$^\dagger$}
\affiliation{QuEra Computing Inc., 1284 Soldiers Field Road, Boston, MA, 02135, USA}

\date{\today}

\begin{abstract}

Quantum spin liquids are elusive but paradigmatic examples of strongly correlated quantum states that are characterized by long-range quantum entanglement. Recently, the direct signatures of a gapped topological $\mathbb{Z}_2$ spin liquid have been observed in a system of Rydberg atoms arrayed on the ruby lattice. Here, we illustrate the concrete realization of a fundamentally different class of spin liquids in a honeycomb array of Rydberg atoms. Exploring the quantum phase diagram of this system using both density-matrix renormalization group and exact diagonalization simulations, several density-wave-ordered phases are characterized and their origins explained. More interestingly, in the regime where third-nearest-neighbor atoms lie within the Rydberg blockade radius, we find a novel ground state---with an emergent $\mathrm{U}(1)\times \mathrm{U}(1)$ local symmetry---formed from superpositions of classical {\it trimer} configurations on the dual triangular lattice. The fidelity of this trimer spin liquid state can be enhanced via dynamical preparation, which we explain by a Rydberg-blockade-based projection mechanism associated with the smooth turnoff of the laser drive. Finally, we discuss the robustness of the trimer spin liquid phase under realistic experimental parameters and demonstrate that our proposal can be readily implemented in current Rydberg atom quantum simulators. 

\end{abstract}


\maketitle

\textit{Introduction.---}Quantum spin liquids are strongly correlated many-body systems that host remarkable phenomena such as emergent gauge fields, long-range entanglement, and fractionalized excitations \cite{Savary2016, Zhou2017}. However, even after 50 years from their original conception as resonating valence bond (RVB) states \cite{Anderson1973}, concrete realizations of these fascinating phases in magnetic insulator materials  are few and far between. Today, advances in neutral-atom quantum simulators have unleashed the potential for realizing highly controllable, coherent quantum many-body systems, which are ideal testbeds for  exploring quantum criticality \cite{Lienhard2018, Keesling2019, PhysRevA.98.023614, PhysRevB.98.205118, Samajdar2020, Chen2022}, probing quantum many-body dynamics \cite{Labuhn2016, Bernien2017, Browaeys2020}, and preparing exotic phases of quantum matter \cite{Ebadi2021, Scholl2021}. In particular, recent experiments on a 219-qubit programmable Rydberg quantum simulator \cite{Semeghini2021} have demonstrated the remarkable realization of a gapped topological phase known as the $\mathbb{Z}_2$ quantum spin liquid \cite{ReadSachdev91,Wen1991,Sachdev92,kitaev2006anyons}.

The zoo of quantum spin liquids \cite{Savary2016, RevModPhys.89.041004,broholm2020quantum}, however, has many other species including, for instance, states where the invariant gauge group is $\mathrm{U}(1)$ instead of $\mathbb{Z}_2$. Such a $\mathrm{U}(1)$ quantum spin liquid is particularly interesting from the perspective of fundamental physics as it hosts an emergent \textit{gapless} excitation, termed a photon in analogy to conventional electromagnetism \cite{hermele2004a}. Unfortunately, experimental efforts to realize such a phase in rare-earth pyrochlore materials \cite{gingras2014quantum} are often complicated by competing microscopic interactions. Standard parton mean-field constructions \cite{Baskaran1988, Wen1991} also do not guarantee the stability of the gapless state under monopole fluctuations \cite{Hermele2004}. These varied considerations highlight the importance of discovering robust candidates for studying the rich physics of $\mathrm{U}(1)$ gauge theories.

\begin{figure*}[t!]
\centering
\includegraphics[width=1.0\textwidth]{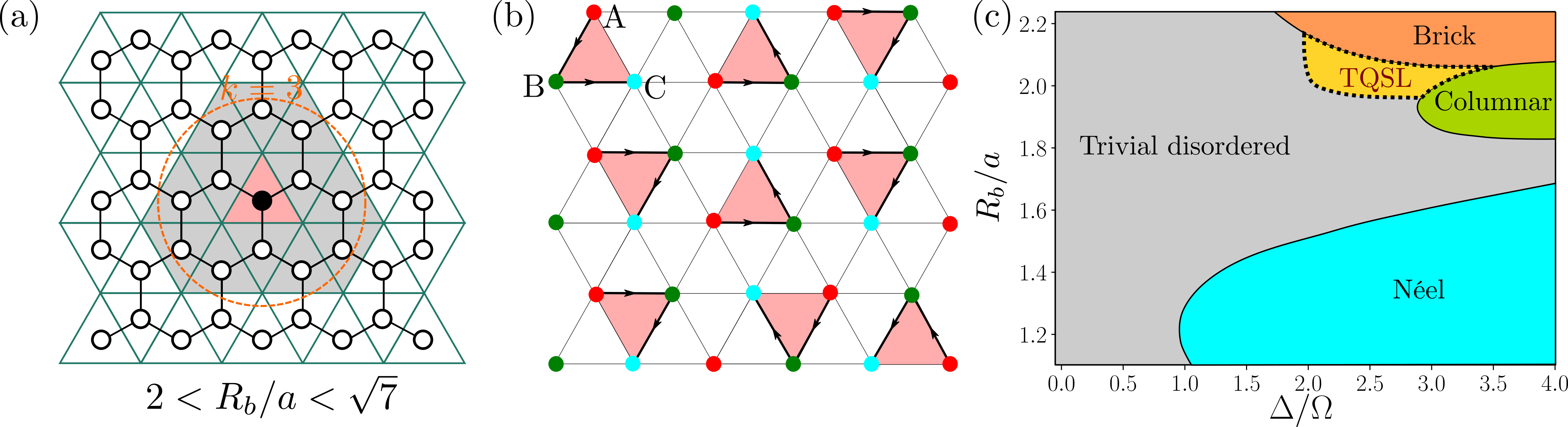}
\caption{\textbf{Trimer quantum spin liquid (TQSL) on a honeycomb lattice of Rydberg atoms.} (a) In the regime where three nearest-neighbor Rydberg atoms are within the blockade radius ($k$\,$=$\,$3$ shell, shown in orange), the blockade-obeying configurations map exactly to the trimer coverings of the triangular lattice with vertices at the centers of the hexagons \cite{Thewes2020}. The number of trimer coverings on the triangular lattice grows exponentially with the system size \cite{Verberkmoes1999}. (b) The TQSL is an equal superposition of exponentially many trimer coverings (one covering shown with filled triangles), and is characterized by a $\mathrm{U}(1)\times \mathrm{U}(1)$ local symmetry due to the tripartite nature of triangular lattice with respect to trimers \cite{Giudice2022}. For the tripartition and the trimer configuration shown, we assign two sets of electric fields directed from A to B and from B to C sublattices (arrows). The two $\mathrm{U}(1)$ degrees of freedom can be related to two conservation laws, as the independent fluxes are equal to the charges $N_A-N_B$ and $N_B-N_C$ enclosed by a closed loop. (c) Quantum phase diagram of Rydberg atoms on a $32 \times 4$ honeycomb lattice retaining three strongest interactions, as obtained by DMRG  \cite{itensor}. The boundaries of the three ordered phases (N\' eel, columnar, and brick) are mapped out by entanglement entropy, energy susceptibility, and fidelity susceptibility peaks (full lines). In addition, a region with a large entanglement entropy is distinguished by fidelity and energy susceptibility (dashed lines) measurements in the regime where third-nearest neighbors are blockaded. The properties of this unordered phase agree with the expected properties of the TQSL state on a finite cylinder.
}
\label{fig:Fig1_PD}
\end{figure*}

In this work, motivated by the recent experimental progress in trapping neutral atoms, we explore the possibility of finding such spin liquids in programmable Rydberg platforms. 
We show that, on the honeycomb lattice, strong van der Waals interactions between Rydberg atoms lead to an effective ``blockade'' constraint \cite{Jaksch2000, Lukin2001, Gaetan2009, Urban2009} that can be mapped to a ``trimer'' constraint on an underlying triangular lattice \cite{Thewes2020}. Related blockade-induced dimer constraints have played a central role in the recent proposals of gapped dimer-RVB $\mathbb{Z}_2$ spin liquids of Rydberg atoms on both the kagome \cite{Samajdar2021, Yan_2022,IGT} and the ruby \cite{Verresen2021} lattice. On the honeycomb lattice, however, the possibility for a novel \emph{trimer} quantum spin liquid (TQSL) state arises owing to the correspondence with trimer---as opposed to dimer---coverings. While such an RVB state of trimers was recently classified as a gapless liquid with an emergent $\mathrm{U}(1)$\,$\times$\,$\mathrm{U}(1)$ local symmetry \cite{Giudice2022}, a microscopic model supporting such a phase and its corresponding physical realizations has yet to be found. Here, we bridge this gap and present conclusive evidence of the realization of TQSLs with honeycomb Rydberg arrays for a range of experimentally relevant parameters.

In order to explore the nature of the ground states, we turn to density-matrix renormalization group (DMRG) \cite{White1992, White1993} simulations and exact diagonalization (ED) of the so-called ``PXP model'' on finite-size clusters \cite{Bloqade}. We find clear signatures of the TQSL phase---including, for example, a high fidelity overlap with the perfect TQSL state---and demonstrate its robustness to real experimental conditions that include long-ranged interactions, relaxed boundary conditions, and experimental state preparation protocols. In addition, we find that the experimental protocol leads to an enhancement of TQSL fidelities compared to the ground state. Our understanding of the underlying mechanism thereof leads us to conjecture a \textit{universal} fidelity enhancement for any state that is a superposition of configurations with the maximum allowed Rydberg excitations (subject to blockade constraints). Finally, we discuss the experimental signatures of the TQSL states and demonstrate that our proposal can be implemented and studied in today's Rydberg atom quantum simulators.

\textit{Trimer model mapping.---}We consider a system of Rydberg atoms arrayed on a honeycomb lattice with the distance between nearest-neighbour sites being $a$. This can be achieved experimentally by placing the neutral atoms in optical tweezers and arranging them using spatial light modulators, with currently attainable system sizes in excess of 200 atoms \cite{Ebadi2021}. The atoms are driven between the ground ($\ket{g}$) and highly excited Rydberg states ($\ket{r}$) by a coherent laser drive with Rabi frequency $\Omega$ and detuning $\Delta$, leading to the Hamiltonian
\begin{equation}\label{eq:RydHam}
    \dfrac{H}{\hbar} = \sum_i \left( \frac{\Omega}{2}  \vert g_i\rangle \langle r_i\vert + \mathrm{h.c.} \right) - \sum_i \Delta n_i  + \sum_{i < j}
    V_{ij}n_i n_j,
\end{equation}
where $\hbar$ is the reduced Planck constant, $i$ denotes the lattice sites, $n_i \equiv |r \rangle_i \langle r |$ counts the occupation of the excited states, and $ V_{ij}$ are the van der Waals interactions between atoms in Rydberg states. The van der Waals interactions fall off with the distance between atoms, $\mathcal{R}$, as $V(\mathcal{R})$\,$=$\,$C_6/\mathcal{R}^6$ and are central to the phenomenon of Rydberg blockade that we utilize. More precisely, of the neighboring atoms lying within a distance $R_b$ (the blockade radius), defined by $V(R_b)$\,$\equiv$\,$\Omega$, only one can be excited to the Rydberg state, leading to the blockade mechanism.

\begin{figure*}[tb]
\centering
\includegraphics[width=1.0\textwidth]{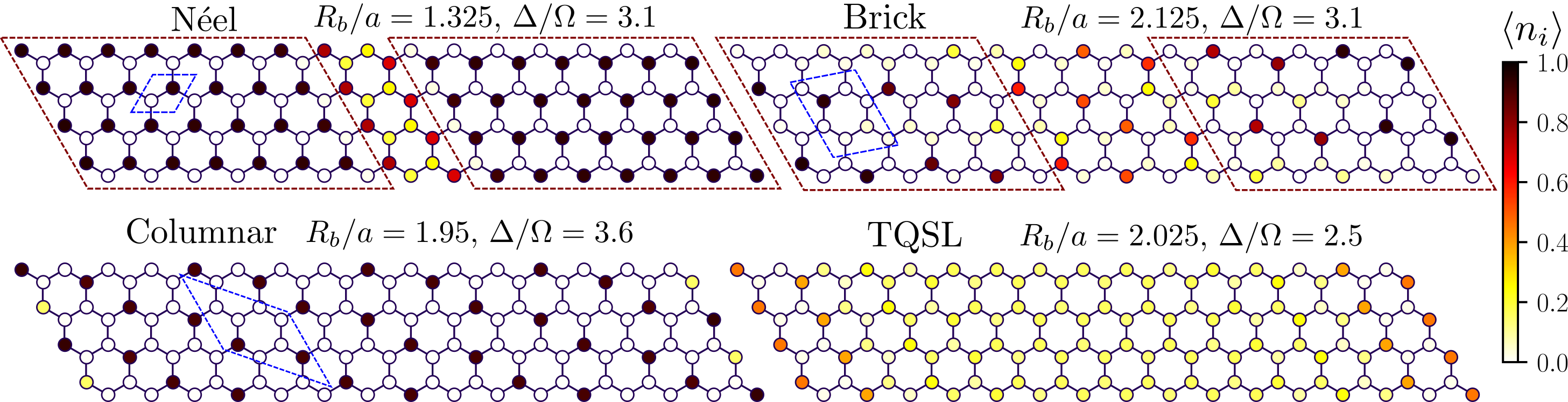}
\caption{\textbf{Phases on the honeycomb lattice.} Rydberg excitation density profiles at representative points of the DMRG phase diagram from Fig.~\ref{fig:Fig1_PD}. While the columnar phase is defectless for this cluster, the N\'eel and brick phases host a domain wall in the middle of the lattice due to their incommensurability  with the open boundary. The two ordered domains on each end of the system are indicated by red boxes, while the primitive unit cells for the ordered phases are delineated in blue. In contrast to the ordered phases, in the TQSL region, a state with no density-wave order and bulk density close to the expected value of $1/6$ is observed, on top of which, density oscillations spread inwards from the boundaries.
}
\label{fig:Fig2_states}
\end{figure*}

Choosing a blockade radius such that the $k$\,$=$\,$3$ nearest neighbors of an atom on the honeycomb lattice are blockaded, the Rydberg blockade constraint becomes identical to a trimer constraint. The trimer in question is a covering of three edges forming a triangle within the triangular lattice built from the vertices placed at the centers of the honeycomb lattice. The constraint enforces that no edge or vertex is shared between trimers \cite{Thewes2020}. This trimer mapping is illustrated in Fig.~\ref{fig:Fig1_PD}(a), where the blockaded neighbors of a central honeycomb atom are shown to match the corresponding forbidden triangular trimers. Thus, the space of maximally filled blockaded configurations---also referred to as the maximum independent set (MIS) subspace \cite{Pichler2018MIS, Ebadi2022}---can be matched to trimer coverings of the dual triangular lattice. The number of covering configurations is found to scale exponentially with the total system size (area) \cite{Verberkmoes1999}, despite the stiffness of individual trimer moves.

While the classical trimer model is interesting in its own right as it was numerically found to avoid ordering for any density \cite{Thewes2020}, we are interested in a trimer model with quantum fluctuations that can be realized in Rydberg quantum simulators. The possibility for novel physics in the quantum model arises by considering a trimer state that is an equal superposition of all of the exponentially many trimer covering configurations (TC):
\begin{equation}\label{eq:TQSL}
    \ket{\mathrm{TQSL}}=\frac{1}{\sqrt{D_{\mathrm{MIS}}}}\sum_{\mathrm{TC}_i}\ket{\mathrm{TC}_i},
\end{equation}
where $D_{\mathrm{MIS}}$ denotes the dimension of the trimer covering subspace (i.e., the number of different trimer coverings), and the sum extends over all the trimer coverings $\mathrm{TC}_i$. Such a state was recently classified by \citet{Giudice2022} as a gapless $\mathrm{U}(1)$\,$\times$\,$\mathrm{U}(1)$ spin liquid. To see the emergence of the two local $\mathrm{U}(1)$ symmetries, one can tripartition the triangular lattice such that the trimers cover one site of each sublattice (A, B, C) and then assign electric fields on A--B and B--C trimer bonds, as shown in Fig.~\ref{fig:Fig1_PD}(b) and argued in Ref.~\onlinecite{Giudice2022}. The total A--B (B--C) flux through a closed loop is then given by the difference between the number of A and B (B and C)  sites enclosed by the loop, showing the presence of two independent $\mathrm{U}(1)$ symmetries (see Supplementary Information \cite{SM} for more details). The gaplessness of the TQSL state then follows from Polyakov's results \cite{Polyakov1977} and was also confirmed numerically by Ref.~\onlinecite{Giudice2022} (see Fig.~8(d) therein). 

While the abovementioned trimer mapping provides a general starting point, the nature of the quantum ground state of the microscopic model has to be carefully investigated to establish the existence of a quantum spin liquid phase, because other possibilities, including trivial disordered or valence bond solid (VBS) states, cannot be excluded \textit{a priori}. Here, we do so by demonstrating that certain ground states of the Hamiltonian~\eqref{eq:RydHam} are adiabatically connected to the perfect TQSL state.

\textit{DMRG phase diagram.---}We first explore the quantum phase diagram of the Rydberg Hamiltonian on the honeycomb lattice using DMRG on long cylinders of finite sizes  \cite{itensor}. We use bond dimensions of up to $1800$ and retain the three strongest Rydberg interactions in Eq.~\eqref{eq:RydHam}, resulting in good convergence. The additional details of the numerical calculations are presented in the Methods section. We use the following quantities to map the boundaries of the different phases: 
\begin{align}\label{eq:quantities}
S_{\mathrm{vN}} & =-\mathrm{Tr}\,(\rho^{}_0 \log \rho^{}_0), \cr
\chi & = -\partial E_0^2/\partial \Delta^2, \\
\mathcal{F} & = 2[1-|\bra{\Psi(\Delta/\Omega)}\ket{\Psi(\Delta/\Omega+\delta)}|]/\delta^2, \notag
\end{align}
where $S_{\mathrm{vN}}$ is the von Neumann entanglement entropy ($\rho_0$ being the reduced density matrix for half of the system), $\chi$ is the energy susceptibility ($E_0$ being the ground state energy), and $\mathcal{F}$ is the fidelity susceptibility ($\ket{\Psi}$ being the ground-state wavefunction). The phase diagram is presented in Fig.~\ref{fig:Fig1_PD}(c).

Among the main features of the phase diagram are the presence of three ordered phases in three different blockade regimes.
Intriguingly, an additional unordered region with a large entanglement entropy is apparent, clearly separated from the trivial disordered phase by a phase transition. The Rydberg density profile in all of the regions is presented in Fig.~\ref{fig:Fig2_states}, while the static structure factors for the different phases are given in the Supplementary Information \cite{SM}. 
The first ordered phase appears in the nearest-neighbor ($k$\,$=$\,$1$) blockade regime, and it corresponds to the N\' eel phase with a staggered order within the honeycomb unit cell. The N\' eel state hosts a domain wall in the middle of the $32 \times 4$ honeycomb cluster employed here due to  two different domains being preferred by the open boundaries on each end.

The next-nearest-neighbor blockade ($k$\,$=$\,$2$) leads to the stabilization of the columnar phase, characterized by the pattern of Rydberg excitations in Fig.~\ref{fig:Fig2_states}. We note, however, that there exists an extensive classical ``string'' degeneracy in this $k$\,$=$\,$2$ blockade regime (see Supplementary Information \cite{SM}). From that classical manifold of states, the columnar pattern is stabilized by both third-neighbor Rydberg interactions and quantum fluctuations. Both of these effects prefer maximal distance between further-neighboring Rydberg excitations (see Ref.~\onlinecite{SM}), a condition satisfied by the columnar state. Incidentally, thermal fluctuations in the classical hard-core boson model on the honeycomb lattice lead to the same ordered phase in this regime \cite{Thewes2020}.

The final ordered phase, which we call the ``brick'' phase \cite{Zhang2021}, appears in the third-neighbor blockade regime ($k$\,$=$\,$3$), with the Rydberg excitations patterned on the so-called brick lattice \cite{Zhang2021} (see Fig.~\ref{fig:Fig2_states}). Similar to the N\' eel state, open boundaries prefer different brick domains, leading to a domain wall in the middle of the cluster considered. The appearance of the brick ground state observed in our simulations is a consequence of quantum fluctuations breaking the exponential degeneracy of valid trimer coverings for the $k$\,$=$\,$3$ blockaded classical model via an order-by-disorder mechanism \cite{Villain1980, Henley1989}. The stabilization of the brick phase, in particular, can be understood by noting that the quantum fluctuations prefer maximally flippable configurations. The brick phase indeed satisfies the maximum flippability condition, as also observed in classical simulations of the hard trimer model relevant to twisted bilayer graphene \cite{Zhang2021}. Additional classical interactions on top of the hard trimer model were shown therein to favor maximal flippability and, in turn, also the  brick state.

\begin{figure}[tb]
\centering
\includegraphics[width=1.0\columnwidth]{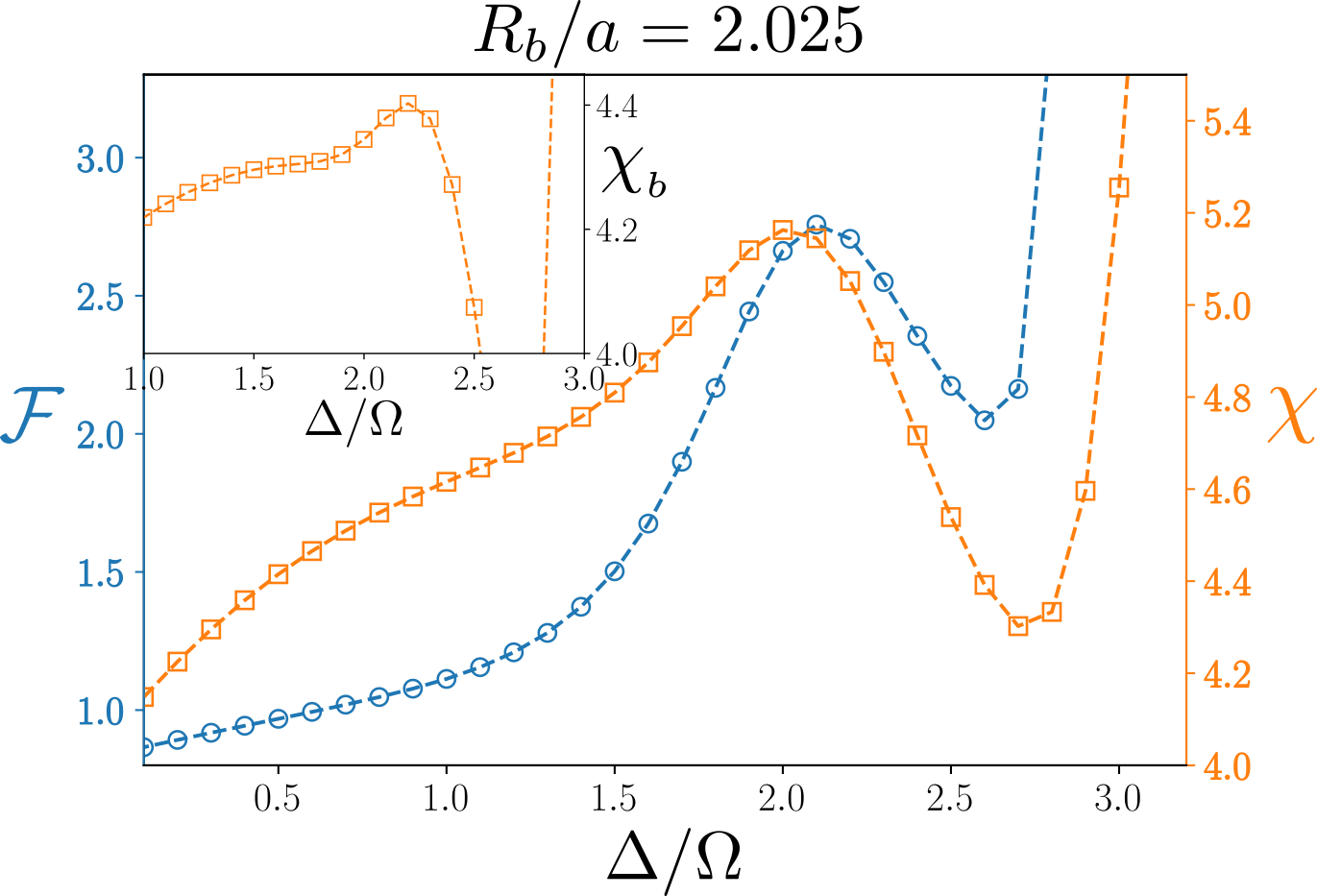}
\caption{\textbf{Transition into the TQSL regime.} On going from the trivial disordered phase to larger $\Delta/\Omega$, clear peaks are visible in the fidelity (evaluated from Eq.~\eqref{eq:quantities} for $\delta=0.1$ on a $32 \times 4$ cluster) and energy susceptibilities that are preserved upon boundary subtraction (between $32 \times 4$ and $24 \times 4$ clusters; inset). The second peak at a larger detuning arises from the transition into the (columnar) ordered phase.}
\label{fig:Fig2b_transition}
\end{figure}

At intermediate detunings, however, an additional region emerges in the $k$\,$=$\,$3$ blockade regime at fillings close to 1/6. This region shows no order and has a high entanglement entropy throughout. On changing the detuning at a fixed blockade radius, the fidelity and energy susceptibilities manifest a clear peak, as presented in Fig.~\ref{fig:Fig2b_transition}, presumably stemming from a nonadiabatic change in the wavefunction compared to the trivial disordered phase. To probe the intrinsic (bulk) nature of the transition, we also calculate the energy susceptibility difference between $32 \times 4$ and $24 \times 4$ clusters ($\chi_{{}_b}$), thus subtracting out the effects of four boundary columns of atoms at each end of the system. The energy susceptibility peak persists after such boundary subtraction with sizable magnitude, pointing to a putative transition in the bulk (inset of Fig.~\ref{fig:Fig2b_transition}). We label this highly entangled region as TQSL in the phase diagram of Fig.~\ref{fig:Fig1_PD} due to its separation from the trivial disordered phase and its appearance in the $k$\,$=$\,$3$ regime with the density of $\approx 1/6$ expected for a TQSL state. In addition, the scaling of the gap with the system size and the enhanced susceptibility towards boundary-induced density oscillations (see Supplementary Information \cite{SM}) are seemingly suggestive of a gapless state in the thermodynamic limit, broadly consistent with the expectations for a TQSL. We note, however, that the state is generically observed as gapped for finite clusters (since the discrete, allowed momenta need not coincide with the gapless point in momentum space) in the absence of twisted boundary conditions or flux insertion \cite{He2017, Ferrari2021, Jin2022}. In the remainder of this paper, we focus on this intriguing TQSL regime identified by our DMRG simulations and demonstrate the existence of a true spin liquid ground state.

\textit{TQSL in the PXP model.---}In order to analyze the existence, characterization, and experimental feasibility of preparation of the TQSL state, we perform large-scale exact diagonalization calculations \cite{Bloqade}. For these simulations, we employ a hard-constraint approximation to the full Rydberg Hamiltonian, known as the PXP model \cite{Sachdev2002}, to enable us to reach large system sizes.
The essence of this approximation is to eliminate  states violating the Rydberg blockade from the Hilbert space.
This is achieved by making the first $k$ Rydberg interactions infinite and discarding the longer-range interactions while projecting the Rabi-oscillation term into the subspace of allowed configurations, leading to the Hamiltonian:
\begin{equation}\label{eq:PXPham}
    \dfrac{H_\mathrm{PXP}}{\hbar}=\frac{\Omega}{2}\sum_i P\sigma^x_i P - \Delta \sum_i n^{}_i,
\end{equation}
where $P$ is the projector onto the blockade subspace and $\sigma^x_i \equiv \left( \vert g_i\rangle \langle r_i\vert + \mathrm{h.c.} \right)$. This allows us to explore the  $k$\,$=$\,$3$ blockade regime with ED \cite{Bloqade, Jinguo2022} on clusters of up to $60$ sites with periodic boundary conditions in both directions, thus also better simulating the \textit{bulk} of a large system. The PXP approximation has been effectively employed to understand a variety of phenomena in Rydberg systems, including quantum scars \cite{Bernien2017, Turner2018, Bluvstein2021}, emergent lattice gauge theories \cite{Suarce2020}, and gapped spin liquids on the ruby lattice \cite{Verresen2021, Semeghini2021}.

\begin{figure*}[t!]
\centering
\includegraphics[width=1.0\textwidth]{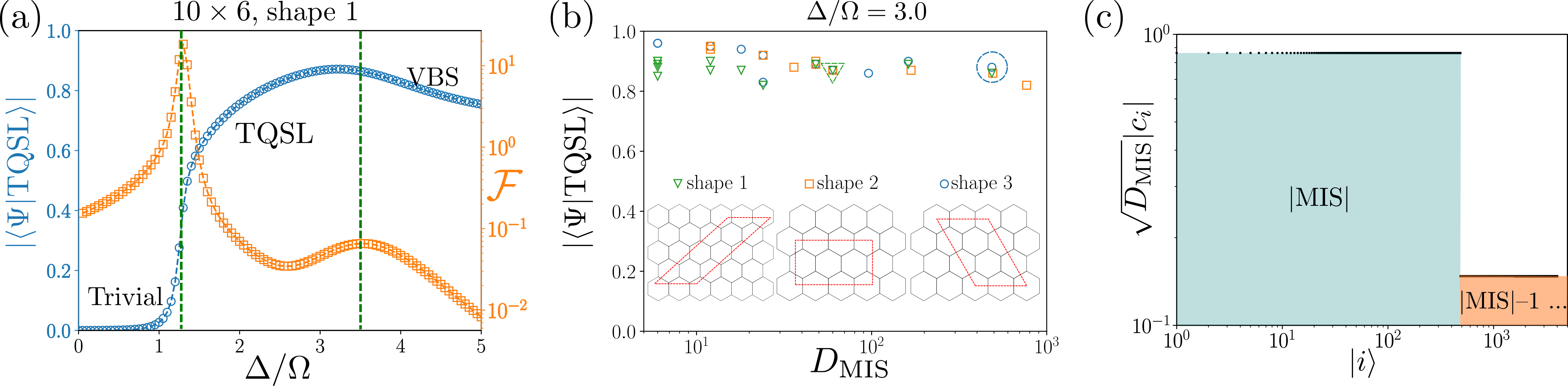}
\caption{\textbf{TQSL in the PXP model.} A robust TQSL is detected in the PXP model \cite{Bloqade, Jinguo2022} with up to third-nearest neighbors blockaded. (a) Overlap of the ground state with the perfect TQSL state reveals three distinct regions in the $10$\,$\times$\,$6$ shape 1 cluster with periodic boundary conditions \cite{SM}, namely, a trivial phase at small $\Delta/\Omega$, a trimer RVB region with high TQSL fidelity in the middle, and a VBS state with decreasing fidelity at high $\Delta/\Omega$. This is further confirmed by considering the fidelity susceptibility [evaluated from Eq.~\eqref{eq:quantities} with $\delta=0.05$], wherein two clear peaks stemming from the trivial--TQSL and TQSL--VBS transitions appear. (b) The TQSL fidelity in the spin-liquid region is preserved upon increasing the effective system size, i.e., the number of trimer coverings (equivalent to the MIS degeneracy, $D_{\mathrm{MIS}}$), irrespective of cluster size or aspect ratio. The inset shows the different cluster shapes employed in simulations. The cluster from (a) is emphasized with a green triangle. (c) The spin-liquid nature of the state with high TQSL overlap is seen by considering the ground-state superposition structure for the typical $6 \times 10$ shape 3 cluster at $\Delta/\Omega=3.0$, circled in (b). All MIS configurations have dominant almost-equal weights and equal phases (see Supplementary Information \cite{SM}), while non-MIS configurations have an order-of-magnitude smaller weights.}
\label{fig:Fig3_fidelity}
\end{figure*}

We map out the ground-state phase diagram of the PXP model as a function of the tuning parameter, $\Delta/\Omega$. The phase boundaries are determined by considering the overlap of the ground state with the perfect TQSL state in  Eq.~\eqref{eq:TQSL} (equal superposition of all MIS states for the given cluster), $|\bra{\Psi}\ket{\mathrm{TQSL}}|$, the fidelity and energy susceptibilities, as well as changes in the low-lying energy spectrum of the model (see Supplementary Information \cite{SM}). To explore the robustness of our predictions, we consider three different honeycomb cluster shapes with several system sizes and aspect ratios for each \cite{SM}. For all the shapes and sizes probed, we find a sizable TQSL region. The TQSL phase is identified by a high fidelity with respect to the perfect TQSL state as well as by explicitly checking that the ground state is predominantly near-equal weight and phase superposition of all trimer configurations  \cite{SM}. An example for a particular cluster is shown in Fig.~\ref{fig:Fig3_fidelity}(a), where the TQSL overlap and the fidelity susceptibility are plotted as a function of the detuning. Three regions, separated by fidelity susceptibility peaks, are observed: a trivial disordered phase for small detunings, a trimer RVB phase with a high TQSL overlap in the intermediate regime, and a phase with decreasing fidelity at high detunings pointing to the formation of a VBS state. In the majority of the other clusters explored, the VBS state is completely absent up to $\Delta/\Omega=5$. This behavior, though driven by quantum fluctuations, is reminiscent of the effect of thermal fluctuations in the classical PXP-equivalent model at finite temperatures, which also lacks order for $k$\,$=$\,$3$ at any density \cite{Thewes2020}. Furthermore, increasing the effective system size, as measured by the number of classically degenerate trimer coverings for a given cluster, leads to no drop in the TQSL fidelity in the spin liquid region, as shown in Fig.~\ref{fig:Fig3_fidelity}(b). Lastly, examination of the structure of the ground states in the RVB phase reveals that they are predominantly equal-weight and equal-phase superposition of all trimer coverings (see Fig.~\ref{fig:Fig3_fidelity}(c) and Supplementary Information \cite{SM}), cementing the TQSL nature of the state.

\begin{figure*}[t!]
\centering
\includegraphics[width=1.0\textwidth]{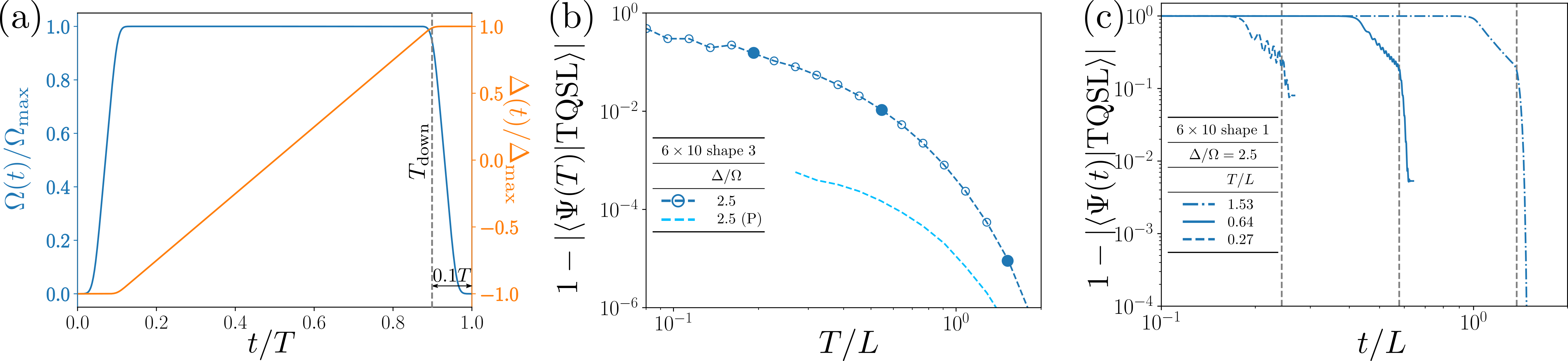}
\caption{\textbf{Adiabatic preparation of the TQSL.} (a) We test the feasibility of preparing the TQSL state with an experimentally relevant adiabatic preparation protocol. The ramp-down time ($T_{\mathrm{down}}=0.9T$) is denoted by a dashed gray line. (b) This results in prepared state fidelities that are several orders-of-magnitude better than the ground-state ones, as seen by plotting the prepared TQSL overlap as a function of the total preparation time (in units of $2\pi/\Omega$), rescaled by the system size ($L=N_xN_y$), for parameters representative of the TQSL phase. The blue dashed line shows the projection of $\ket{\Psi{(0.9T)}}$ onto the MIS subspace. (c) TQSL overlap during the quasi-adiabatic sweep presented for the filled data points in (b), showing that the gain compared to the ground state stems mostly from the off-ramp part of the protocol, to the left of the gray dashed lines denoting $T_{\mathrm{down}}$. The off ramp effectively acts as a projector to the MIS subspace \cite{SM} in a mechanism that is expected to be a general feature of Rydberg systems.}
\label{fig:Fig3b_adiabatic}
\end{figure*}

\textit{Dynamical preparation of TQSL states.---}We now explore the feasibility of preparing the TQSL state with an experimentally relevant quasi-adiabatic protocol, illustrated in Fig.~\ref{fig:Fig3b_adiabatic}(a). The protocol, of total time $T$, consists of starting from an initial state where all atoms are in $\ket{g}$ and increasing $\Omega$ to a desired value at a fixed large negative detuning in the first segment of duration $0.1T$. This is followed by increasing $\Delta$ to its desired final value, and then finally, an $\Omega$ off ramp  of length $0.1T$ at a fixed detuning. The pulses are then smoothed with a Gaussian kernel to eliminate short timescale effects.

The results for the TQSL overlap of the state at the end of the ramp as a function of the total time $T$ are showcased in Fig.~\ref{fig:Fig3b_adiabatic}(b). The obtained fidelities are, in most cases \cite{SM}, several orders of magnitude above the ground-state fidelity. 
It also appears that the fidelities can approach  arbitrarily close to one with increasing $T$.
The prepared fidelity depends only weakly on detuning within the TQSL phase, while it drops in the trivial and VBS phases \cite{SM}. This remarkable fidelity enhancement points to an important role that the quasi-adiabatic preparation protocol might play in the preparation of spin liquid states in general and is consistent with the recent results reported in simulations of the $\mathbb{Z}_2$ spin liquid on the ruby lattice \cite{Semeghini2021, Giudici2022}.

Here, we are able to assign the origin of the fidelity enhancement to the off-ramp part of the pulse. As shown in Fig.~\ref{fig:Fig3b_adiabatic}(c), the fidelity reached during the constant-$\Omega$ part of the pulse is of the order of the ground-state fidelity. In the off-ramp part (last $0.1T$ of time), a significant enhancement is seen. This is in agreement with the experimental observations on the ruby lattice~\footnote{Private communication, Harvard atom array team.} and is shown to be valid for both honeycomb and ruby lattices \cite{SM}. We explain this off-ramp fidelity enhancement by a projection mechanism. First, we note that the time-dependent Hamiltonian during the off ramp can be thought of as a sequence of Hamiltonians with ever-increasing values of the ratio $\Delta/\Omega$. This leads to greater penalties for state admixtures with less than the maximum allowed number of Rydberg excitations (non-MIS configurations), thus leading to an effective projection to the MIS subspace. We test this hypothesis by comparing the TQSL overlap of the final prepared state to that of the state obtained by projecting $\ket{\Psi{(0.9T)}}$ on to the MIS subspace. We find that while the fidelity of the projection is always higher than that of the prepared one, the two approach each other for long total preparation times (Fig.~\ref{fig:Fig3b_adiabatic}(b) and \cite{SM}). While such a mechanism appears to be connected to the PXP model's details, we show in the next section that it applies more generally to Rydberg systems in a slightly modified form.

\begin{figure*}[thb]
\centering
\includegraphics[width=1.0\textwidth]{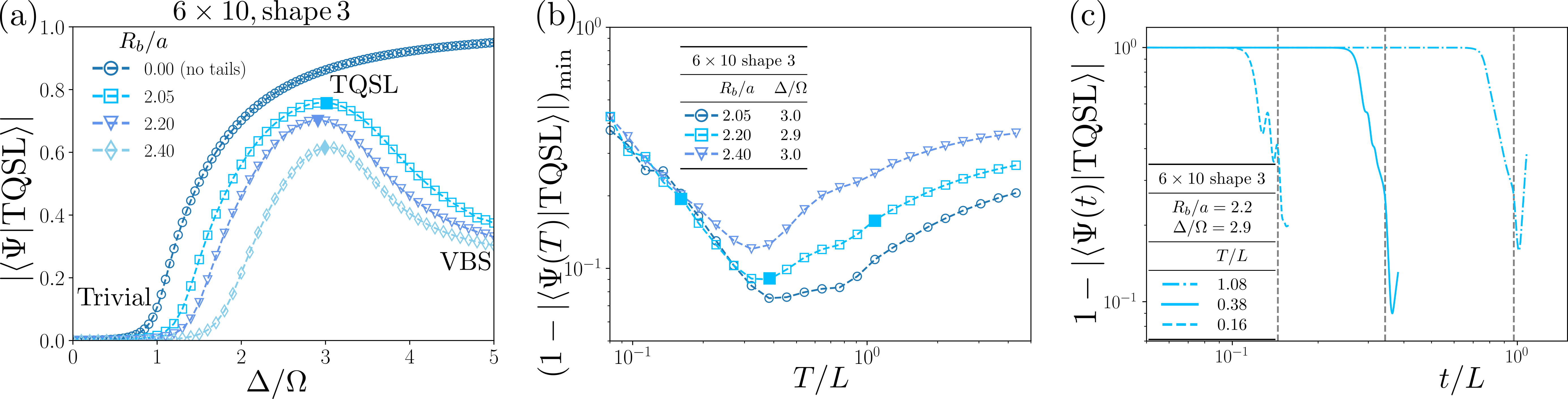}
\caption{\textbf{Effect of interaction tails.} To probe the robustness of the TQSL, we add the interaction tails up to $\mathcal{R}=3a$ to the PXP Hamiltonian. (a) The ground-state overlap with the perfect TQSL state as a function of $\Delta/\Omega$ for different interaction strengths ($R_b/a$) shows that the spin liquid survives the long-ranged interactions on the typical cluster with the large MIS degeneracy, though with a smaller parameter region for the TSQL phase and lower fidelities. (b) The adiabatic preparation protocol at parameters corresponding to the peak ground-state fidelity in the TQSL region, denoted by filled symbols in (a), still shows maximum preparation fidelities significantly above the ground state ones. In contrast to the system without tails, an optimal total preparation time exists, showing the significance of semi-adiabatic effects. (c) The TQSL overlap during the adiabatic sweep continues to show a sizable gain during the off ramp, with an additional nonmonotonic feature. The behavior during the off ramp in the presence of interaction tails can still be explained by the universal off-ramp projection mechanism \cite{SM}.
}
\label{fig:Fig4_tails}
\end{figure*}

\textit{Robustness of the TQSL state.---}Given the theoretically expected RVB nature of the TQSL state, it is important to probe its robustness to perturbations stemming from the ``tails'' of the van der Waals interactions, which break the classical degeneracy of the trimer configurations. These (experimentally relevant) interaction tails discarded in the pure PXP model of Eq.~\eqref{eq:PXPham} are always present in real atomic systems, so their inclusion is necessary in a realistic model. Note that the PXP approximation of the hard blockade was already relaxed completely in the DMRG studies above. Proceeding further, we now add the interaction tails up to $\mathcal{R}$\,$=$\,$3a$ to the PXP Hamiltonian of Eq.~\eqref{eq:PXPham}. The strength of the interaction tails is controlled by the dimensionless parameter $R_b/a$ that we explore in the realistic range between $2$ and $\sqrt{7}$ for the $k$\,$=$\,$3$ regime.

The resulting ground-state fidelities for several values  of $R_b/a$ are shown for a typical case in Figs.~\ref{fig:Fig4_tails}(a), with additional clusters presented in the Supplementary Information \cite{SM}. Without tails, this cluster does not show a VBS phase up to $\Delta/\Omega$\,$=$\,$5.0$, thus presenting an extended TQSL region with the wavefunction character shown in Fig.~\ref{fig:Fig3_fidelity}(c). Taking the tails into account, a sizable spin-liquid region, manifested as a high TQSL fidelity plateau, survives to large $R_b/a$ at intermediate detunings. The size of the region and the maximum TQSL fidelity reduces upon increasing the strength of the interaction's tails. In addition, this region is now followed by the VBS plateau with fidelities independent of the interaction tails' strengths.

Next, we explore the quasi-adiabatic preparation protocol of Fig.~\ref{fig:Fig3b_adiabatic}(a) for the TQSL state in the presence of long-ranged interaction tails. The optimal TQSL overlap during the preparation protocol is shown in Fig.~\ref{fig:Fig4_tails}(b) for parameters corresponding to peaks of the ground-state fidelity in the TQSL phase [solid symbols in Fig.~\ref{fig:Fig4_tails}(a)]. We observe that the preparation fidelity---despite being lower than that for the pure PXP case and falling off with increasing $R_b/a$---still significantly outperforms the ground-state fidelity. Unlike the pure honeycomb PXP case, there now exists some optimal value of the total preparation time that is linear in the system size \cite{SM}, similar to the case of the pure PXP model on the ruby lattice \cite{Giudici2022}.

In order to gain further insight into the fidelity enhancement observed with dynamical state preparation, we also consider the fidelity as a function of the preparation time, as showcased in Fig.~\ref{fig:Fig4_tails}(c). We find that in the first $0.9T$ segment of the protocol (denoted by gray dashed lines), the TQSL overlap from this dynamical preparation is similar to the one obtained from the ground state [see Fig.~\ref{fig:Fig4_tails}(a)]; however, the $\Omega$ off-ramp (the last $0.1T$) leads to a significant fidelity enhancement over the ground state. Compared to the case without interaction tails, an additional upturn in $1- |\bra{\Psi(t)}\ket{\mathrm{TQSL}}|$ is consistently observed at the end of the ramp, leading us to consider the optimal protocol fidelity at intermediate times instead of the final prepared fidelity in Fig.~\ref{fig:Fig4_tails}(b). This upturn can also be explained within the off-ramp projection mechanism. The semi-adiabatic protocol suppresses the destabilizing effect of the interaction tails for intermediate total preparation times, leading to a state before the off ramp (at $0.9T$) that has equal MIS weights and phases, but still large admixtures of non-MIS configurations. Then, at the start of the off ramp, an effective projection to the $k$\,$=$\,$3$ MIS subspace takes place. However, once $\Omega$ drops to a value such that the next shell becomes effectively blockaded, i.e., $V_{k+1}(R_b/a)=\Omega(t^*)$, the projection to the MIS subspace for $k$\,$=$\,$4$ is the effective description of the off-ramp Hamiltonian evolution. Therefore, if one wishes to optimize for $k$\,$=$\,$3$ ground states, the off ramp should be sharply cut off before $t^*$. This picture is independent of the PXP-type approximations and generalizes well to the preparation of entangled quantum states or to the optimization algorithms \cite{Ebadi2022, Nguyen2022} arising from blockade physics in Rydberg atom simulators.

\textit{Discussion.---}Thus far, we have shown how a highly entangled TQSL phase can emerge in a honeycomb lattice of Rydberg atoms and presented evidence for its existence and robustness on finite-sized clusters.
Exploring the experimentally accessible preparation protocols, we also uncovered an off-ramp fidelity enhancement mechanism potentially relevant to a broad range of quantum state preparation tasks in Rydberg platforms. We now turn to the question of the experimental characterization of this novel TQSL state.

The spin liquid state that we report here is directly accessible in current-generation Rydberg atom simulators that can realize the relevant lattice geometry, achieve the necessary parameter regimes, and employ the quasi-adiabatic preparation protocol. From our simulations, the parameter range to search for the TQSL phase in experiments corresponds to $R_b/a\approx 2.0$--$2.4$ and $\Delta_{\max}/\Omega_{\max}\approx 1.0$--$4.0$. We consider $^{87}$Rb atoms and laser coupling a hyperfine ground state to a $70S_{1/2}$ Rydberg state; a realistic choice of $\Omega_{\max} = 4.0 \times 2\pi \, \mathrm{MHz}$ and $C_6=8.6 \times 10^5 \times 2\pi \, \mathrm{MHz} \, \mu\mathrm{m}^6$ leads to a honeycomb lattice constant of $a \approx 3.2$--$3.9 \, \mu\mathrm{m}$, therefore easily accommodating $L$\,$>$\,$200$ atoms in a $100\times 100 \, \mu\mathrm{m}$ array. The typical preparation times of $\sim 3$--$5 \, \mu\mathrm{s}$ correspond to $T/L \sim 0.1$, which, though currently less than optimal, should still lead to sizable TQSL fidelities and are comparable to the ones employed in the ruby-lattice experiments preparing a $\mathbb{Z}_2$ spin liquid phase \cite{Semeghini2021}. Further improvements to experimental coherence times by increasing the laser power and moving to larger intermediate-state detunings can extend $T/L$ to the optimal preparation time in experiments.  

The main experimental signature to characterize such a state would be obtained by testing its trimer character and the associated $\mathrm{U}(1)$\,$\times$\,$\mathrm{U}(1)$ symmetry upon sampling the wavefunction in the experimentally accessible $Z$-basis. For each snapshot of the TQSL state, a corresponding trimer structure with fluxes can be assigned as presented in Fig.~\ref{fig:Fig1_PD}. One can then test for the $\mathrm{U}(1)$\,$\times$\,$\mathrm{U}(1)$ conservation law by evaluating the flux enclosed in a closed loop and comparing it to the theoretical expectations based on the occupation of different sublattices within the loop (see Supplementary Information \cite{SM}). Additionally, the resonance between different trimer configurations can be explored using $X$-basis measurements. These measurements would require a global basis rotation applied in the experimental protocol for $X$-loop measurements, akin to the measurements performed for the ruby lattice \cite{Verresen2021, Semeghini2021}. The relevant $X$-loop operators to be probed are those that flip between different trimer configurations; these can be found by considering the difference between two irregular breathing  honeycomb lattices \cite{Villain1980b, Coppersmith1982, Verberkmoes1999, Zhang2021} describing the underlying trimer configurations, as sketched in Fig.~\ref{fig:FigS1a_trimers} of the Supplementary Information \cite{SM}. Such $X$-loop operators are expected to decay exponentially with the perimeter of the loop only in the TQSL phase, akin to the ruby-lattice $X$-loop operators. 

Other useful probes of the TQSL phase may be more indirect. For instance, the TQSL should be featureless in the bulk, thus setting it apart from proximate ordered phases in measurements of the static structure factor. However, it should also be distinguishable from the trivial disordered state via energy susceptibility measurements that exhibit a peak at the trivial--TQSL transition \cite{SM}; note that the energy susceptibility can be experimentally extracted from the total density in the $Z$-basis, as $\chi=\partial \langle n \rangle / \partial \Delta$. 

Our study opens up several new research directions. Theoretically, it would be important to understand whether a $\mathrm{U}(1)$\,$\times$\,$\mathrm{U}(1)$ spin liquid state can be stabilized in (2+1) dimensions by coupling the gauge field to gapless fermionic matter and to analytically demonstrate the irrelevance of perturbations (in the renormalization-group sense) about the spin-liquid fixed point \cite{Hermele2004}.
Numerically, a question for future work would be to establish the TQSL as a stable phase of matter in the thermodynamic limit, perhaps using methods such as infinite DMRG \cite{iDMRG} or quantum Monte Carlo, which can provides new insights beyond exact diagonalization calculations that inevitably suffers from finite-size effects. 
Experimentally, the preparation and characterization of the TQSL state would pave a new path towards exploring its physics, including the robustness of the state, its gauge field dynamics, and the fractionalization of excitations. For example, in order to probe the excitations of the TQSL, Rydberg spectroscopy consisting of time-dependent correlator measurements can be employed using locally addressable Rydberg arrays \cite{Knap2013, Baez2020, Bluvstein2022}. Furthermore, it would be interesting to investigate the nature of the quantum critical points leading out of the TQSL phase as well as the associated nonequilibrium quantum many-body dynamics and potential dynamical phase transitions. Finally, the generic understanding of fidelity enhancements developed in this work can be useful for not only adapting the preparation protocol to obtaining strongly correlated states, but also solving hard optimization problems on Rydberg atom simulators.

\section*{Methods}
\label{sec:Methods}

\textit{Exact diagonalization.}---Exact diagonalization simulations of the PXP Hamiltonian were performed using the Bloqade \cite{Bloqade} and Generic Tensor Networks \cite{Jinguo2022} packages with periodic boundary conditions on a torus. The Hamiltonians in the blockaded subspace were generated by Bloqade's routines and diagonalized using the Lanczos scheme. The perfect TQSL state was generated by finding all MIS configurations and calculating the MIS degeneracy using Generic Tensor Networks. Three distinct shapes were explored with a variety of aspect ratios and system sizes of up to 60 sites. These cluster shapes are shown in 
Fig.~\ref{fig:Fig3_fidelity}(b), which presents an example of a $N_x =4, N_y=4$ shape 1 cluster, a $N_x =6, N_y=2$ shape 2 cluster, and a $N_x =4, N_y=3$ shape 3 cluster. 

\begin{figure}[htb]
\vspace*{-0.0cm}
\centering
\includegraphics[width=0.8\columnwidth]{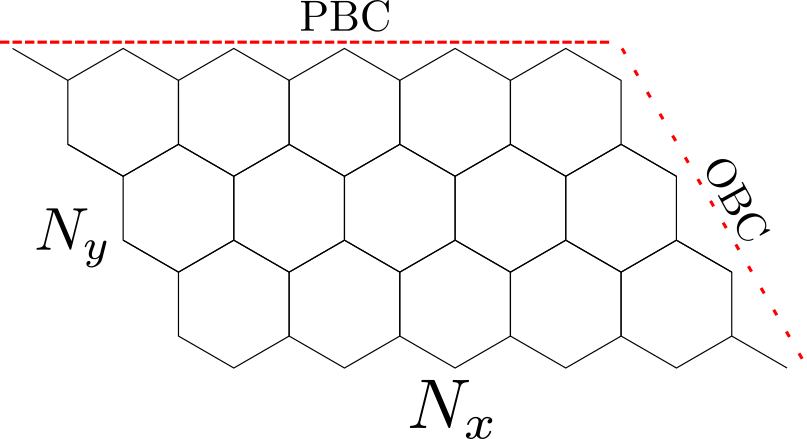}
\vskip -0.cm
\caption{\textbf{Cluster shape used in DMRG simulations.} DMRG was performed on cylindrical boundary conditions with small transverse cylinder sizes.
}
\label{fig:FigM1_shapes}
\vskip -0.cm
\end{figure}

\textit{State preparation.---}The quasi-adiabatic state preparation simulations of the PXP model without tails were executed using Bloqade's \cite{Bloqade} ODE-solver-based routines with the same clusters and boundary conditions as reported for exact diagonalization and with the preparation protocol from Fig.~\ref{fig:Fig3b_adiabatic}(a) with a Gaussian kernel radius set at $T/100$. The time step used for simulating the dynamics was $5\times 10^{-4}$ (in units of $2\pi/\Omega$), leading to excellent convergence as manifested by a TQSL overlap changing by at most $10^{-7}$ upon further decreasing the time step to $10^{-4}$. The preparation with tails was executed with Bloqade-generated Hamiltonians and pulses with Krylov-subspace-based evolution routines. The time step used for Trotterization was $10^{-3}$, which achieved a similar level of convergence as the ODE-solver-based methods. The two methods tested against each other for the Hamiltonian without tails have shown excellent agreement.

\textit{DMRG.---}The DMRG calculations were performed using the ITensor package \cite{itensor}. The geometry studied here consisted of a long cylinder (with open boundary conditions along the shorter edge and periodic along the longer) shown in Fig.~\ref{fig:FigM1_shapes} for a $N_x=12, N_y=4$ system, with the phase diagram of Fig.~\ref{fig:Fig1_PD} constructed for a $32 \times 4$ system. Phase diagrams for $24 \times 4$, $26 \times 4$, and $32 \times 3$ cylinders, as well as for a $32\times 4$ system with the addition of interaction tails up to fifth-nearest neighbors ($\mathcal{R}$\,$=$\,$3a$) were also fully reconstructed, with the phase boundaries qualitatively matching the ones for the $32 \times 4$ system, including the TQSL region. Other aspect ratios were employed for probing the properties of the TQSL regime, with $N_x=16$--$32$ and $N_y=3$--$6$. The sweep protocol employed was based on typically $\mathcal{O}(200)$ sweeps with bond dimensions in the range $400$--$1800$, depending on the transverse system size ($N_y$). Initially, many sweeps were performed at a relatively small bond dimension ($<100$), with the bond dimension being progressively increased in later stages. A gradually decreasing noise term was added until the final stages of the sweep to prevent the DMRG from being stuck in local minima. The protocol achieved good convergence with typical discarded weights below $10^{-10}$ and the relative change  in the ground state energy after the final increase in the bond dimension being below $10^{-6}$.

\textit{Acknowledgements.---}We acknowledge fruitful conversations with Sergio Cantu, Jinguo Liu, Pedro Lopes, and Xiu-Zhe (Roger) Luo. We thank Alex Keesling, Mikhail Lukin, Hannes Pichler, and Giulia Semeghini for carefully reading the manuscript and providing useful feedback. R.S. is supported by the Princeton Quantum Initiative Postdoctoral Fellowship.

\textit{Author contributions.---}F.L. proposed this work. M.K., R.S., and F.L. carried out the numerical and theoretical analysis. All authors contributed extensively to the interpretation of the data, discussions, and the preparation of this manuscript. All work was supervised by R.S., S.-T.W, and F.L.

\textit{Competing interests.---}The authors declare no competing interests.

\textit{Materials and Correspondence.---}All correspondence and material requests should be addressed to R.~S., S.-T.W., and F.~L.

\textit{Data availability.---}All data supporting the results of this study are available within the paper and its Supplementary Information or from the corresponding authors upon reasonable request.

\textit{Note added.---}After the completion of this work, we became aware of Ref.~\onlinecite{Sheng2022}, which considers the ordered phases of the honeycomb Rydberg array. In contrast to that work, we explore the mapping to the trimer model and, crucially, the existence of the trimer quantum spin liquid phase. While the ordered phases we find are in agreement with Ref.~\onlinecite{Sheng2022}, we, in addition, provide a concrete understanding of their stabilization.

\bibliography{HC.bib}

\begin{thebibliography}{68}%
\makeatletter
\providecommand \@ifxundefined [1]{%
 \@ifx{#1\undefined}
}%
\providecommand \@ifnum [1]{%
 \ifnum #1\expandafter \@firstoftwo
 \else \expandafter \@secondoftwo
 \fi
}%
\providecommand \@ifx [1]{%
 \ifx #1\expandafter \@firstoftwo
 \else \expandafter \@secondoftwo
 \fi
}%
\providecommand \natexlab [1]{#1}%
\providecommand \enquote  [1]{``#1''}%
\providecommand \bibnamefont  [1]{#1}%
\providecommand \bibfnamefont [1]{#1}%
\providecommand \citenamefont [1]{#1}%
\providecommand \href@noop [0]{\@secondoftwo}%
\providecommand \href [0]{\begingroup \@sanitize@url \@href}%
\providecommand \@href[1]{\@@startlink{#1}\@@href}%
\providecommand \@@href[1]{\endgroup#1\@@endlink}%
\providecommand \@sanitize@url [0]{\catcode `\\12\catcode `\$12\catcode
  `\&12\catcode `\#12\catcode `\^12\catcode `\_12\catcode `\%12\relax}%
\providecommand \@@startlink[1]{}%
\providecommand \@@endlink[0]{}%
\providecommand \url  [0]{\begingroup\@sanitize@url \@url }%
\providecommand \@url [1]{\endgroup\@href {#1}{\urlprefix }}%
\providecommand \urlprefix  [0]{URL }%
\providecommand \Eprint [0]{\href }%
\providecommand \doibase [0]{http://dx.doi.org/}%
\providecommand \selectlanguage [0]{\@gobble}%
\providecommand \bibinfo  [0]{\@secondoftwo}%
\providecommand \bibfield  [0]{\@secondoftwo}%
\providecommand \translation [1]{[#1]}%
\providecommand \BibitemOpen [0]{}%
\providecommand \bibitemStop [0]{}%
\providecommand \bibitemNoStop [0]{.\EOS\space}%
\providecommand \EOS [0]{\spacefactor3000\relax}%
\providecommand \BibitemShut  [1]{\csname bibitem#1\endcsname}%
\let\auto@bib@innerbib\@empty
\bibitem [{\citenamefont {Savary}\ and\ \citenamefont
  {Balents}(2016)}]{Savary2016}%
  \BibitemOpen
  \bibfield  {author} {\bibinfo {author} {\bibfnamefont {Lucile}\ \bibnamefont
  {Savary}}\ and\ \bibinfo {author} {\bibfnamefont {Leon}\ \bibnamefont
  {Balents}},\ }\bibfield  {title} {\enquote {\bibinfo {title} {Quantum spin
  liquids: a review},}\ }\href {\doibase 10.1088/0034-4885/80/1/016502}
  {\bibfield  {journal} {\bibinfo  {journal} {Rep. Prog. Phys.}\ }\textbf
  {\bibinfo {volume} {80}},\ \bibinfo {pages} {016502} (\bibinfo {year}
  {2016})}\BibitemShut {NoStop}%
\bibitem [{\citenamefont {Zhou}\ \emph {et~al.}(2017)\citenamefont {Zhou},
  \citenamefont {Kanoda},\ and\ \citenamefont {Ng}}]{Zhou2017}%
  \BibitemOpen
  \bibfield  {author} {\bibinfo {author} {\bibfnamefont {Yi}~\bibnamefont
  {Zhou}}, \bibinfo {author} {\bibfnamefont {Kazushi}\ \bibnamefont {Kanoda}},
  \ and\ \bibinfo {author} {\bibfnamefont {Tai-Kai}\ \bibnamefont {Ng}},\
  }\bibfield  {title} {\enquote {\bibinfo {title} {Quantum spin liquid
  states},}\ }\href {\doibase 10.1103/RevModPhys.89.025003} {\bibfield
  {journal} {\bibinfo  {journal} {Rev. Mod. Phys.}\ }\textbf {\bibinfo {volume}
  {89}},\ \bibinfo {pages} {025003} (\bibinfo {year} {2017})}\BibitemShut
  {NoStop}%
\bibitem [{\citenamefont {Anderson}(1973)}]{Anderson1973}%
  \BibitemOpen
  \bibfield  {author} {\bibinfo {author} {\bibfnamefont {P.W.}\ \bibnamefont
  {Anderson}},\ }\bibfield  {title} {\enquote {\bibinfo {title} {Resonating
  valence bonds: A new kind of insulator?}}\ }\href {\doibase
  https://doi.org/10.1016/0025-5408(73)90167-0} {\bibfield  {journal} {\bibinfo
   {journal} {Materials Research Bulletin}\ }\textbf {\bibinfo {volume} {8}},\
  \bibinfo {pages} {153--160} (\bibinfo {year} {1973})}\BibitemShut {NoStop}%
\bibitem [{\citenamefont {Lienhard}\ \emph {et~al.}(2018)\citenamefont
  {Lienhard}, \citenamefont {de~L\'es\'eleuc}, \citenamefont {Barredo},
  \citenamefont {Lahaye}, \citenamefont {Browaeys}, \citenamefont {Schuler},
  \citenamefont {Henry},\ and\ \citenamefont {L\"auchli}}]{Lienhard2018}%
  \BibitemOpen
  \bibfield  {author} {\bibinfo {author} {\bibfnamefont {Vincent}\ \bibnamefont
  {Lienhard}}, \bibinfo {author} {\bibfnamefont {Sylvain}\ \bibnamefont
  {de~L\'es\'eleuc}}, \bibinfo {author} {\bibfnamefont {Daniel}\ \bibnamefont
  {Barredo}}, \bibinfo {author} {\bibfnamefont {Thierry}\ \bibnamefont
  {Lahaye}}, \bibinfo {author} {\bibfnamefont {Antoine}\ \bibnamefont
  {Browaeys}}, \bibinfo {author} {\bibfnamefont {Michael}\ \bibnamefont
  {Schuler}}, \bibinfo {author} {\bibfnamefont {Louis-Paul}\ \bibnamefont
  {Henry}}, \ and\ \bibinfo {author} {\bibfnamefont {Andreas~M.}\ \bibnamefont
  {L\"auchli}},\ }\bibfield  {title} {\enquote {\bibinfo {title} {{Observing
  the space- and time-dependent growth of correlations in dynamically tuned
  synthetic Ising models with antiferromagnetic interactions}},}\ }\href
  {\doibase 10.1103/PhysRevX.8.021070} {\bibfield  {journal} {\bibinfo
  {journal} {Phys. Rev. X}\ }\textbf {\bibinfo {volume} {8}},\ \bibinfo {pages}
  {021070} (\bibinfo {year} {2018})}\BibitemShut {NoStop}%
\bibitem [{\citenamefont {Keesling}\ \emph {et~al.}(2019)\citenamefont
  {Keesling}, \citenamefont {Omran}, \citenamefont {Levine}, \citenamefont
  {Bernien}, \citenamefont {Pichler}, \citenamefont {Choi}, \citenamefont
  {Samajdar}, \citenamefont {Schwartz}, \citenamefont {Silvi}, \citenamefont
  {Sachdev}, \citenamefont {Zoller}, \citenamefont {Endres}, \citenamefont
  {Greiner}, \citenamefont {Vuleti{\'{c}}},\ and\ \citenamefont
  {Lukin}}]{Keesling2019}%
  \BibitemOpen
  \bibfield  {author} {\bibinfo {author} {\bibfnamefont {Alexander}\
  \bibnamefont {Keesling}}, \bibinfo {author} {\bibfnamefont {Ahmed}\
  \bibnamefont {Omran}}, \bibinfo {author} {\bibfnamefont {Harry}\ \bibnamefont
  {Levine}}, \bibinfo {author} {\bibfnamefont {Hannes}\ \bibnamefont
  {Bernien}}, \bibinfo {author} {\bibfnamefont {Hannes}\ \bibnamefont
  {Pichler}}, \bibinfo {author} {\bibfnamefont {Soonwon}\ \bibnamefont {Choi}},
  \bibinfo {author} {\bibfnamefont {Rhine}\ \bibnamefont {Samajdar}}, \bibinfo
  {author} {\bibfnamefont {Sylvain}\ \bibnamefont {Schwartz}}, \bibinfo
  {author} {\bibfnamefont {Pietro}\ \bibnamefont {Silvi}}, \bibinfo {author}
  {\bibfnamefont {Subir}\ \bibnamefont {Sachdev}}, \bibinfo {author}
  {\bibfnamefont {Peter}\ \bibnamefont {Zoller}}, \bibinfo {author}
  {\bibfnamefont {Manuel}\ \bibnamefont {Endres}}, \bibinfo {author}
  {\bibfnamefont {Markus}\ \bibnamefont {Greiner}}, \bibinfo {author}
  {\bibfnamefont {Vladan}\ \bibnamefont {Vuleti{\'{c}}}}, \ and\ \bibinfo
  {author} {\bibfnamefont {Mikhail~D.}\ \bibnamefont {Lukin}},\ }\bibfield
  {title} {\enquote {\bibinfo {title} {{Quantum Kibble--Zurek mechanism and
  critical dynamics on a programmable Rydberg simulator}},}\ }\href {\doibase
  10.1038/s41586-019-1070-1} {\bibfield  {journal} {\bibinfo  {journal}
  {Nature}\ }\textbf {\bibinfo {volume} {568}},\ \bibinfo {pages} {207--211}
  (\bibinfo {year} {2019})}\BibitemShut {NoStop}%
\bibitem [{\citenamefont {Samajdar}\ \emph {et~al.}(2018)\citenamefont
  {Samajdar}, \citenamefont {Choi}, \citenamefont {Pichler}, \citenamefont
  {Lukin},\ and\ \citenamefont {Sachdev}}]{PhysRevA.98.023614}%
  \BibitemOpen
  \bibfield  {author} {\bibinfo {author} {\bibfnamefont {Rhine}\ \bibnamefont
  {Samajdar}}, \bibinfo {author} {\bibfnamefont {Soonwon}\ \bibnamefont
  {Choi}}, \bibinfo {author} {\bibfnamefont {Hannes}\ \bibnamefont {Pichler}},
  \bibinfo {author} {\bibfnamefont {Mikhail~D.}\ \bibnamefont {Lukin}}, \ and\
  \bibinfo {author} {\bibfnamefont {Subir}\ \bibnamefont {Sachdev}},\
  }\bibfield  {title} {\enquote {\bibinfo {title} {{Numerical study of the
  chiral ${\mathbb{Z}}_{3}$ quantum phase transition in one spatial
  dimension}},}\ }\href {\doibase 10.1103/PhysRevA.98.023614} {\bibfield
  {journal} {\bibinfo  {journal} {Phys. Rev. A}\ }\textbf {\bibinfo {volume}
  {98}},\ \bibinfo {pages} {023614} (\bibinfo {year} {2018})}\BibitemShut
  {NoStop}%
\bibitem [{\citenamefont {Whitsitt}\ \emph {et~al.}(2018)\citenamefont
  {Whitsitt}, \citenamefont {Samajdar},\ and\ \citenamefont
  {Sachdev}}]{PhysRevB.98.205118}%
  \BibitemOpen
  \bibfield  {author} {\bibinfo {author} {\bibfnamefont {Seth}\ \bibnamefont
  {Whitsitt}}, \bibinfo {author} {\bibfnamefont {Rhine}\ \bibnamefont
  {Samajdar}}, \ and\ \bibinfo {author} {\bibfnamefont {Subir}\ \bibnamefont
  {Sachdev}},\ }\bibfield  {title} {\enquote {\bibinfo {title} {Quantum field
  theory for the chiral clock transition in one spatial dimension},}\ }\href
  {\doibase 10.1103/PhysRevB.98.205118} {\bibfield  {journal} {\bibinfo
  {journal} {Phys. Rev. B}\ }\textbf {\bibinfo {volume} {98}},\ \bibinfo
  {pages} {205118} (\bibinfo {year} {2018})}\BibitemShut {NoStop}%
\bibitem [{\citenamefont {Samajdar}\ \emph {et~al.}(2020)\citenamefont
  {Samajdar}, \citenamefont {Ho}, \citenamefont {Pichler}, \citenamefont
  {Lukin},\ and\ \citenamefont {Sachdev}}]{Samajdar2020}%
  \BibitemOpen
  \bibfield  {author} {\bibinfo {author} {\bibfnamefont {Rhine}\ \bibnamefont
  {Samajdar}}, \bibinfo {author} {\bibfnamefont {Wen~Wei}\ \bibnamefont {Ho}},
  \bibinfo {author} {\bibfnamefont {Hannes}\ \bibnamefont {Pichler}}, \bibinfo
  {author} {\bibfnamefont {Mikhail~D.}\ \bibnamefont {Lukin}}, \ and\ \bibinfo
  {author} {\bibfnamefont {Subir}\ \bibnamefont {Sachdev}},\ }\bibfield
  {title} {\enquote {\bibinfo {title} {{Complex density wave orders and quantum
  phase transitions in a model of square-lattice Rydberg atom arrays}},}\
  }\href {\doibase 10.1103/PhysRevLett.124.103601} {\bibfield  {journal}
  {\bibinfo  {journal} {Phys. Rev. Lett.}\ }\textbf {\bibinfo {volume} {124}},\
  \bibinfo {pages} {103601} (\bibinfo {year} {2020})}\BibitemShut {NoStop}%
\bibitem [{\citenamefont {Chen}\ \emph {et~al.}(2022)\citenamefont {Chen},
  \citenamefont {Bornet}, \citenamefont {Bintz}, \citenamefont {Emperauger},
  \citenamefont {Leclerc}, \citenamefont {Liu}, \citenamefont {Scholl},
  \citenamefont {Barredo}, \citenamefont {Hauschild}, \citenamefont
  {Chatterjee}, \citenamefont {Schuler}, \citenamefont {Laeuchli},
  \citenamefont {Zaletel}, \citenamefont {Lahaye}, \citenamefont {Yao},\ and\
  \citenamefont {Browaeys}}]{Chen2022}%
  \BibitemOpen
  \bibfield  {author} {\bibinfo {author} {\bibfnamefont {Cheng}\ \bibnamefont
  {Chen}}, \bibinfo {author} {\bibfnamefont {Guillaume}\ \bibnamefont
  {Bornet}}, \bibinfo {author} {\bibfnamefont {Marcus}\ \bibnamefont {Bintz}},
  \bibinfo {author} {\bibfnamefont {Gabriel}\ \bibnamefont {Emperauger}},
  \bibinfo {author} {\bibfnamefont {Lucas}\ \bibnamefont {Leclerc}}, \bibinfo
  {author} {\bibfnamefont {Vincent~S.}\ \bibnamefont {Liu}}, \bibinfo {author}
  {\bibfnamefont {Pascal}\ \bibnamefont {Scholl}}, \bibinfo {author}
  {\bibfnamefont {Daniel}\ \bibnamefont {Barredo}}, \bibinfo {author}
  {\bibfnamefont {Johannes}\ \bibnamefont {Hauschild}}, \bibinfo {author}
  {\bibfnamefont {Shubhayu}\ \bibnamefont {Chatterjee}}, \bibinfo {author}
  {\bibfnamefont {Michael}\ \bibnamefont {Schuler}}, \bibinfo {author}
  {\bibfnamefont {Andreas~M.}\ \bibnamefont {Laeuchli}}, \bibinfo {author}
  {\bibfnamefont {Michael~P.}\ \bibnamefont {Zaletel}}, \bibinfo {author}
  {\bibfnamefont {Thierry}\ \bibnamefont {Lahaye}}, \bibinfo {author}
  {\bibfnamefont {Norman~Y.}\ \bibnamefont {Yao}}, \ and\ \bibinfo {author}
  {\bibfnamefont {Antoine}\ \bibnamefont {Browaeys}},\ }\href@noop {} {\enquote
  {\bibinfo {title} {{Continuous symmetry breaking in a two-dimensional Rydberg
  array}},}\ } (\bibinfo {year} {2022}),\ \Eprint
  {http://arxiv.org/abs/2207.12930} {arXiv:2207.12930 [cond-mat.quant-gas]}
  \BibitemShut {NoStop}%
\bibitem [{\citenamefont {Labuhn}\ \emph {et~al.}(2016)\citenamefont {Labuhn},
  \citenamefont {Barredo}, \citenamefont {Ravets}, \citenamefont
  {de~L{\'e}s{\'e}leuc}, \citenamefont {Macr{\`i}}, \citenamefont {Lahaye},\
  and\ \citenamefont {Browaeys}}]{Labuhn2016}%
  \BibitemOpen
  \bibfield  {author} {\bibinfo {author} {\bibfnamefont {Henning}\ \bibnamefont
  {Labuhn}}, \bibinfo {author} {\bibfnamefont {Daniel}\ \bibnamefont
  {Barredo}}, \bibinfo {author} {\bibfnamefont {Sylvain}\ \bibnamefont
  {Ravets}}, \bibinfo {author} {\bibfnamefont {Sylvain}\ \bibnamefont
  {de~L{\'e}s{\'e}leuc}}, \bibinfo {author} {\bibfnamefont {Tommaso}\
  \bibnamefont {Macr{\`i}}}, \bibinfo {author} {\bibfnamefont {Thierry}\
  \bibnamefont {Lahaye}}, \ and\ \bibinfo {author} {\bibfnamefont {Antoine}\
  \bibnamefont {Browaeys}},\ }\bibfield  {title} {\enquote {\bibinfo {title}
  {{Tunable two-dimensional arrays of single Rydberg atoms for realizing
  quantum Ising models}},}\ }\href {\doibase 10.1038/nature18274} {\bibfield
  {journal} {\bibinfo  {journal} {Nature}\ }\textbf {\bibinfo {volume} {534}},\
  \bibinfo {pages} {667--670} (\bibinfo {year} {2016})}\BibitemShut {NoStop}%
\bibitem [{\citenamefont {Bernien}\ \emph {et~al.}(2017)\citenamefont
  {Bernien}, \citenamefont {Schwartz}, \citenamefont {Keesling}, \citenamefont
  {Levine}, \citenamefont {Omran}, \citenamefont {Pichler}, \citenamefont
  {Choi}, \citenamefont {Zibrov}, \citenamefont {Endres}, \citenamefont
  {Greiner}, \citenamefont {Vuleti{\'{c}}},\ and\ \citenamefont
  {Lukin}}]{Bernien2017}%
  \BibitemOpen
  \bibfield  {author} {\bibinfo {author} {\bibfnamefont {Hannes}\ \bibnamefont
  {Bernien}}, \bibinfo {author} {\bibfnamefont {Sylvain}\ \bibnamefont
  {Schwartz}}, \bibinfo {author} {\bibfnamefont {Alexander}\ \bibnamefont
  {Keesling}}, \bibinfo {author} {\bibfnamefont {Harry}\ \bibnamefont
  {Levine}}, \bibinfo {author} {\bibfnamefont {Ahmed}\ \bibnamefont {Omran}},
  \bibinfo {author} {\bibfnamefont {Hannes}\ \bibnamefont {Pichler}}, \bibinfo
  {author} {\bibfnamefont {Soonwon}\ \bibnamefont {Choi}}, \bibinfo {author}
  {\bibfnamefont {Alexander~S.}\ \bibnamefont {Zibrov}}, \bibinfo {author}
  {\bibfnamefont {Manuel}\ \bibnamefont {Endres}}, \bibinfo {author}
  {\bibfnamefont {Markus}\ \bibnamefont {Greiner}}, \bibinfo {author}
  {\bibfnamefont {Vladan}\ \bibnamefont {Vuleti{\'{c}}}}, \ and\ \bibinfo
  {author} {\bibfnamefont {Mikhail~D.}\ \bibnamefont {Lukin}},\ }\bibfield
  {title} {\enquote {\bibinfo {title} {Probing many-body dynamics on a 51-atom
  quantum simulator},}\ }\href {\doibase 10.1038/nature24622} {\bibfield
  {journal} {\bibinfo  {journal} {Nature}\ }\textbf {\bibinfo {volume} {551}},\
  \bibinfo {pages} {579--584} (\bibinfo {year} {2017})}\BibitemShut {NoStop}%
\bibitem [{\citenamefont {Browaeys}\ and\ \citenamefont
  {Lahaye}(2020)}]{Browaeys2020}%
  \BibitemOpen
  \bibfield  {author} {\bibinfo {author} {\bibfnamefont {Antoine}\ \bibnamefont
  {Browaeys}}\ and\ \bibinfo {author} {\bibfnamefont {Thierry}\ \bibnamefont
  {Lahaye}},\ }\bibfield  {title} {\enquote {\bibinfo {title} {{Many-body
  physics with individually controlled Rydberg atoms}},}\ }\href {\doibase
  10.1038/s41567-019-0733-z} {\bibfield  {journal} {\bibinfo  {journal} {Nat.
  Phys.}\ }\textbf {\bibinfo {volume} {16}},\ \bibinfo {pages} {132--142}
  (\bibinfo {year} {2020})}\BibitemShut {NoStop}%
\bibitem [{\citenamefont {Ebadi}\ \emph {et~al.}(2021)\citenamefont {Ebadi},
  \citenamefont {Wang}, \citenamefont {Levine}, \citenamefont {Keesling},
  \citenamefont {Semeghini}, \citenamefont {Omran}, \citenamefont {Bluvstein},
  \citenamefont {Samajdar}, \citenamefont {Pichler}, \citenamefont {Ho},
  \citenamefont {Choi}, \citenamefont {Sachdev}, \citenamefont {Greiner},
  \citenamefont {Vuleti{\'{c}}},\ and\ \citenamefont {Lukin}}]{Ebadi2021}%
  \BibitemOpen
  \bibfield  {author} {\bibinfo {author} {\bibfnamefont {Sepehr}\ \bibnamefont
  {Ebadi}}, \bibinfo {author} {\bibfnamefont {Tout~T.}\ \bibnamefont {Wang}},
  \bibinfo {author} {\bibfnamefont {Harry}\ \bibnamefont {Levine}}, \bibinfo
  {author} {\bibfnamefont {Alexander}\ \bibnamefont {Keesling}}, \bibinfo
  {author} {\bibfnamefont {Giulia}\ \bibnamefont {Semeghini}}, \bibinfo
  {author} {\bibfnamefont {Ahmed}\ \bibnamefont {Omran}}, \bibinfo {author}
  {\bibfnamefont {Dolev}\ \bibnamefont {Bluvstein}}, \bibinfo {author}
  {\bibfnamefont {Rhine}\ \bibnamefont {Samajdar}}, \bibinfo {author}
  {\bibfnamefont {Hannes}\ \bibnamefont {Pichler}}, \bibinfo {author}
  {\bibfnamefont {Wen~Wei}\ \bibnamefont {Ho}}, \bibinfo {author}
  {\bibfnamefont {Soonwon}\ \bibnamefont {Choi}}, \bibinfo {author}
  {\bibfnamefont {Subir}\ \bibnamefont {Sachdev}}, \bibinfo {author}
  {\bibfnamefont {Markus}\ \bibnamefont {Greiner}}, \bibinfo {author}
  {\bibfnamefont {Vladan}\ \bibnamefont {Vuleti{\'{c}}}}, \ and\ \bibinfo
  {author} {\bibfnamefont {Mikhail~D.}\ \bibnamefont {Lukin}},\ }\bibfield
  {title} {\enquote {\bibinfo {title} {Quantum phases of matter on a 256-atom
  programmable quantum simulator},}\ }\href {\doibase
  10.1038/s41586-021-03582-4} {\bibfield  {journal} {\bibinfo  {journal}
  {Nature}\ }\textbf {\bibinfo {volume} {595}},\ \bibinfo {pages} {227--232}
  (\bibinfo {year} {2021})}\BibitemShut {NoStop}%
\bibitem [{\citenamefont {Scholl}\ \emph {et~al.}(2021)\citenamefont {Scholl},
  \citenamefont {Schuler}, \citenamefont {Williams}, \citenamefont
  {Eberharter}, \citenamefont {Barredo}, \citenamefont {Schymik}, \citenamefont
  {Lienhard}, \citenamefont {Henry}, \citenamefont {Lang}, \citenamefont
  {Lahaye}, \citenamefont {La\"uchli},\ and\ \citenamefont
  {Browaeys}}]{Scholl2021}%
  \BibitemOpen
  \bibfield  {author} {\bibinfo {author} {\bibfnamefont {Pascal}\ \bibnamefont
  {Scholl}}, \bibinfo {author} {\bibfnamefont {Michael}\ \bibnamefont
  {Schuler}}, \bibinfo {author} {\bibfnamefont {Hannah~J}\ \bibnamefont
  {Williams}}, \bibinfo {author} {\bibfnamefont {Alexander~A}\ \bibnamefont
  {Eberharter}}, \bibinfo {author} {\bibfnamefont {Daniel}\ \bibnamefont
  {Barredo}}, \bibinfo {author} {\bibfnamefont {Kai-Niklas}\ \bibnamefont
  {Schymik}}, \bibinfo {author} {\bibfnamefont {Vincent}\ \bibnamefont
  {Lienhard}}, \bibinfo {author} {\bibfnamefont {Louis-Paul}\ \bibnamefont
  {Henry}}, \bibinfo {author} {\bibfnamefont {Thomas~C}\ \bibnamefont {Lang}},
  \bibinfo {author} {\bibfnamefont {Thierry}\ \bibnamefont {Lahaye}}, \bibinfo
  {author} {\bibfnamefont {Andreas~M.}\ \bibnamefont {La\"uchli}}, \ and\
  \bibinfo {author} {\bibfnamefont {Antoine}\ \bibnamefont {Browaeys}},\
  }\bibfield  {title} {\enquote {\bibinfo {title} {{Quantum simulation of 2D
  antiferromagnets with hundreds of Rydberg atoms}},}\ }\href {\doibase
  https://doi.org/10.1038/s41586-021-03585-1} {\bibfield  {journal} {\bibinfo
  {journal} {Nature}\ }\textbf {\bibinfo {volume} {595}},\ \bibinfo {pages}
  {233--238} (\bibinfo {year} {2021})}\BibitemShut {NoStop}%
\bibitem [{\citenamefont {Semeghini}\ \emph {et~al.}(2021)\citenamefont
  {Semeghini}, \citenamefont {Levine}, \citenamefont {Keesling}, \citenamefont
  {Ebadi}, \citenamefont {Wang}, \citenamefont {Bluvstein}, \citenamefont
  {Verresen}, \citenamefont {Pichler}, \citenamefont {Kalinowski},
  \citenamefont {Samajdar}, \citenamefont {Omran}, \citenamefont {Sachdev},
  \citenamefont {Vishwanath}, \citenamefont {Greiner}, \citenamefont
  {Vuletić},\ and\ \citenamefont {Lukin}}]{Semeghini2021}%
  \BibitemOpen
  \bibfield  {author} {\bibinfo {author} {\bibfnamefont {G.}~\bibnamefont
  {Semeghini}}, \bibinfo {author} {\bibfnamefont {H.}~\bibnamefont {Levine}},
  \bibinfo {author} {\bibfnamefont {A.}~\bibnamefont {Keesling}}, \bibinfo
  {author} {\bibfnamefont {S.}~\bibnamefont {Ebadi}}, \bibinfo {author}
  {\bibfnamefont {T.~T.}\ \bibnamefont {Wang}}, \bibinfo {author}
  {\bibfnamefont {D.}~\bibnamefont {Bluvstein}}, \bibinfo {author}
  {\bibfnamefont {R.}~\bibnamefont {Verresen}}, \bibinfo {author}
  {\bibfnamefont {H.}~\bibnamefont {Pichler}}, \bibinfo {author} {\bibfnamefont
  {M.}~\bibnamefont {Kalinowski}}, \bibinfo {author} {\bibfnamefont
  {R.}~\bibnamefont {Samajdar}}, \bibinfo {author} {\bibfnamefont
  {A.}~\bibnamefont {Omran}}, \bibinfo {author} {\bibfnamefont
  {S.}~\bibnamefont {Sachdev}}, \bibinfo {author} {\bibfnamefont
  {A.}~\bibnamefont {Vishwanath}}, \bibinfo {author} {\bibfnamefont
  {M.}~\bibnamefont {Greiner}}, \bibinfo {author} {\bibfnamefont
  {V.}~\bibnamefont {Vuletić}}, \ and\ \bibinfo {author} {\bibfnamefont
  {M.~D.}\ \bibnamefont {Lukin}},\ }\bibfield  {title} {\enquote {\bibinfo
  {title} {{Probing topological spin liquids on a programmable quantum
  simulator}},}\ }\href {\doibase 10.1126/science.abi8794} {\bibfield
  {journal} {\bibinfo  {journal} {Science}\ }\textbf {\bibinfo {volume}
  {374}},\ \bibinfo {pages} {1242--1247} (\bibinfo {year} {2021})}\BibitemShut
  {NoStop}%
\bibitem [{\citenamefont {Read}\ and\ \citenamefont
  {Sachdev}(1991)}]{ReadSachdev91}%
  \BibitemOpen
  \bibfield  {author} {\bibinfo {author} {\bibfnamefont {N.}~\bibnamefont
  {Read}}\ and\ \bibinfo {author} {\bibfnamefont {Subir}\ \bibnamefont
  {Sachdev}},\ }\bibfield  {title} {\enquote {\bibinfo {title} {{Large-N
  expansion for frustrated quantum antiferromagnets}},}\ }\href {\doibase
  10.1103/PhysRevLett.66.1773} {\bibfield  {journal} {\bibinfo  {journal}
  {Phys. Rev. Lett.}\ }\textbf {\bibinfo {volume} {66}},\ \bibinfo {pages}
  {1773--1776} (\bibinfo {year} {1991})}\BibitemShut {NoStop}%
\bibitem [{\citenamefont {Wen}(1991)}]{Wen1991}%
  \BibitemOpen
  \bibfield  {author} {\bibinfo {author} {\bibfnamefont {X.~G.}\ \bibnamefont
  {Wen}},\ }\bibfield  {title} {\enquote {\bibinfo {title} {Mean-field theory
  of spin-liquid states with finite energy gap and topological orders},}\
  }\href {\doibase 10.1103/PhysRevB.44.2664} {\bibfield  {journal} {\bibinfo
  {journal} {Phys. Rev. B}\ }\textbf {\bibinfo {volume} {44}},\ \bibinfo
  {pages} {2664--2672} (\bibinfo {year} {1991})}\BibitemShut {NoStop}%
\bibitem [{\citenamefont {Sachdev}(1992)}]{Sachdev92}%
  \BibitemOpen
  \bibfield  {author} {\bibinfo {author} {\bibfnamefont {Subir}\ \bibnamefont
  {Sachdev}},\ }\bibfield  {title} {\enquote {\bibinfo {title} {{Kagom{\'e}-
  and triangular-lattice Heisenberg antiferromagnets: Ordering from quantum
  fluctuations and quantum-disordered ground states with unconfined bosonic
  spinons}},}\ }\href {\doibase 10.1103/PhysRevB.45.12377} {\bibfield
  {journal} {\bibinfo  {journal} {Phys. Rev. B}\ }\textbf {\bibinfo {volume}
  {45}},\ \bibinfo {pages} {12377--12396} (\bibinfo {year} {1992})}\BibitemShut
  {NoStop}%
\bibitem [{\citenamefont {Kitaev}(2006)}]{kitaev2006anyons}%
  \BibitemOpen
  \bibfield  {author} {\bibinfo {author} {\bibfnamefont {Alexei}\ \bibnamefont
  {Kitaev}},\ }\bibfield  {title} {\enquote {\bibinfo {title} {Anyons in an
  exactly solved model and beyond},}\ }\href {\doibase
  10.1016/j.aop.2005.10.005} {\bibfield  {journal} {\bibinfo  {journal} {Ann.
  Phys.}\ }\textbf {\bibinfo {volume} {321}},\ \bibinfo {pages} {2--111}
  (\bibinfo {year} {2006})}\BibitemShut {NoStop}%
\bibitem [{\citenamefont {Wen}(2017)}]{RevModPhys.89.041004}%
  \BibitemOpen
  \bibfield  {author} {\bibinfo {author} {\bibfnamefont {Xiao-Gang}\
  \bibnamefont {Wen}},\ }\bibfield  {title} {\enquote {\bibinfo {title}
  {{Colloquium: Zoo of quantum-topological phases of matter}},}\ }\href
  {\doibase 10.1103/RevModPhys.89.041004} {\bibfield  {journal} {\bibinfo
  {journal} {Rev. Mod. Phys.}\ }\textbf {\bibinfo {volume} {89}},\ \bibinfo
  {pages} {041004} (\bibinfo {year} {2017})}\BibitemShut {NoStop}%
\bibitem [{\citenamefont {Broholm}\ \emph {et~al.}(2020)\citenamefont
  {Broholm}, \citenamefont {Cava}, \citenamefont {Kivelson}, \citenamefont
  {Nocera}, \citenamefont {Norman},\ and\ \citenamefont
  {Senthil}}]{broholm2020quantum}%
  \BibitemOpen
  \bibfield  {author} {\bibinfo {author} {\bibfnamefont {C}~\bibnamefont
  {Broholm}}, \bibinfo {author} {\bibfnamefont {R~J}\ \bibnamefont {Cava}},
  \bibinfo {author} {\bibfnamefont {S~A}\ \bibnamefont {Kivelson}}, \bibinfo
  {author} {\bibfnamefont {D~G}\ \bibnamefont {Nocera}}, \bibinfo {author}
  {\bibfnamefont {M~R}\ \bibnamefont {Norman}}, \ and\ \bibinfo {author}
  {\bibfnamefont {T}~\bibnamefont {Senthil}},\ }\bibfield  {title} {\enquote
  {\bibinfo {title} {Quantum spin liquids},}\ }\href {\doibase
  10.1126/science.aay0668} {\bibfield  {journal} {\bibinfo  {journal}
  {Science}\ }\textbf {\bibinfo {volume} {367}},\ \bibinfo {pages} {eaay0668}
  (\bibinfo {year} {2020})}\BibitemShut {NoStop}%
\bibitem [{\citenamefont {Hermele}\ \emph
  {et~al.}(2004{\natexlab{a}})\citenamefont {Hermele}, \citenamefont {Fisher},\
  and\ \citenamefont {Balents}}]{hermele2004a}%
  \BibitemOpen
  \bibfield  {author} {\bibinfo {author} {\bibfnamefont {Michael}\ \bibnamefont
  {Hermele}}, \bibinfo {author} {\bibfnamefont {Matthew P.~A.}\ \bibnamefont
  {Fisher}}, \ and\ \bibinfo {author} {\bibfnamefont {Leon}\ \bibnamefont
  {Balents}},\ }\bibfield  {title} {\enquote {\bibinfo {title} {{Pyrochlore
  photons: The $U(1)$ spin liquid in a $S=\frac{1}{2}$ three-dimensional
  frustrated magnet}},}\ }\href {\doibase 10.1103/PhysRevB.69.064404}
  {\bibfield  {journal} {\bibinfo  {journal} {Phys. Rev. B}\ }\textbf {\bibinfo
  {volume} {69}},\ \bibinfo {pages} {064404} (\bibinfo {year}
  {2004}{\natexlab{a}})}\BibitemShut {NoStop}%
\bibitem [{\citenamefont {Gingras}\ and\ \citenamefont
  {McClarty}(2014)}]{gingras2014quantum}%
  \BibitemOpen
  \bibfield  {author} {\bibinfo {author} {\bibfnamefont {Michel J~P}\
  \bibnamefont {Gingras}}\ and\ \bibinfo {author} {\bibfnamefont {Paul~A}\
  \bibnamefont {McClarty}},\ }\bibfield  {title} {\enquote {\bibinfo {title}
  {Quantum spin ice: a search for gapless quantum spin liquids in pyrochlore
  magnets},}\ }\href {\doibase 10.1088/0034-4885/77/5/056501} {\bibfield
  {journal} {\bibinfo  {journal} {Rep. Prog. Phys.}\ }\textbf {\bibinfo
  {volume} {77}},\ \bibinfo {pages} {056501} (\bibinfo {year}
  {2014})}\BibitemShut {NoStop}%
\bibitem [{\citenamefont {Baskaran}\ and\ \citenamefont
  {Anderson}(1988)}]{Baskaran1988}%
  \BibitemOpen
  \bibfield  {author} {\bibinfo {author} {\bibfnamefont {G.}~\bibnamefont
  {Baskaran}}\ and\ \bibinfo {author} {\bibfnamefont {P.~W.}\ \bibnamefont
  {Anderson}},\ }\bibfield  {title} {\enquote {\bibinfo {title} {{Gauge theory
  of high-temperature superconductors and strongly correlated Fermi
  systems}},}\ }\href {\doibase 10.1103/PhysRevB.37.580} {\bibfield  {journal}
  {\bibinfo  {journal} {Phys. Rev. B}\ }\textbf {\bibinfo {volume} {37}},\
  \bibinfo {pages} {580--583} (\bibinfo {year} {1988})}\BibitemShut {NoStop}%
\bibitem [{\citenamefont {Hermele}\ \emph
  {et~al.}(2004{\natexlab{b}})\citenamefont {Hermele}, \citenamefont {Senthil},
  \citenamefont {Fisher}, \citenamefont {Lee}, \citenamefont {Nagaosa},\ and\
  \citenamefont {Wen}}]{Hermele2004}%
  \BibitemOpen
  \bibfield  {author} {\bibinfo {author} {\bibfnamefont {Michael}\ \bibnamefont
  {Hermele}}, \bibinfo {author} {\bibfnamefont {T.}~\bibnamefont {Senthil}},
  \bibinfo {author} {\bibfnamefont {Matthew P.~A.}\ \bibnamefont {Fisher}},
  \bibinfo {author} {\bibfnamefont {Patrick~A.}\ \bibnamefont {Lee}}, \bibinfo
  {author} {\bibfnamefont {Naoto}\ \bibnamefont {Nagaosa}}, \ and\ \bibinfo
  {author} {\bibfnamefont {Xiao-Gang}\ \bibnamefont {Wen}},\ }\bibfield
  {title} {\enquote {\bibinfo {title} {{Stability of $U(1)$ spin liquids in two
  dimensions}},}\ }\href {\doibase 10.1103/PhysRevB.70.214437} {\bibfield
  {journal} {\bibinfo  {journal} {Phys. Rev. B}\ }\textbf {\bibinfo {volume}
  {70}},\ \bibinfo {pages} {214437} (\bibinfo {year}
  {2004}{\natexlab{b}})}\BibitemShut {NoStop}%
\bibitem [{\citenamefont {Thewes}\ and\ \citenamefont
  {Fernandes}(2020)}]{Thewes2020}%
  \BibitemOpen
  \bibfield  {author} {\bibinfo {author} {\bibfnamefont {Filipe~C.}\
  \bibnamefont {Thewes}}\ and\ \bibinfo {author} {\bibfnamefont {Heitor C.~M.}\
  \bibnamefont {Fernandes}},\ }\bibfield  {title} {\enquote {\bibinfo {title}
  {Phase transitions in hard-core lattice gases on the honeycomb lattice},}\
  }\href {\doibase 10.1103/PhysRevE.101.062138} {\bibfield  {journal} {\bibinfo
   {journal} {Phys. Rev. E}\ }\textbf {\bibinfo {volume} {101}},\ \bibinfo
  {pages} {062138} (\bibinfo {year} {2020})}\BibitemShut {NoStop}%
\bibitem [{\citenamefont {Verberkmoes}\ and\ \citenamefont
  {Nienhuis}(1999)}]{Verberkmoes1999}%
  \BibitemOpen
  \bibfield  {author} {\bibinfo {author} {\bibfnamefont {Alain}\ \bibnamefont
  {Verberkmoes}}\ and\ \bibinfo {author} {\bibfnamefont {Bernard}\ \bibnamefont
  {Nienhuis}},\ }\bibfield  {title} {\enquote {\bibinfo {title} {Triangular
  trimers on the triangular lattice: An exact solution},}\ }\href {\doibase
  10.1103/PhysRevLett.83.3986} {\bibfield  {journal} {\bibinfo  {journal}
  {Phys. Rev. Lett.}\ }\textbf {\bibinfo {volume} {83}},\ \bibinfo {pages}
  {3986--3989} (\bibinfo {year} {1999})}\BibitemShut {NoStop}%
\bibitem [{\citenamefont {Giudice}\ \emph {et~al.}(2022)\citenamefont
  {Giudice}, \citenamefont {Surace}, \citenamefont {Pichler},\ and\
  \citenamefont {Giudici}}]{Giudice2022}%
  \BibitemOpen
  \bibfield  {author} {\bibinfo {author} {\bibfnamefont {Giacomo}\ \bibnamefont
  {Giudice}}, \bibinfo {author} {\bibfnamefont {Federica~Maria}\ \bibnamefont
  {Surace}}, \bibinfo {author} {\bibfnamefont {Hannes}\ \bibnamefont
  {Pichler}}, \ and\ \bibinfo {author} {\bibfnamefont {Giuliano}\ \bibnamefont
  {Giudici}},\ }\href@noop {} {\enquote {\bibinfo {title} {{Trimer states with
  $\mathbb{Z}_3$ topological order in Rydberg atom arrays}},}\ } (\bibinfo
  {year} {2022}),\ \Eprint {http://arxiv.org/abs/2205.10387} {arXiv:2205.10387
  [quant-ph]} \BibitemShut {NoStop}%
\bibitem [{\citenamefont {Fishman}\ \emph {et~al.}(2020)\citenamefont
  {Fishman}, \citenamefont {White},\ and\ \citenamefont
  {Stoudenmire}}]{itensor}%
  \BibitemOpen
  \bibfield  {author} {\bibinfo {author} {\bibfnamefont {Matthew}\ \bibnamefont
  {Fishman}}, \bibinfo {author} {\bibfnamefont {Steven~R.}\ \bibnamefont
  {White}}, \ and\ \bibinfo {author} {\bibfnamefont {E.~Miles}\ \bibnamefont
  {Stoudenmire}},\ }\href@noop {} {\enquote {\bibinfo {title} {The
  \mbox{ITensor} software library for tensor network calculations},}\ }
  (\bibinfo {year} {2020}),\ \Eprint {http://arxiv.org/abs/2007.14822}
  {arXiv:2007.14822 [cs.MS]} \BibitemShut {NoStop}%
\bibitem [{\citenamefont {Jaksch}\ \emph {et~al.}(2000)\citenamefont {Jaksch},
  \citenamefont {Cirac}, \citenamefont {Zoller}, \citenamefont {Rolston},
  \citenamefont {C\^ot\'e},\ and\ \citenamefont {Lukin}}]{Jaksch2000}%
  \BibitemOpen
  \bibfield  {author} {\bibinfo {author} {\bibfnamefont {D.}~\bibnamefont
  {Jaksch}}, \bibinfo {author} {\bibfnamefont {J.~I.}\ \bibnamefont {Cirac}},
  \bibinfo {author} {\bibfnamefont {P.}~\bibnamefont {Zoller}}, \bibinfo
  {author} {\bibfnamefont {S.~L.}\ \bibnamefont {Rolston}}, \bibinfo {author}
  {\bibfnamefont {R.}~\bibnamefont {C\^ot\'e}}, \ and\ \bibinfo {author}
  {\bibfnamefont {M.~D.}\ \bibnamefont {Lukin}},\ }\bibfield  {title} {\enquote
  {\bibinfo {title} {Fast quantum gates for neutral atoms},}\ }\href {\doibase
  10.1103/PhysRevLett.85.2208} {\bibfield  {journal} {\bibinfo  {journal}
  {Phys. Rev. Lett.}\ }\textbf {\bibinfo {volume} {85}},\ \bibinfo {pages}
  {2208--2211} (\bibinfo {year} {2000})}\BibitemShut {NoStop}%
\bibitem [{\citenamefont {Lukin}\ \emph {et~al.}(2001)\citenamefont {Lukin},
  \citenamefont {Fleischhauer}, \citenamefont {Cote}, \citenamefont {Duan},
  \citenamefont {Jaksch}, \citenamefont {Cirac},\ and\ \citenamefont
  {Zoller}}]{Lukin2001}%
  \BibitemOpen
  \bibfield  {author} {\bibinfo {author} {\bibfnamefont {M.~D.}\ \bibnamefont
  {Lukin}}, \bibinfo {author} {\bibfnamefont {M.}~\bibnamefont {Fleischhauer}},
  \bibinfo {author} {\bibfnamefont {R.}~\bibnamefont {Cote}}, \bibinfo {author}
  {\bibfnamefont {L.~M.}\ \bibnamefont {Duan}}, \bibinfo {author}
  {\bibfnamefont {D.}~\bibnamefont {Jaksch}}, \bibinfo {author} {\bibfnamefont
  {J.~I.}\ \bibnamefont {Cirac}}, \ and\ \bibinfo {author} {\bibfnamefont
  {P.}~\bibnamefont {Zoller}},\ }\bibfield  {title} {\enquote {\bibinfo {title}
  {Dipole blockade and quantum information processing in mesoscopic atomic
  ensembles},}\ }\href {\doibase 10.1103/PhysRevLett.87.037901} {\bibfield
  {journal} {\bibinfo  {journal} {Phys. Rev. Lett.}\ }\textbf {\bibinfo
  {volume} {87}},\ \bibinfo {pages} {037901} (\bibinfo {year}
  {2001})}\BibitemShut {NoStop}%
\bibitem [{\citenamefont {Ga{\"e}tan}\ \emph {et~al.}(2009)\citenamefont
  {Ga{\"e}tan}, \citenamefont {Miroshnychenko}, \citenamefont {Wilk},
  \citenamefont {Chotia}, \citenamefont {Viteau}, \citenamefont {Comparat},
  \citenamefont {Pillet}, \citenamefont {Browaeys},\ and\ \citenamefont
  {Grangier}}]{Gaetan2009}%
  \BibitemOpen
  \bibfield  {author} {\bibinfo {author} {\bibfnamefont {Alpha}\ \bibnamefont
  {Ga{\"e}tan}}, \bibinfo {author} {\bibfnamefont {Yevhen}\ \bibnamefont
  {Miroshnychenko}}, \bibinfo {author} {\bibfnamefont {Tatjana}\ \bibnamefont
  {Wilk}}, \bibinfo {author} {\bibfnamefont {Amodsen}\ \bibnamefont {Chotia}},
  \bibinfo {author} {\bibfnamefont {Matthieu}\ \bibnamefont {Viteau}}, \bibinfo
  {author} {\bibfnamefont {Daniel}\ \bibnamefont {Comparat}}, \bibinfo {author}
  {\bibfnamefont {Pierre}\ \bibnamefont {Pillet}}, \bibinfo {author}
  {\bibfnamefont {Antoine}\ \bibnamefont {Browaeys}}, \ and\ \bibinfo {author}
  {\bibfnamefont {Philippe}\ \bibnamefont {Grangier}},\ }\bibfield  {title}
  {\enquote {\bibinfo {title} {{Observation of collective excitation of two
  individual atoms in the Rydberg blockade regime}},}\ }\href {\doibase
  10.1038/nphys1183} {\bibfield  {journal} {\bibinfo  {journal} {Nat. Phys.}\
  }\textbf {\bibinfo {volume} {5}},\ \bibinfo {pages} {115--118} (\bibinfo
  {year} {2009})}\BibitemShut {NoStop}%
\bibitem [{\citenamefont {Urban}\ \emph {et~al.}(2009)\citenamefont {Urban},
  \citenamefont {Johnson}, \citenamefont {Henage}, \citenamefont {Isenhower},
  \citenamefont {Yavuz}, \citenamefont {Walker},\ and\ \citenamefont
  {Saffman}}]{Urban2009}%
  \BibitemOpen
  \bibfield  {author} {\bibinfo {author} {\bibfnamefont {E.}~\bibnamefont
  {Urban}}, \bibinfo {author} {\bibfnamefont {T.~A.}\ \bibnamefont {Johnson}},
  \bibinfo {author} {\bibfnamefont {T.}~\bibnamefont {Henage}}, \bibinfo
  {author} {\bibfnamefont {L.}~\bibnamefont {Isenhower}}, \bibinfo {author}
  {\bibfnamefont {D.~D.}\ \bibnamefont {Yavuz}}, \bibinfo {author}
  {\bibfnamefont {T.~G.}\ \bibnamefont {Walker}}, \ and\ \bibinfo {author}
  {\bibfnamefont {M.}~\bibnamefont {Saffman}},\ }\bibfield  {title} {\enquote
  {\bibinfo {title} {{Observation of Rydberg blockade between two atoms}},}\
  }\href {\doibase 10.1038/nphys1178} {\bibfield  {journal} {\bibinfo
  {journal} {Nat. Phys.}\ }\textbf {\bibinfo {volume} {5}},\ \bibinfo {pages}
  {110--114} (\bibinfo {year} {2009})}\BibitemShut {NoStop}%
\bibitem [{\citenamefont {Samajdar}\ \emph {et~al.}(2021)\citenamefont
  {Samajdar}, \citenamefont {Ho}, \citenamefont {Pichler}, \citenamefont
  {Lukin},\ and\ \citenamefont {Sachdev}}]{Samajdar2021}%
  \BibitemOpen
  \bibfield  {author} {\bibinfo {author} {\bibfnamefont {Rhine}\ \bibnamefont
  {Samajdar}}, \bibinfo {author} {\bibfnamefont {Wen~Wei}\ \bibnamefont {Ho}},
  \bibinfo {author} {\bibfnamefont {Hannes}\ \bibnamefont {Pichler}}, \bibinfo
  {author} {\bibfnamefont {Mikhail~D.}\ \bibnamefont {Lukin}}, \ and\ \bibinfo
  {author} {\bibfnamefont {Subir}\ \bibnamefont {Sachdev}},\ }\bibfield
  {title} {\enquote {\bibinfo {title} {{Quantum phases of Rydberg atoms on a
  kagome lattice}},}\ }\href {\doibase 10.1073/pnas.2015785118} {\bibfield
  {journal} {\bibinfo  {journal} {Proc. Natl. Acad. Sci. U.S.A.}\ }\textbf
  {\bibinfo {volume} {118}},\ \bibinfo {pages} {e2015785118} (\bibinfo {year}
  {2021})}\BibitemShut {NoStop}%
\bibitem [{\citenamefont {Yan}\ \emph {et~al.}(2022)\citenamefont {Yan},
  \citenamefont {Samajdar}, \citenamefont {Wang}, \citenamefont {Sachdev},\
  and\ \citenamefont {Meng}}]{Yan_2022}%
  \BibitemOpen
  \bibfield  {author} {\bibinfo {author} {\bibfnamefont {Zheng}\ \bibnamefont
  {Yan}}, \bibinfo {author} {\bibfnamefont {Rhine}\ \bibnamefont {Samajdar}},
  \bibinfo {author} {\bibfnamefont {Yan-Cheng}\ \bibnamefont {Wang}}, \bibinfo
  {author} {\bibfnamefont {Subir}\ \bibnamefont {Sachdev}}, \ and\ \bibinfo
  {author} {\bibfnamefont {Zi~Yang}\ \bibnamefont {Meng}},\ }\bibfield  {title}
  {\enquote {\bibinfo {title} {Triangular lattice quantum dimer model with
  variable dimer density},}\ }\href {\doibase 10.1038/s41467-022-33431-5}
  {\bibfield  {journal} {\bibinfo  {journal} {Nat. Commun.}\ }\textbf {\bibinfo
  {volume} {13}},\ \bibinfo {pages} {5799} (\bibinfo {year}
  {2022})}\BibitemShut {NoStop}%
\bibitem [{\citenamefont {Samajdar}\ \emph {et~al.}(2022)\citenamefont
  {Samajdar}, \citenamefont {Joshi}, \citenamefont {Teng},\ and\ \citenamefont
  {Sachdev}}]{IGT}%
  \BibitemOpen
  \bibfield  {author} {\bibinfo {author} {\bibfnamefont {Rhine}\ \bibnamefont
  {Samajdar}}, \bibinfo {author} {\bibfnamefont {Darshan~G.}\ \bibnamefont
  {Joshi}}, \bibinfo {author} {\bibfnamefont {Yanting}\ \bibnamefont {Teng}}, \
  and\ \bibinfo {author} {\bibfnamefont {Subir}\ \bibnamefont {Sachdev}},\
  }\href@noop {} {\enquote {\bibinfo {title} {{Emergent $\mathbb{Z}_2$ gauge
  theories and topological excitations in Rydberg atom arrays}},}\ } (\bibinfo
  {year} {2022}),\ \Eprint {http://arxiv.org/abs/2204.00632} {arXiv:2204.00632
  [cond-mat.quant-gas]} \BibitemShut {NoStop}%
\bibitem [{\citenamefont {Verresen}\ \emph {et~al.}(2021)\citenamefont
  {Verresen}, \citenamefont {Lukin},\ and\ \citenamefont
  {Vishwanath}}]{Verresen2021}%
  \BibitemOpen
  \bibfield  {author} {\bibinfo {author} {\bibfnamefont {Ruben}\ \bibnamefont
  {Verresen}}, \bibinfo {author} {\bibfnamefont {Mikhail~D.}\ \bibnamefont
  {Lukin}}, \ and\ \bibinfo {author} {\bibfnamefont {Ashvin}\ \bibnamefont
  {Vishwanath}},\ }\bibfield  {title} {\enquote {\bibinfo {title} {{Prediction
  of toric code topological order from Rydberg blockade}},}\ }\href {\doibase
  10.1103/PhysRevX.11.031005} {\bibfield  {journal} {\bibinfo  {journal} {Phys.
  Rev. X}\ }\textbf {\bibinfo {volume} {11}},\ \bibinfo {pages} {031005}
  (\bibinfo {year} {2021})}\BibitemShut {NoStop}%
\bibitem [{\citenamefont {White}(1992)}]{White1992}%
  \BibitemOpen
  \bibfield  {author} {\bibinfo {author} {\bibfnamefont {Steven~R.}\
  \bibnamefont {White}},\ }\bibfield  {title} {\enquote {\bibinfo {title}
  {Density matrix formulation for quantum renormalization groups},}\ }\href
  {\doibase 10.1103/PhysRevLett.69.2863} {\bibfield  {journal} {\bibinfo
  {journal} {Phys. Rev. Lett.}\ }\textbf {\bibinfo {volume} {69}},\ \bibinfo
  {pages} {2863--2866} (\bibinfo {year} {1992})}\BibitemShut {NoStop}%
\bibitem [{\citenamefont {White}(1993)}]{White1993}%
  \BibitemOpen
  \bibfield  {author} {\bibinfo {author} {\bibfnamefont {Steven~R.}\
  \bibnamefont {White}},\ }\bibfield  {title} {\enquote {\bibinfo {title}
  {Density-matrix algorithms for quantum renormalization groups},}\ }\href
  {\doibase 10.1103/PhysRevB.48.10345} {\bibfield  {journal} {\bibinfo
  {journal} {Phys. Rev. B}\ }\textbf {\bibinfo {volume} {48}},\ \bibinfo
  {pages} {10345--10356} (\bibinfo {year} {1993})}\BibitemShut {NoStop}%
\bibitem [{Blo(2022)}]{Bloqade}%
  \BibitemOpen
  \href {https://github.com/QuEraComputing/Bloqade.jl} {\enquote {\bibinfo
  {title} {Bloqade, a package for the quantum computation and quantum dynamics
  based on neutral-atom architectures},}\ } (\bibinfo {year}
  {2022})\BibitemShut {NoStop}%
\bibitem [{\citenamefont {{Pichler}}\ \emph {et~al.}(2018)\citenamefont
  {{Pichler}}, \citenamefont {{Wang}}, \citenamefont {{Zhou}}, \citenamefont
  {{Choi}},\ and\ \citenamefont {{Lukin}}}]{Pichler2018MIS}%
  \BibitemOpen
  \bibfield  {author} {\bibinfo {author} {\bibfnamefont {Hannes}\ \bibnamefont
  {{Pichler}}}, \bibinfo {author} {\bibfnamefont {Sheng-Tao}\ \bibnamefont
  {{Wang}}}, \bibinfo {author} {\bibfnamefont {Leo}\ \bibnamefont {{Zhou}}},
  \bibinfo {author} {\bibfnamefont {Soonwon}\ \bibnamefont {{Choi}}}, \ and\
  \bibinfo {author} {\bibfnamefont {Mikhail~D.}\ \bibnamefont {{Lukin}}},\
  }\bibfield  {title} {\enquote {\bibinfo {title} {{Quantum optimization for
  maximum independent set using Rydberg atom arrays}},}\ }\href@noop {}
  {\bibfield  {journal} {\bibinfo  {journal} {arXiv e-prints}\ } (\bibinfo
  {year} {2018})},\ \Eprint {http://arxiv.org/abs/1808.10816} {arXiv:1808.10816
  [quant-ph]} \BibitemShut {NoStop}%
\bibitem [{\citenamefont {Ebadi}\ \emph {et~al.}(2022)\citenamefont {Ebadi},
  \citenamefont {Keesling}, \citenamefont {Cain}, \citenamefont {Wang},
  \citenamefont {Levine}, \citenamefont {Bluvstein}, \citenamefont {Semeghini},
  \citenamefont {Omran}, \citenamefont {Liu}, \citenamefont {Samajdar},
  \citenamefont {Luo}, \citenamefont {Nash}, \citenamefont {Gao}, \citenamefont
  {Barak}, \citenamefont {Farhi}, \citenamefont {Sachdev}, \citenamefont
  {Gemelke}, \citenamefont {Zhou}, \citenamefont {Choi}, \citenamefont
  {Pichler}, \citenamefont {Wang}, \citenamefont {Greiner}, \citenamefont
  {Vuletić},\ and\ \citenamefont {Lukin}}]{Ebadi2022}%
  \BibitemOpen
  \bibfield  {author} {\bibinfo {author} {\bibfnamefont {S.}~\bibnamefont
  {Ebadi}}, \bibinfo {author} {\bibfnamefont {A.}~\bibnamefont {Keesling}},
  \bibinfo {author} {\bibfnamefont {M.}~\bibnamefont {Cain}}, \bibinfo {author}
  {\bibfnamefont {T.~T.}\ \bibnamefont {Wang}}, \bibinfo {author}
  {\bibfnamefont {H.}~\bibnamefont {Levine}}, \bibinfo {author} {\bibfnamefont
  {D.}~\bibnamefont {Bluvstein}}, \bibinfo {author} {\bibfnamefont
  {G.}~\bibnamefont {Semeghini}}, \bibinfo {author} {\bibfnamefont
  {A.}~\bibnamefont {Omran}}, \bibinfo {author} {\bibfnamefont {J.-G.}\
  \bibnamefont {Liu}}, \bibinfo {author} {\bibfnamefont {R.}~\bibnamefont
  {Samajdar}}, \bibinfo {author} {\bibfnamefont {X.-Z.}\ \bibnamefont {Luo}},
  \bibinfo {author} {\bibfnamefont {B.}~\bibnamefont {Nash}}, \bibinfo {author}
  {\bibfnamefont {X.}~\bibnamefont {Gao}}, \bibinfo {author} {\bibfnamefont
  {B.}~\bibnamefont {Barak}}, \bibinfo {author} {\bibfnamefont
  {E.}~\bibnamefont {Farhi}}, \bibinfo {author} {\bibfnamefont
  {S.}~\bibnamefont {Sachdev}}, \bibinfo {author} {\bibfnamefont
  {N.}~\bibnamefont {Gemelke}}, \bibinfo {author} {\bibfnamefont
  {L.}~\bibnamefont {Zhou}}, \bibinfo {author} {\bibfnamefont {S.}~\bibnamefont
  {Choi}}, \bibinfo {author} {\bibfnamefont {H.}~\bibnamefont {Pichler}},
  \bibinfo {author} {\bibfnamefont {S.-T.}\ \bibnamefont {Wang}}, \bibinfo
  {author} {\bibfnamefont {M.}~\bibnamefont {Greiner}}, \bibinfo {author}
  {\bibfnamefont {V.}~\bibnamefont {Vuletić}}, \ and\ \bibinfo {author}
  {\bibfnamefont {M.~D.}\ \bibnamefont {Lukin}},\ }\bibfield  {title} {\enquote
  {\bibinfo {title} {{Quantum optimization of maximum independent set using
  Rydberg atom arrays}},}\ }\href {\doibase 10.1126/science.abo6587} {\bibfield
   {journal} {\bibinfo  {journal} {Science}\ }\textbf {\bibinfo {volume}
  {376}},\ \bibinfo {pages} {1209--1215} (\bibinfo {year} {2022})}\BibitemShut
  {NoStop}%
\bibitem [{SM()}]{SM}%
  \BibitemOpen
  \bibinfo {note} {See Supplementary Information below}\BibitemShut {NoStop}%
\bibitem [{\citenamefont {Polyakov}(1977)}]{Polyakov1977}%
  \BibitemOpen
  \bibfield  {author} {\bibinfo {author} {\bibfnamefont {A.M.}\ \bibnamefont
  {Polyakov}},\ }\bibfield  {title} {\enquote {\bibinfo {title} {Quark
  confinement and topology of gauge theories},}\ }\href {\doibase
  https://doi.org/10.1016/0550-3213(77)90086-4} {\bibfield  {journal} {\bibinfo
   {journal} {Nucl. Phys. B}\ }\textbf {\bibinfo {volume} {120}},\ \bibinfo
  {pages} {429--458} (\bibinfo {year} {1977})}\BibitemShut {NoStop}%
\bibitem [{\citenamefont {Zhang}\ \emph {et~al.}(2021)\citenamefont {Zhang},
  \citenamefont {Zhang}, \citenamefont {Fu},\ and\ \citenamefont
  {Kim}}]{Zhang2021}%
  \BibitemOpen
  \bibfield  {author} {\bibinfo {author} {\bibfnamefont {Kevin}\ \bibnamefont
  {Zhang}}, \bibinfo {author} {\bibfnamefont {Yang}\ \bibnamefont {Zhang}},
  \bibinfo {author} {\bibfnamefont {Liang}\ \bibnamefont {Fu}}, \ and\ \bibinfo
  {author} {\bibfnamefont {Eun-Ah}\ \bibnamefont {Kim}},\ }\href@noop {}
  {\enquote {\bibinfo {title} {{Fractional correlated insulating states at $n
  \pm 1/3$ filled magic angle twisted bilayer graphene}},}\ } (\bibinfo {year}
  {2021}),\ \Eprint {http://arxiv.org/abs/2105.13371} {arXiv:2105.13371
  [cond-mat.mes-hall]} \BibitemShut {NoStop}%
\bibitem [{\citenamefont {{Villain, J.}}\ \emph {et~al.}(1980)\citenamefont
  {{Villain, J.}}, \citenamefont {{Bidaux, R.}}, \citenamefont {{Carton,
  J.-P.}},\ and\ \citenamefont {{Conte, R.}}}]{Villain1980}%
  \BibitemOpen
  \bibfield  {author} {\bibinfo {author} {\bibnamefont {{Villain, J.}}},
  \bibinfo {author} {\bibnamefont {{Bidaux, R.}}}, \bibinfo {author}
  {\bibnamefont {{Carton, J.-P.}}}, \ and\ \bibinfo {author} {\bibnamefont
  {{Conte, R.}}},\ }\bibfield  {title} {\enquote {\bibinfo {title} {Order as an
  effect of disorder},}\ }\href {\doibase 10.1051/jphys:0198000410110126300}
  {\bibfield  {journal} {\bibinfo  {journal} {J. Phys. France}\ }\textbf
  {\bibinfo {volume} {41}},\ \bibinfo {pages} {1263--1272} (\bibinfo {year}
  {1980})}\BibitemShut {NoStop}%
\bibitem [{\citenamefont {Henley}(1989)}]{Henley1989}%
  \BibitemOpen
  \bibfield  {author} {\bibinfo {author} {\bibfnamefont {Christopher~L.}\
  \bibnamefont {Henley}},\ }\bibfield  {title} {\enquote {\bibinfo {title}
  {Ordering due to disorder in a frustrated vector antiferromagnet},}\ }\href
  {\doibase 10.1103/PhysRevLett.62.2056} {\bibfield  {journal} {\bibinfo
  {journal} {Phys. Rev. Lett.}\ }\textbf {\bibinfo {volume} {62}},\ \bibinfo
  {pages} {2056--2059} (\bibinfo {year} {1989})}\BibitemShut {NoStop}%
\bibitem [{\citenamefont {He}\ \emph {et~al.}(2017)\citenamefont {He},
  \citenamefont {Zaletel}, \citenamefont {Oshikawa},\ and\ \citenamefont
  {Pollmann}}]{He2017}%
  \BibitemOpen
  \bibfield  {author} {\bibinfo {author} {\bibfnamefont {Yin-Chen}\
  \bibnamefont {He}}, \bibinfo {author} {\bibfnamefont {Michael~P.}\
  \bibnamefont {Zaletel}}, \bibinfo {author} {\bibfnamefont {Masaki}\
  \bibnamefont {Oshikawa}}, \ and\ \bibinfo {author} {\bibfnamefont {Frank}\
  \bibnamefont {Pollmann}},\ }\bibfield  {title} {\enquote {\bibinfo {title}
  {{Signatures of Dirac cones in a DMRG study of the kagome Heisenberg
  model}},}\ }\href {\doibase 10.1103/PhysRevX.7.031020} {\bibfield  {journal}
  {\bibinfo  {journal} {Phys. Rev. X}\ }\textbf {\bibinfo {volume} {7}},\
  \bibinfo {pages} {031020} (\bibinfo {year} {2017})}\BibitemShut {NoStop}%
\bibitem [{\citenamefont {Ferrari}\ \emph {et~al.}(2021)\citenamefont
  {Ferrari}, \citenamefont {Parola},\ and\ \citenamefont
  {Becca}}]{Ferrari2021}%
  \BibitemOpen
  \bibfield  {author} {\bibinfo {author} {\bibfnamefont {Francesco}\
  \bibnamefont {Ferrari}}, \bibinfo {author} {\bibfnamefont {Alberto}\
  \bibnamefont {Parola}}, \ and\ \bibinfo {author} {\bibfnamefont {Federico}\
  \bibnamefont {Becca}},\ }\bibfield  {title} {\enquote {\bibinfo {title}
  {Gapless spin liquids in disguise},}\ }\href {\doibase
  10.1103/PhysRevB.103.195140} {\bibfield  {journal} {\bibinfo  {journal}
  {Phys. Rev. B}\ }\textbf {\bibinfo {volume} {103}},\ \bibinfo {pages}
  {195140} (\bibinfo {year} {2021})}\BibitemShut {NoStop}%
\bibitem [{\citenamefont {Jin}\ \emph {et~al.}(2022)\citenamefont {Jin},
  \citenamefont {Natori},\ and\ \citenamefont {Knolle}}]{Jin2022}%
  \BibitemOpen
  \bibfield  {author} {\bibinfo {author} {\bibfnamefont {Hui-Ke}\ \bibnamefont
  {Jin}}, \bibinfo {author} {\bibfnamefont {W.~M.~H.}\ \bibnamefont {Natori}},
  \ and\ \bibinfo {author} {\bibfnamefont {Johannes}\ \bibnamefont {Knolle}},\
  }\href@noop {} {\enquote {\bibinfo {title} {{Twisting the Dirac cones of the
  $SU(4)$ spin-orbital liquid on the honeycomb lattice}},}\ } (\bibinfo {year}
  {2022}),\ \Eprint {http://arxiv.org/abs/2207.02239} {arXiv:2207.02239
  [cond-mat.str-el]} \BibitemShut {NoStop}%
\bibitem [{\citenamefont {Sachdev}\ \emph {et~al.}(2002)\citenamefont
  {Sachdev}, \citenamefont {Sengupta},\ and\ \citenamefont
  {Girvin}}]{Sachdev2002}%
  \BibitemOpen
  \bibfield  {author} {\bibinfo {author} {\bibfnamefont {Subir}\ \bibnamefont
  {Sachdev}}, \bibinfo {author} {\bibfnamefont {K.}~\bibnamefont {Sengupta}}, \
  and\ \bibinfo {author} {\bibfnamefont {S.~M.}\ \bibnamefont {Girvin}},\
  }\bibfield  {title} {\enquote {\bibinfo {title} {Mott insulators in strong
  electric fields},}\ }\href {\doibase 10.1103/PhysRevB.66.075128} {\bibfield
  {journal} {\bibinfo  {journal} {Phys. Rev. B}\ }\textbf {\bibinfo {volume}
  {66}},\ \bibinfo {pages} {075128} (\bibinfo {year} {2002})}\BibitemShut
  {NoStop}%
\bibitem [{\citenamefont {Liu}\ \emph {et~al.}(2022)\citenamefont {Liu},
  \citenamefont {Gao}, \citenamefont {Cain}, \citenamefont {Lukin},\ and\
  \citenamefont {Wang}}]{Jinguo2022}%
  \BibitemOpen
  \bibfield  {author} {\bibinfo {author} {\bibfnamefont {Jin-Guo}\ \bibnamefont
  {Liu}}, \bibinfo {author} {\bibfnamefont {Xun}\ \bibnamefont {Gao}}, \bibinfo
  {author} {\bibfnamefont {Madelyn}\ \bibnamefont {Cain}}, \bibinfo {author}
  {\bibfnamefont {Mikhail~D.}\ \bibnamefont {Lukin}}, \ and\ \bibinfo {author}
  {\bibfnamefont {Sheng-Tao}\ \bibnamefont {Wang}},\ }\href@noop {} {\enquote
  {\bibinfo {title} {Computing solution space properties of combinatorial
  optimization problems via generic tensor networks},}\ } (\bibinfo {year}
  {2022}),\ \Eprint {http://arxiv.org/abs/2205.03718} {arXiv:2205.03718
  [cond-mat.stat-mech]} \BibitemShut {NoStop}%
\bibitem [{\citenamefont {Turner}\ \emph {et~al.}(2018)\citenamefont {Turner},
  \citenamefont {Michailidis}, \citenamefont {Abanin}, \citenamefont {Serbyn},\
  and\ \citenamefont {Papi{\'{c}}}}]{Turner2018}%
  \BibitemOpen
  \bibfield  {author} {\bibinfo {author} {\bibfnamefont {C.~J.}\ \bibnamefont
  {Turner}}, \bibinfo {author} {\bibfnamefont {A.~A.}\ \bibnamefont
  {Michailidis}}, \bibinfo {author} {\bibfnamefont {D.~A.}\ \bibnamefont
  {Abanin}}, \bibinfo {author} {\bibfnamefont {M.}~\bibnamefont {Serbyn}}, \
  and\ \bibinfo {author} {\bibfnamefont {Z.}~\bibnamefont {Papi{\'{c}}}},\
  }\bibfield  {title} {\enquote {\bibinfo {title} {Weak ergodicity breaking
  from quantum many-body scars},}\ }\href {\doibase 10.1038/s41567-018-0137-5}
  {\bibfield  {journal} {\bibinfo  {journal} {Nat. Phys.}\ }\textbf {\bibinfo
  {volume} {14}},\ \bibinfo {pages} {745--749} (\bibinfo {year}
  {2018})}\BibitemShut {NoStop}%
\bibitem [{\citenamefont {Bluvstein}\ \emph {et~al.}(2021)\citenamefont
  {Bluvstein}, \citenamefont {Omran}, \citenamefont {Levine}, \citenamefont
  {Keesling}, \citenamefont {Semeghini}, \citenamefont {Ebadi}, \citenamefont
  {Wang}, \citenamefont {Michailidis}, \citenamefont {Maskara}, \citenamefont
  {Ho}, \citenamefont {Choi}, \citenamefont {Serbyn}, \citenamefont {Greiner},
  \citenamefont {Vuletić},\ and\ \citenamefont {Lukin}}]{Bluvstein2021}%
  \BibitemOpen
  \bibfield  {author} {\bibinfo {author} {\bibfnamefont {D.}~\bibnamefont
  {Bluvstein}}, \bibinfo {author} {\bibfnamefont {A.}~\bibnamefont {Omran}},
  \bibinfo {author} {\bibfnamefont {H.}~\bibnamefont {Levine}}, \bibinfo
  {author} {\bibfnamefont {A.}~\bibnamefont {Keesling}}, \bibinfo {author}
  {\bibfnamefont {G.}~\bibnamefont {Semeghini}}, \bibinfo {author}
  {\bibfnamefont {S.}~\bibnamefont {Ebadi}}, \bibinfo {author} {\bibfnamefont
  {T.~T.}\ \bibnamefont {Wang}}, \bibinfo {author} {\bibfnamefont {A.~A.}\
  \bibnamefont {Michailidis}}, \bibinfo {author} {\bibfnamefont
  {N.}~\bibnamefont {Maskara}}, \bibinfo {author} {\bibfnamefont {W.~W.}\
  \bibnamefont {Ho}}, \bibinfo {author} {\bibfnamefont {S.}~\bibnamefont
  {Choi}}, \bibinfo {author} {\bibfnamefont {M.}~\bibnamefont {Serbyn}},
  \bibinfo {author} {\bibfnamefont {M.}~\bibnamefont {Greiner}}, \bibinfo
  {author} {\bibfnamefont {V.}~\bibnamefont {Vuletić}}, \ and\ \bibinfo
  {author} {\bibfnamefont {M.~D.}\ \bibnamefont {Lukin}},\ }\bibfield  {title}
  {\enquote {\bibinfo {title} {{Controlling quantum many-body dynamics in
  driven Rydberg atom arrays}},}\ }\href {\doibase 10.1126/science.abg2530}
  {\bibfield  {journal} {\bibinfo  {journal} {Science}\ }\textbf {\bibinfo
  {volume} {371}},\ \bibinfo {pages} {1355--1359} (\bibinfo {year}
  {2021})}\BibitemShut {NoStop}%
\bibitem [{\citenamefont {Surace}\ \emph {et~al.}(2020)\citenamefont {Surace},
  \citenamefont {Mazza}, \citenamefont {Giudici}, \citenamefont {Lerose},
  \citenamefont {Gambassi},\ and\ \citenamefont {Dalmonte}}]{Suarce2020}%
  \BibitemOpen
  \bibfield  {author} {\bibinfo {author} {\bibfnamefont {Federica~M.}\
  \bibnamefont {Surace}}, \bibinfo {author} {\bibfnamefont {Paolo~P.}\
  \bibnamefont {Mazza}}, \bibinfo {author} {\bibfnamefont {Giuliano}\
  \bibnamefont {Giudici}}, \bibinfo {author} {\bibfnamefont {Alessio}\
  \bibnamefont {Lerose}}, \bibinfo {author} {\bibfnamefont {Andrea}\
  \bibnamefont {Gambassi}}, \ and\ \bibinfo {author} {\bibfnamefont {Marcello}\
  \bibnamefont {Dalmonte}},\ }\bibfield  {title} {\enquote {\bibinfo {title}
  {{Lattice gauge theories and string dynamics in Rydberg atom quantum
  simulators}},}\ }\href {\doibase 10.1103/PhysRevX.10.021041} {\bibfield
  {journal} {\bibinfo  {journal} {Phys. Rev. X}\ }\textbf {\bibinfo {volume}
  {10}},\ \bibinfo {pages} {021041} (\bibinfo {year} {2020})}\BibitemShut
  {NoStop}%
\bibitem [{\citenamefont {Giudici}\ \emph {et~al.}(2022)\citenamefont
  {Giudici}, \citenamefont {Lukin},\ and\ \citenamefont
  {Pichler}}]{Giudici2022}%
  \BibitemOpen
  \bibfield  {author} {\bibinfo {author} {\bibfnamefont {Giuliano}\
  \bibnamefont {Giudici}}, \bibinfo {author} {\bibfnamefont {Mikhail~D.}\
  \bibnamefont {Lukin}}, \ and\ \bibinfo {author} {\bibfnamefont {Hannes}\
  \bibnamefont {Pichler}},\ }\bibfield  {title} {\enquote {\bibinfo {title}
  {{Dynamical preparation of quantum spin liquids in Rydberg atom arrays}},}\
  }\href {\doibase 10.1103/PhysRevLett.129.090401} {\bibfield  {journal}
  {\bibinfo  {journal} {Phys. Rev. Lett.}\ }\textbf {\bibinfo {volume} {129}},\
  \bibinfo {pages} {090401} (\bibinfo {year} {2022})}\BibitemShut {NoStop}%
\bibitem [{Note1()}]{Note1}%
  \BibitemOpen
  \bibinfo {note} {Private communication, Harvard atom array team.}\BibitemShut
  {Stop}%
\bibitem [{\citenamefont {Nguyen}\ \emph {et~al.}(2022)\citenamefont {Nguyen},
  \citenamefont {Liu}, \citenamefont {Wurtz}, \citenamefont {Lukin},
  \citenamefont {Wang},\ and\ \citenamefont {Pichler}}]{Nguyen2022}%
  \BibitemOpen
  \bibfield  {author} {\bibinfo {author} {\bibfnamefont {Minh-Thi}\
  \bibnamefont {Nguyen}}, \bibinfo {author} {\bibfnamefont {Jin-Guo}\
  \bibnamefont {Liu}}, \bibinfo {author} {\bibfnamefont {Jonathan}\
  \bibnamefont {Wurtz}}, \bibinfo {author} {\bibfnamefont {Mikhail~D.}\
  \bibnamefont {Lukin}}, \bibinfo {author} {\bibfnamefont {Sheng-Tao}\
  \bibnamefont {Wang}}, \ and\ \bibinfo {author} {\bibfnamefont {Hannes}\
  \bibnamefont {Pichler}},\ }\href@noop {} {\enquote {\bibinfo {title}
  {{Quantum optimization with arbitrary connectivity using Rydberg atom
  arrays}},}\ } (\bibinfo {year} {2022}),\ \Eprint
  {http://arxiv.org/abs/2209.03965} {arXiv:2209.03965 [quant-ph]} \BibitemShut
  {NoStop}%
\bibitem [{\citenamefont {Villain}(1980)}]{Villain1980b}%
  \BibitemOpen
  \bibfield  {author} {\bibinfo {author} {\bibfnamefont {Jacques}\ \bibnamefont
  {Villain}},\ }\bibfield  {title} {\enquote {\bibinfo {title}
  {Commensurate-incommensurate transition of krypton monolayers on graphite: A
  low temperature theory},}\ }\href {\doibase
  https://doi.org/10.1016/0039-6028(80)90115-6} {\bibfield  {journal} {\bibinfo
   {journal} {Surface Science}\ }\textbf {\bibinfo {volume} {97}},\ \bibinfo
  {pages} {219--242} (\bibinfo {year} {1980})}\BibitemShut {NoStop}%
\bibitem [{\citenamefont {Coppersmith}\ \emph {et~al.}(1982)\citenamefont
  {Coppersmith}, \citenamefont {Fisher}, \citenamefont {Halperin},
  \citenamefont {Lee},\ and\ \citenamefont {Brinkman}}]{Coppersmith1982}%
  \BibitemOpen
  \bibfield  {author} {\bibinfo {author} {\bibfnamefont {S.~N.}\ \bibnamefont
  {Coppersmith}}, \bibinfo {author} {\bibfnamefont {Daniel~S.}\ \bibnamefont
  {Fisher}}, \bibinfo {author} {\bibfnamefont {B.~I.}\ \bibnamefont
  {Halperin}}, \bibinfo {author} {\bibfnamefont {P.~A.}\ \bibnamefont {Lee}}, \
  and\ \bibinfo {author} {\bibfnamefont {W.~F.}\ \bibnamefont {Brinkman}},\
  }\bibfield  {title} {\enquote {\bibinfo {title} {Dislocations and the
  commensurate-incommensurate transition in two dimensions},}\ }\href {\doibase
  10.1103/PhysRevB.25.349} {\bibfield  {journal} {\bibinfo  {journal} {Phys.
  Rev. B}\ }\textbf {\bibinfo {volume} {25}},\ \bibinfo {pages} {349--363}
  (\bibinfo {year} {1982})}\BibitemShut {NoStop}%
\bibitem [{\citenamefont {McCulloch}(2008)}]{iDMRG}%
  \BibitemOpen
  \bibfield  {author} {\bibinfo {author} {\bibfnamefont {I.~P.}\ \bibnamefont
  {McCulloch}},\ }\href@noop {} {\enquote {\bibinfo {title} {Infinite size
  density matrix renormalization group, revisited},}\ } (\bibinfo {year}
  {2008}),\ \Eprint {http://arxiv.org/abs/0804.2509} {arXiv:0804.2509
  [cond-mat.str-el]} \BibitemShut {NoStop}%
\bibitem [{\citenamefont {Knap}\ \emph {et~al.}(2013)\citenamefont {Knap},
  \citenamefont {Kantian}, \citenamefont {Giamarchi}, \citenamefont {Bloch},
  \citenamefont {Lukin},\ and\ \citenamefont {Demler}}]{Knap2013}%
  \BibitemOpen
  \bibfield  {author} {\bibinfo {author} {\bibfnamefont {Michael}\ \bibnamefont
  {Knap}}, \bibinfo {author} {\bibfnamefont {Adrian}\ \bibnamefont {Kantian}},
  \bibinfo {author} {\bibfnamefont {Thierry}\ \bibnamefont {Giamarchi}},
  \bibinfo {author} {\bibfnamefont {Immanuel}\ \bibnamefont {Bloch}}, \bibinfo
  {author} {\bibfnamefont {Mikhail~D.}\ \bibnamefont {Lukin}}, \ and\ \bibinfo
  {author} {\bibfnamefont {Eugene}\ \bibnamefont {Demler}},\ }\bibfield
  {title} {\enquote {\bibinfo {title} {{Probing real-space and time-resolved
  correlation functions with many-body Ramsey interferometry}},}\ }\href
  {\doibase 10.1103/PhysRevLett.111.147205} {\bibfield  {journal} {\bibinfo
  {journal} {Phys. Rev. Lett.}\ }\textbf {\bibinfo {volume} {111}},\ \bibinfo
  {pages} {147205} (\bibinfo {year} {2013})}\BibitemShut {NoStop}%
\bibitem [{\citenamefont {Baez}\ \emph {et~al.}(2020)\citenamefont {Baez},
  \citenamefont {Goihl}, \citenamefont {Haferkamp}, \citenamefont
  {Bermejo-Vega}, \citenamefont {Gluza},\ and\ \citenamefont
  {Eisert}}]{Baez2020}%
  \BibitemOpen
  \bibfield  {author} {\bibinfo {author} {\bibfnamefont {Maria~Laura}\
  \bibnamefont {Baez}}, \bibinfo {author} {\bibfnamefont {Marcel}\ \bibnamefont
  {Goihl}}, \bibinfo {author} {\bibfnamefont {Jonas}\ \bibnamefont
  {Haferkamp}}, \bibinfo {author} {\bibfnamefont {Juani}\ \bibnamefont
  {Bermejo-Vega}}, \bibinfo {author} {\bibfnamefont {Marek}\ \bibnamefont
  {Gluza}}, \ and\ \bibinfo {author} {\bibfnamefont {Jens}\ \bibnamefont
  {Eisert}},\ }\bibfield  {title} {\enquote {\bibinfo {title} {Dynamical
  structure factors of dynamical quantum simulators},}\ }\href {\doibase
  10.1073/pnas.2006103117} {\bibfield  {journal} {\bibinfo  {journal} {Proc.
  Natl. Acad. Sci. U.S.A.}\ }\textbf {\bibinfo {volume} {117}},\ \bibinfo
  {pages} {26123--26134} (\bibinfo {year} {2020})}\BibitemShut {NoStop}%
\bibitem [{\citenamefont {Bluvstein}\ \emph {et~al.}(2022)\citenamefont
  {Bluvstein}, \citenamefont {Levine}, \citenamefont {Semeghini}, \citenamefont
  {Wang}, \citenamefont {Ebadi}, \citenamefont {Kalinowski}, \citenamefont
  {Keesling}, \citenamefont {Maskara}, \citenamefont {Pichler}, \citenamefont
  {Greiner}, \citenamefont {Vuleti{\'{c}}},\ and\ \citenamefont
  {Lukin}}]{Bluvstein2022}%
  \BibitemOpen
  \bibfield  {author} {\bibinfo {author} {\bibfnamefont {Dolev}\ \bibnamefont
  {Bluvstein}}, \bibinfo {author} {\bibfnamefont {Harry}\ \bibnamefont
  {Levine}}, \bibinfo {author} {\bibfnamefont {Giulia}\ \bibnamefont
  {Semeghini}}, \bibinfo {author} {\bibfnamefont {Tout~T.}\ \bibnamefont
  {Wang}}, \bibinfo {author} {\bibfnamefont {Sepehr}\ \bibnamefont {Ebadi}},
  \bibinfo {author} {\bibfnamefont {Marcin}\ \bibnamefont {Kalinowski}},
  \bibinfo {author} {\bibfnamefont {Alexander}\ \bibnamefont {Keesling}},
  \bibinfo {author} {\bibfnamefont {Nishad}\ \bibnamefont {Maskara}}, \bibinfo
  {author} {\bibfnamefont {Hannes}\ \bibnamefont {Pichler}}, \bibinfo {author}
  {\bibfnamefont {Markus}\ \bibnamefont {Greiner}}, \bibinfo {author}
  {\bibfnamefont {Vladan}\ \bibnamefont {Vuleti{\'{c}}}}, \ and\ \bibinfo
  {author} {\bibfnamefont {Mikhail~D.}\ \bibnamefont {Lukin}},\ }\bibfield
  {title} {\enquote {\bibinfo {title} {{A quantum processor based on coherent
  transport of entangled atom arrays}},}\ }\href {\doibase
  10.1038/s41586-022-04592-6} {\bibfield  {journal} {\bibinfo  {journal}
  {Nature}\ }\textbf {\bibinfo {volume} {604}},\ \bibinfo {pages} {451--456}
  (\bibinfo {year} {2022})}\BibitemShut {NoStop}%
\bibitem [{\citenamefont {Yang}\ and\ \citenamefont {Xu}(2022)}]{Sheng2022}%
  \BibitemOpen
  \bibfield  {author} {\bibinfo {author} {\bibfnamefont {Sheng}\ \bibnamefont
  {Yang}}\ and\ \bibinfo {author} {\bibfnamefont {Jing-Bo}\ \bibnamefont
  {Xu}},\ }\bibfield  {title} {\enquote {\bibinfo {title}
  {{Density-wave-ordered phases of Rydberg atoms on a honeycomb lattice}},}\
  }\href {\doibase 10.1103/PhysRevE.106.034121} {\bibfield  {journal} {\bibinfo
   {journal} {Phys. Rev. E}\ }\textbf {\bibinfo {volume} {106}},\ \bibinfo
  {pages} {034121} (\bibinfo {year} {2022})}\BibitemShut {NoStop}%
\bibitem [{\citenamefont {Shibata}\ \emph {et~al.}(1996)\citenamefont
  {Shibata}, \citenamefont {Ueda}, \citenamefont {Nishino},\ and\ \citenamefont
  {Ishii}}]{Shibata1996}%
  \BibitemOpen
  \bibfield  {author} {\bibinfo {author} {\bibfnamefont {Naokazu}\ \bibnamefont
  {Shibata}}, \bibinfo {author} {\bibfnamefont {Kazuo}\ \bibnamefont {Ueda}},
  \bibinfo {author} {\bibfnamefont {Tomotoshi}\ \bibnamefont {Nishino}}, \ and\
  \bibinfo {author} {\bibfnamefont {Chikara}\ \bibnamefont {Ishii}},\
  }\bibfield  {title} {\enquote {\bibinfo {title} {{Friedel oscillations in the
  one-dimensional Kondo lattice model}},}\ }\href {\doibase
  10.1103/PhysRevB.54.13495} {\bibfield  {journal} {\bibinfo  {journal} {Phys.
  Rev. B}\ }\textbf {\bibinfo {volume} {54}},\ \bibinfo {pages} {13495--13498}
  (\bibinfo {year} {1996})}\BibitemShut {NoStop}%
\bibitem [{\citenamefont {Bed\"urftig}\ \emph {et~al.}(1998)\citenamefont
  {Bed\"urftig}, \citenamefont {Brendel}, \citenamefont {Frahm},\ and\
  \citenamefont {Noack}}]{Bedurftig1998}%
  \BibitemOpen
  \bibfield  {author} {\bibinfo {author} {\bibfnamefont {G.}~\bibnamefont
  {Bed\"urftig}}, \bibinfo {author} {\bibfnamefont {B.}~\bibnamefont
  {Brendel}}, \bibinfo {author} {\bibfnamefont {H.}~\bibnamefont {Frahm}}, \
  and\ \bibinfo {author} {\bibfnamefont {R.~M.}\ \bibnamefont {Noack}},\
  }\bibfield  {title} {\enquote {\bibinfo {title} {{Friedel oscillations in the
  open Hubbard chain}},}\ }\href {\doibase 10.1103/PhysRevB.58.10225}
  {\bibfield  {journal} {\bibinfo  {journal} {Phys. Rev. B}\ }\textbf {\bibinfo
  {volume} {58}},\ \bibinfo {pages} {10225--10235} (\bibinfo {year}
  {1998})}\BibitemShut {NoStop}%
\bibitem [{\citenamefont {Feldmeier}\ \emph {et~al.}(2019)\citenamefont
  {Feldmeier}, \citenamefont {Pollmann},\ and\ \citenamefont
  {Knap}}]{Feldmeier2019}%
  \BibitemOpen
  \bibfield  {author} {\bibinfo {author} {\bibfnamefont {Johannes}\
  \bibnamefont {Feldmeier}}, \bibinfo {author} {\bibfnamefont {Frank}\
  \bibnamefont {Pollmann}}, \ and\ \bibinfo {author} {\bibfnamefont {Michael}\
  \bibnamefont {Knap}},\ }\bibfield  {title} {\enquote {\bibinfo {title}
  {Emergent glassy dynamics in a quantum dimer model},}\ }\href {\doibase
  10.1103/PhysRevLett.123.040601} {\bibfield  {journal} {\bibinfo  {journal}
  {Phys. Rev. Lett.}\ }\textbf {\bibinfo {volume} {123}},\ \bibinfo {pages}
  {040601} (\bibinfo {year} {2019})}\BibitemShut {NoStop}%
\end{thebibliography}%

\clearpage
\onecolumngrid
\begin{center}
{\bf Supplementary Information: ``Trimer quantum spin liquid in a honeycomb array of Rydberg atoms''}
\end{center}
\vspace{.5cm}
\linespread{1.025}

In this Supplementary Information, we discuss in more detail the experimental detection of the TQSL state and present additional results from both DMRG and ED calculations supporting the TQSL character of the state and its robustness on a wide range of geometries. In \cref{sec:A}, we provide details of the trimer mapping relevant for the experimental detection of the TQSL;  \cref{sec:B} and \cref{sec:C} present additional DMRG and ED results, respectively; \cref{sec:D} includes additional data on the dynamical preparation of the TQSL state; finally, \cref{sec:E} shows the numerical results for the second-neighbor-blockaded PXP model for comparison.

\section{Trimer model mapping and experimental signatures of the TQSL}
\label{sec:A}

\begin{figure}[htb]
\vspace*{-0.0cm}
\centering
\includegraphics[width=1.0\textwidth]{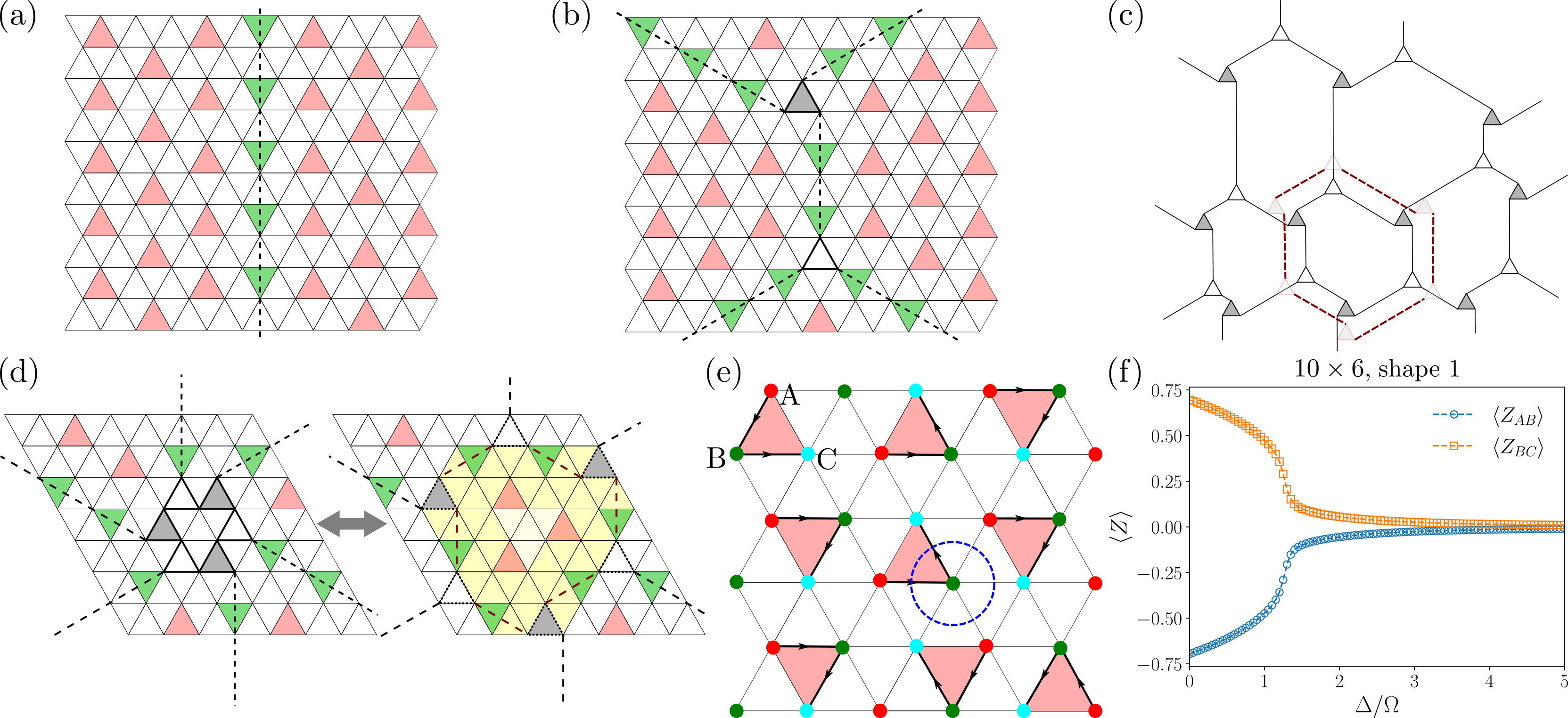}
\vskip -0.cm
\caption{\textbf{Trimer model mapping and experimental probes of the TQSL.} (a) To classify the trimer degeneracy on the triangular lattice \cite{Verberkmoes1999, Zhang2021, Villain1980b} one can start from an ordered trimer state and note that flipping a line of trimers still leads to a valid trimer covering. (b) The same is true for a confluence of three flipped lines ending with a filled (shaded) or empty (unshaded) trimer. (c) All the trimer configurations reachable from the original ordered state can then be described by any irregular honeycomb tiling of the plane. The number of such configurations grows exponentially with the system size. This can be seen by noting that each of the irregular honeycombs has a ``breathing'' degree of freedom---it can contract and expand while keeping its aspect ratio. Such a move is local (shown in red), changing only the configuration in the vicinity of the breathing irregular honeycomb, making it a potentially useful experimental probe of the resonance between different trimer configurations. (d) This experimental probe can be explicitly constructed for a given breathing move as it is equivalent to a product of $\sigma_{x}$ operators needed to expand the loop. An example of a simple breathing move is shown with configurations before and after the move differing only within the yellow shell around the perimeter of the loop. The moves needed to expand the honeycomb, in this case, lead to the many-body $X$-loop operator that is a product of 24 individual honeycomb $\sigma_{x}$ operators. (e) The main experimental probe that definitively distinguishes between trivial and TQSL phases entails checking the two $\mathrm{U}(1)$ conservation laws \cite{Giudice2022} relating the electric field (arrows) flux with enclosed charges (number of different sublattice sites) along an arbitrary closed loop (example shown in blue). (f) Two $Z$-loop operators evaluated for the loop shown in blue in (e) for the $10 \times 6$ shape 1 cluster from Fig.~\ref{fig:Fig3_fidelity}(a) of the main text. The two $\mathrm{U}(1)$ conservation laws are violated in the trivial phase, but are approximately satisfied in both the TQSL and VBS phases.
}
\label{fig:FigS1a_trimers}
\end{figure}

We start by presenting the details of the trimer model mapping relevant to the experimental detection of the TQSL phase. This actually turns out to be closely related to the question of classifying the degenerate trimer configurations. It is known \cite{Verberkmoes1999, Zhang2021, Villain1980b} that all the trimer configurations can be represented by a corresponding irregular honeycomb lattice that tiles the plane, as shown in Fig.~\ref{fig:FigS1a_trimers}(a--c). The exponential degeneracy of trimers is then directly related to the ``breathing'' degree of freedom for each of the irregular honeycombs: the freedom to contract and expand while keeping the aspect ratio fixed. Such a move is local, as it only involves changes in the immediate vicinity of the breathing honeycomb. This has an important consequence for the construction of experimental probes of the TQSL state. In particular, a local operator comprised of spin flips ($\prod_i\sigma^x_i$) that corresponds to a particular breathing move will, in general, permute within the classes of states in the MIS subspace. Thus, measuring the expectation value of different breathing operators, explicitly constructed from irregular honeycomb mappings, can distinguish between the TQSL and ordered trimer phases. The breathing operators are analogous to the $X$-loops for the ruby lattice \cite{Semeghini2021}.  They also naturally require measurements in the $X$-basis, necessitating a global basis rotation performed after a quench to noninteracting atom regimes or employing hyperfine mapping and single-qubit rotations \cite{Bluvstein2022}. 

The $X$-loop breathing operators are expected to decay rapidly with increasing loop and system size due to the structure of the trimer subspace and monomer fluctuations. The fluctuation-induced decay is expected to scale exponentially with the loop perimeter only in the TQSL phase, in complete analogy to the ruby-lattice $X$-loops \cite{Verresen2021, Semeghini2021}. The perimeter-law scaling is showcased for a breathing move in Fig.~\ref{fig:FigS1a_trimers}(d), where any difference between two loop-connected configurations is limited to the yellow-shaded region of two triangular layers around the loop's perimeter. The additional prefactor for the $X$-loop's intensity is related to the trimer subspace's fractionalization into sectors described by topologically distinct irregular honeycomb lattices \cite{Villain1980b}. This prefactor is, in the worst case, proportional to the inverse of the number of topological classes, scaling as $1/L^2$ for system size $L$ \cite{Villain1980b}. The constant nature of the prefactor for a given system size and shape still allows for probing  the perimeter-law scaling of the $X$-loops, and thus, the TQSL phase.

A second set of operators can be used to experimentally distinguish between the trivial and TQSL states, allowing for a definite experimental detection of the TQSL phase in conjunction with the breathing operators. These operators check the two $\mathrm{U}(1)$ conservation laws \cite{Giudice2022} by evaluating the electric flux and conserved charges for an arbitrary closed loop, as illustrated in Fig.~\ref{fig:FigS1a_trimers}(d). For a given snapshot of the honeycomb lattice, a related triangular lattice can be defined, tripartitioned, and covered with trimers. Electric fields related to the associated $\mathrm{U}(1)$ gauge degrees of freedom can then be assigned on the trimers, as described in the main text, following which, their flux ($\Phi$) can be evaluated and compared to the enclosed charge. For A--B electric fields, $Z_{AB}=\Phi_{AB}-(N_A-N_B)$ equals 0 for the perfect trimer covering, and likewise for B--C. Thus, an expectation value close to $0$ of these $Z$-basis-accessible operators on arbitrary closed loops distinguishes between trivial and trimer states. These operators are direct analogs of the closed $Z$-loops \cite{Semeghini2021} employed for detection of the $\mathbb{Z}_2$ topological phase on the ruby lattice.

As an example, we numerically evaluate the $Z$-loop operators across the phase diagram of the $10 \times 6$ shape 1 cluster, showcased earlier in Fig.~\ref{fig:Fig3_fidelity}(a) of the main text. In particular, $Z_{AB}$ and $Z_{BC}$ are calculated for the $Z$-loop presented in Fig.~\ref{fig:FigS1a_trimers}(d) and shown in Fig.~\ref{fig:FigS1a_trimers}(f). The expectation values for both of the loop operators are high in the trivial phase and approach the expected value of zero in the TQSL phase. The second transition between the TQSL and the VBS present on this cluster is not distinguishable from $Z$-loop measurements alone. 

\section{Additional DMRG results}
\label{sec:B}

\subsection{Structure factors of the phases}
\begin{figure}[htb]
\vspace*{-0.0cm}
\centering
\includegraphics[width=0.9\columnwidth]{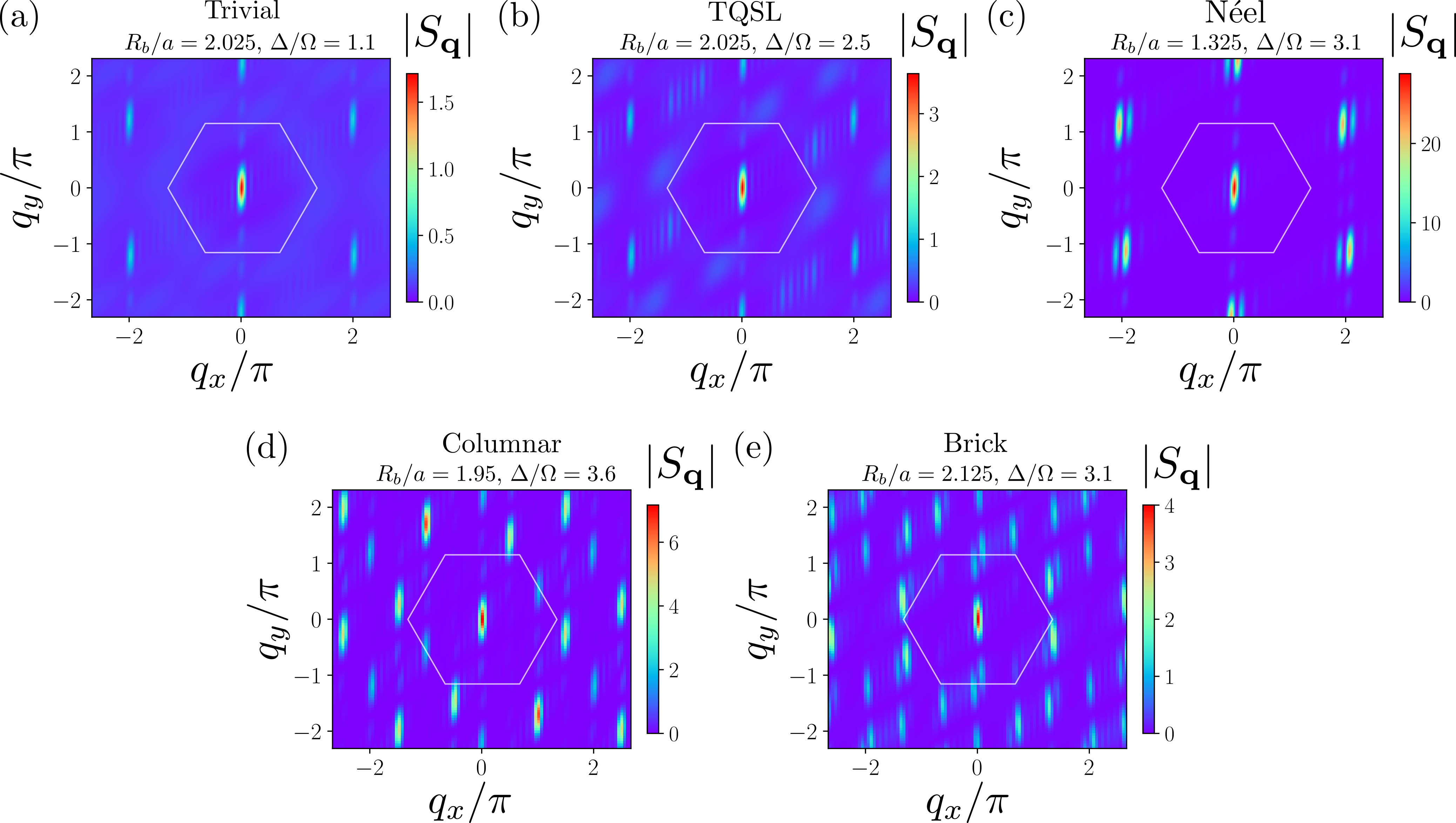}
\vskip -0.cm
\caption{\textbf{Structure factors of the phases.} Static structure factors for different regions of the phase diagram, as obtained by DMRG, are shown. The first Brillouin zone is denoted by a white hexagon. (a) The structure factor for the trivial disordered phase displays only a trivial peak at the $\Gamma$ point. (b) The same is true in the TQSL regime (indicating the absence of order) along with additional broad features stemming from the trimer constraint being fulfilled. (c) For the N\'eel state---where the order is present within the honeycomb unit cell and does not break the translational symmetry of the lattice---sharp Bragg peaks are visible in the second Brillouin zone. In contrast, the columnar (d) and brick (e) phases break translational symmetry and present additional nontrivial ordering peaks.
}
\label{fig:FigS2_dmrgSF}
\vskip -0.cm
\end{figure}

The static structure factor quantifies the correlations between Rydberg excitations in the various phases and presents a direct experimental classification tool for the density-wave-ordered ones. The structure factors were calculated as the Fourier transform of density-density correlations:
\begin{equation} \label{eq:SF}
    S_{\mathbf{q}}=\frac{1}{L}\sum_{i,j} e^{-i \mathbf{q}\cdot \left(\mathbf{r}_i-\mathbf{r}_j\right)}\langle n_i n_j\rangle,
\end{equation}
where $\mathbf{r}_i$ denote the positions of the atoms. The results are presented in Fig.~\ref{fig:FigS2_dmrgSF} for representative points in the different phase regions with the same parameters as in Fig.~\ref{fig:Fig2_states} of the main text. For all the structure factors shown, one has to take into account the effect of finite-size clusters with open boundary conditions as well as the choice of one of the degenerate ordered configurations made by DMRG due to boundary conditions that explicitly break the point group symmetries of the lattice. In the experimental scenario, with a larger system, different ordering patterns may occur in different snapshots, averaging out to a symmetric structure factor.

From Fig.~\ref{fig:FigS2_dmrgSF}, we see that the disordered phase, as expected, exhibits only a trivial peak at the $\Gamma$ point of the first Brillouin zone defined by the reciprocal lattice vectors, $\mathbf{b}_1$\,$=$\,$2\pi(-1, -1/\sqrt{3})$ and $\mathbf{b}_2$\,$=$\,$2\pi(0, 2/\sqrt{3})$. This is similar to the structure factor in the TQSL regime, which too shows a complete absence of sharp ordering peaks. On top of the expected $\Gamma$-point peak, broad features can also be seen in the structure factor, stemming from the short-range correlations induced by the strong $k$\,$=$\,$3$ trimer constraint. 

In contrast, the structure factor of the N\'eel phase displays strong Bragg peaks in the second Brillouin zone because the N\'eel ordering occurs \textit{within} the unit cell of the honeycomb lattice. The doubling of the peaks in the second Brillouin zone is a consequence of the two domains present in our numerically obtained N\'eel state. The columnar and the brick phases, on the other hand, break the translational symmetry of the lattice and can be characterized by considering the structure factor peaks inside the first Brillouin zone. For the columnar phase, the main ordering peaks occur at $\pm (\mathbf{b}_1/2+\mathbf{b}_2)$ with two additional sets of secondary peaks, the first appearing at $\pm (3\mathbf{b}_1/4+\mathbf{b}_2/2)$ and $\pm (-\mathbf{b}_1/4+\mathbf{b}_2)$, while the second is at $\pm \mathbf{b}_1/2$. All of these peaks are expected from the real-space structure of the columnar phase. Finally, the main ordering peaks in the brick phase are at $\pm (2\mathbf{b}_1/3+\mathbf{b}_2/2)$ and $\pm 2\mathbf{b}_1/3$, with secondary peaks at $\pm (\mathbf{b}_1/3+\mathbf{b}_2)$, $\pm (\mathbf{b}_1/3+\mathbf{b}_2/2)$, and $\pm\mathbf{b}_2/2$. The brick-phase peaks are all consistent with the real-space ordering observed and show doubling due to the presence of two distinct domains. 

\subsection{Gap and density fluctuations in the TQSL regime \label{sec:gap_density}}

\begin{figure}[tb]
\vspace*{-0.0cm}
\centering
\includegraphics[width=1.0\columnwidth]{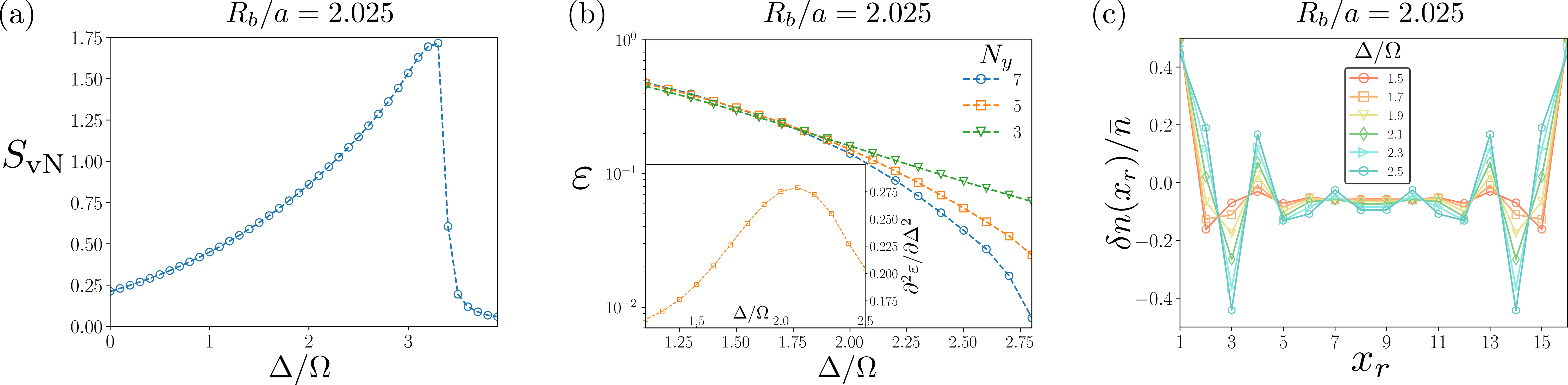}
\vskip -0.cm
\caption{\textbf{Properties of the TQSL region in DMRG.} (a) In contrast to the ordered phases, the TQSL region does not show a peak in von Neumann entanglement entropy and is characterized by a large entanglement entropy. (b) 
Going from trivial disordered phase to the TQSL region, one detects a strong decay of the gap as a function of $\Delta/\Omega$ on the one hand, as well as a change in the behavior with increasing transverse system size. While in the trivial phase, the gap is approximately constant for different system sizes, in the TQSL region, the gap generally drops with increasing system size, which might hint at a gapless state in the thermodynamic limit. Inset:
The second derivative of the gap ($N_y=5$) also displays a peak akin to the energy and fidelity susceptibilities. (c) Similarly, on a plot of the column density fluctuations, a strong susceptibility towards fluctuations is seen  on going from the trivial phase ($\Delta/\Omega<1.9$) to larger $\Delta/\Omega$. Although these fluctuations decay faster than the expected Friedel oscillations for a truly gapless state, the result is broadly consistent with a gapless state in the thermodynamic limit.
}
\label{fig:FigS3_dmrgTQSL}
\vskip -0.cm
\end{figure}

In this section, we present some additional properties of the TQSL region observed with DMRG that hint at the possibly gapless character of the state in the thermodynamic limit. Figure~\ref{fig:FigS3_dmrgTQSL}(a) shows the entanglement entropy, which attains large values in the TQSL region, with the only phase transition discernible being that into the ordered (columnar) phase. In Fig.~\ref{fig:FigS3_dmrgTQSL}(b), we show the energy gap, $\varepsilon$, as a function of the detuning for several system sizes across the trivial and TQSL regions. The gap is generically expected to be nonzero in a finite system like the one studied. In general, a gapless spin liquid on finite clusters without specially engineered boundary conditions and flux insertion (inaccessible in this case) becomes gapped  \cite{He2017, Ferrari2021, Jin2022}. Still, a qualitative distinction is apparent between the trivial phase, where the gap is independent of the transverse size, and the TQSL region, where the gap sharply drops with increasing transverse size, alluding to a potentially gapless state in the limit of infinite transverse size. True extrapolation to the thermodynamic limit is presently not viable with this dataset of small transverse sizes and increasing the transverse size is, unfortunately, exponentially costly with respect to the DMRG bond dimension. Interestingly, the second derivative of the gap with respect to the detuning also exhibits a peak consistent with those in the energy and fidelity susceptibilities, as shown in the inset of  Fig.~\ref{fig:FigS3_dmrgTQSL}(b).

Another interesting feature of the ground state in the TQSL regime is seen in the column-averaged density oscillations presented in Fig.~\ref{fig:FigS3_dmrgTQSL}(c). The system, for a given $N_x$, hosts $N_x/2$ columns, and the relative density oscillations across the columns compared to the average density are shown. Going from the trivial phase at $\Delta/\Omega=1.5$ to $\Delta/\Omega=2.5$, one notices a stark increase in the density oscillations' magnitude in both the bulk and at the boundaries. These density oscillations are induced by the open boundaries \cite{Shibata1996, Bedurftig1998} and in a true gapless state, they correspond to the Friedel oscillations expected to decay with a power law. In the small transverse-size systems we employ, an exponential (rather than power-law) decay is observed in the TQSL regime. However, as the gap is reduced, the oscillations are enhanced in the bulk. The possibility of an increase in susceptibility towards boundary-induced oscillations being a signature of the TQSL state is compounded by the observation of the same feature in the cylindrical clusters studied with ED and presented in Sec.~\ref{sec:OBC_ED}.

\section{Exact diagonalization of the \texorpdfstring{$k$\,$=$\,$3$}{k=3} PXP model}
\label{sec:C}

\subsection{Dependence on system size and shape \label{sec:GS_size}}

\begin{figure}[htb]
\vspace*{-0.0cm}
\centering
\includegraphics[width=1.0\columnwidth]{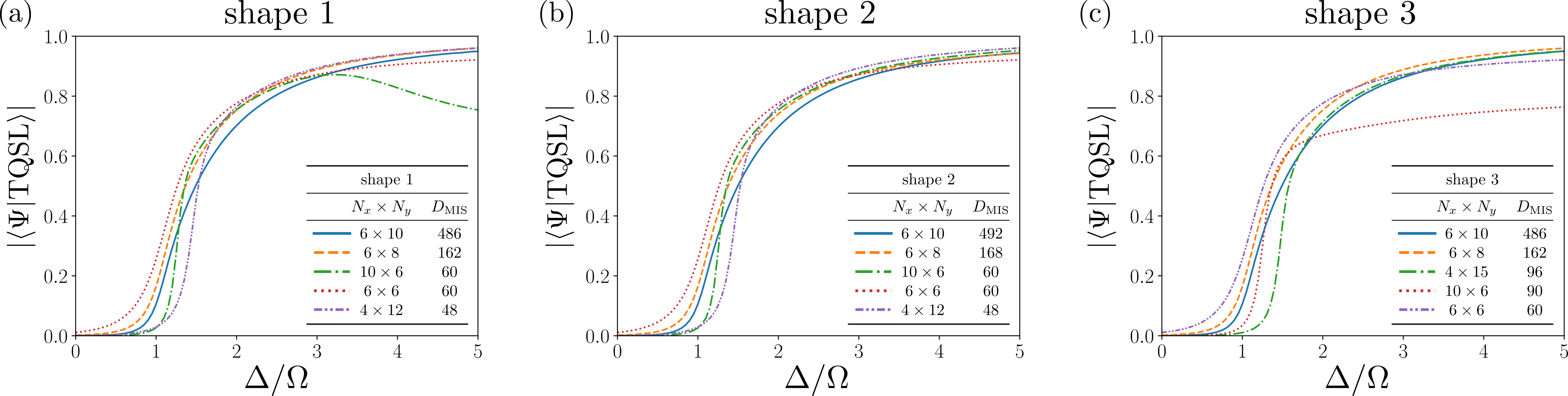}
\vskip -0.cm
\caption{\textbf{Ground-state fidelity for various system shapes and sizes.} The ground-state TQSL overlap  is plotted as a function of $\Delta/\Omega$ for five system sizes with the largest MIS degeneracies for each of the probed shapes. One sees that in all but two of the clusters, a TQSL state is reached with high fidelity at $\Delta/\Omega \gtrsim 1.5$, with the VBS phase absent up to the largest detunings probed. The two remaining  $10 \times 6$ clusters have relatively small MIS degeneracies (60 and 90, respectively), leading to a less robust TQSL state, and show the presence of the VBS state at larger $\Delta/\Omega$.
}
\label{fig:FigS4_shapefidelity}
\vskip -0.cm
\end{figure}

A crucial part of establishing the presence of the TQSL state in ED simulations is to probe different system sizes, aspect ratios, and shapes. The most important takeaway from such an analysis was presented in Fig.~\ref{fig:Fig3_fidelity}(b), where it was shown that the large TQSL overlap persists upon increasing the effective system size parameter---the MIS degeneracy---for a given cluster.  Our ED results point to a significantly better ground-state spin liquid fidelity and fidelity retention with increasing MIS degeneracy, than observed in the ED simulations of the PXP model on ruby-lattice clusters for the $\mathbb{Z}_2$ spin liquid \cite{Giudici2022}. Here, in Fig.~\ref{fig:FigS4_shapefidelity}, we show the TQSL overlap as a function of the detuning for a range of clusters with the largest MIS degeneracies.

The typical result valid for the clusters with MIS degeneracies above $\mathcal{O}(100)$ as well as for most of the smaller MIS degeneracy clusters is the presence of two phases only: trivial at small detunings and TQSL at large ones, as exemplified by the high TQSL fidelity for the largest detunings probed. The only deviation to this trend occurs for clusters with a large system size ($10 \times 6$) but comparatively small ($<100$) MIS degeneracy. This destabilizes the TQSL state at large detunings and leads to the appearance of the VBS phase. This is seen clearly for the $10 \times 6$ shape 1  cluster, where the TQSL region that survives is evident in the overlap. On the $10 \times 6$ shape 3  cluster, the TQSL region is still present, as detected by the changing superposition structure of the ground state (see Sec.~\ref{sec:GS_structure}), but only in a diminished detuning range ($1.3$--$1.9$ $\Delta/\Omega$), making it hard to establish two transitions in the overlap and the fidelity susceptibility.

\subsection{Structure of the ground states and low-energy spectrum \label{sec:GS_structure}}

\begin{figure}[htb]
\vspace*{-0.0cm}
\centering
\includegraphics[width=0.75\columnwidth]{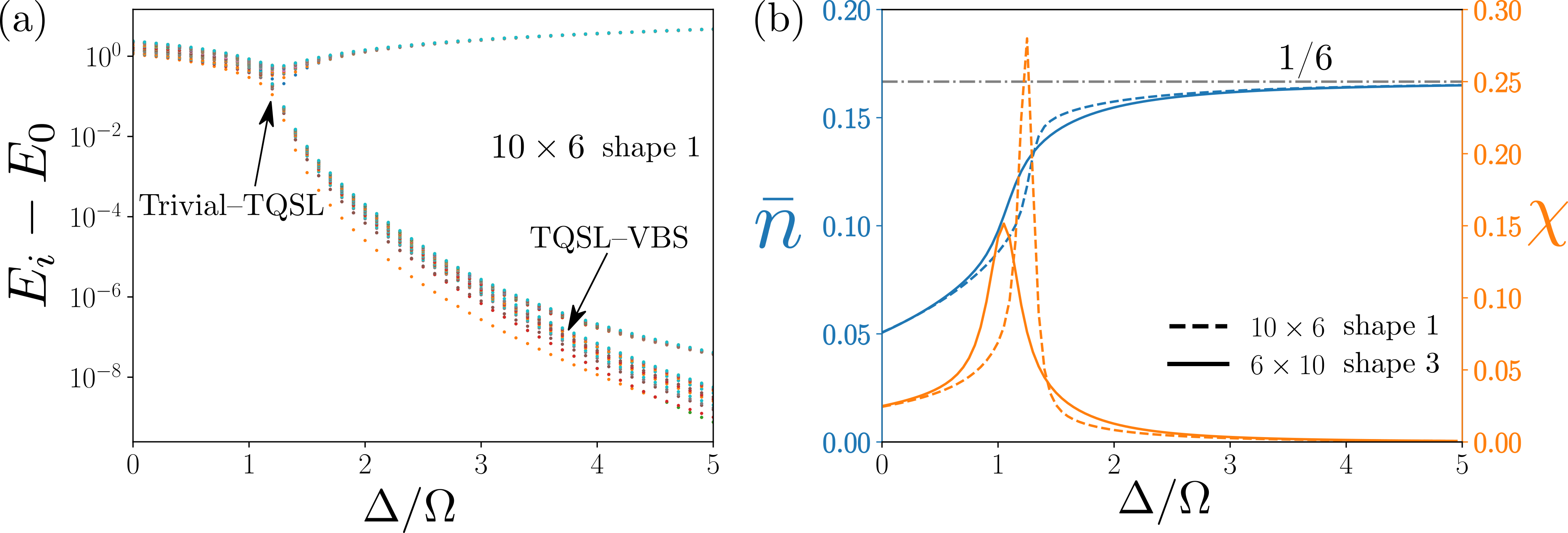}
\vskip -0.cm
\caption{\textbf{Exact-diagonalization signatures of transitions into the TQSL.} Besides the signatures in TQSL overlap and fidelity susceptibility, transitions into the TQSL can be followed by (a) evolution of the low-lying energy spectrum where the character of the spectrum exhibits changes at both trivial--TQSL and TQSL--VBS transitions, and (b) the density or the energy susceptibility, which only detect the first transition, similar to Ref.~\onlinecite{Verresen2021}.
}
\label{fig:FigS5_spectrum}
\vskip -0.cm
\end{figure}

We now establish additional signatures showing the presence of the TQSL phase in the ED simulations and provide details of the ground-state wavefunctions obtained. In Fig \ref{fig:FigS5_spectrum}(a), the lowest 100 eigenenergies for the $10 \times 6$ shape 1  cluster are shown. The spectrum displays two transition points, as evidenced by the qualitative changes in energy splittings appearing at the same positions as peaks in the fidelity susceptibility in Fig.~\ref{fig:Fig3_fidelity}(a). The first qualitative change can be related to the MIS manifold becoming dominant for the low-energy eigenstate. This is seen in the spectrum as the appearance of a large energy splitting $\propto \Delta$ between two groups of states corresponding to predominantly $|\mathrm{MIS}|$ and $|\mathrm{MIS}|-1$ superpositions. The energy splittings within the groupings are determined by $\Omega$-induced hoppings between different states with the same number of Rydberg excitations and decay as a high power of $\Omega/\Delta$. The second transition can be connected to level crossings and energy splittings within the MIS manifold itself, signaling the onset of the VBS order. The trivial--TQSL transition can also be detected in the (experimentally accessible) energy susceptibility, as shown in Fig.~\ref{fig:FigS5_spectrum}(b) for two representative clusters, or equivalently, the average density as a function of detuning. The second (TQSL--VBS) transition, however, is hard to detect from the density, as both states have an average density very close to $1/6$ as expected for trimer coverings.

\begin{figure}[tb]
\vspace*{-0.0cm}
\centering
\includegraphics[width=1.0\columnwidth]{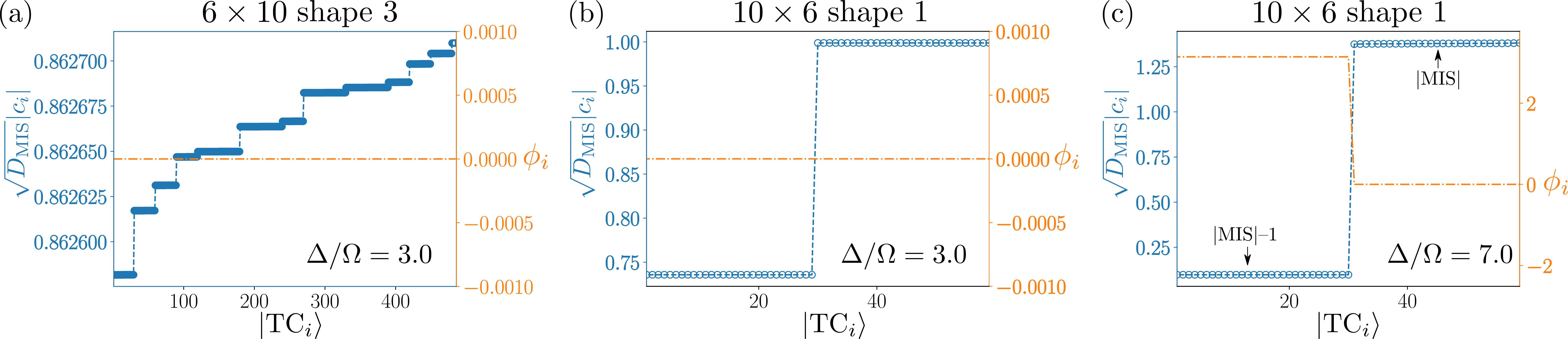}
\vskip -0.cm
\caption{\textbf{Structure of the TQSL wavefunction.} The weights and phases of different configurations in the ground-state wavefunction on two representative clusters. (a) An almost  perfect equal-superposition of all MIS states describes the $6 \times 10$ shape 3 ground state in the TQSL region. The remaining weight is distributed among configurations with monomers with at least an order-of-magnitude smaller coefficients (see Fig.~\ref{fig:Fig3_fidelity}(c) in the main text). This is representative of all the shapes showing the perfect TQSL state in Fig.~\ref{fig:FigS4_shapefidelity}. (b) For the $10 \times 6$ shape 1 cluster in the TQSL region, the ground state is still well described by a superposition of all the MIS states with equal phases, but with a more pronounced weight asymmetry. (c) In contrast, deep in the VBS region of the $10 \times 6$ shape 1 cluster, the ground state is predominantly a superposition of only 29 out of 60 MIS states, with the next-highest contribution coming from the $|\mathrm{MIS}|-1$ subspace.} 
\label{fig:FigS6_MISvsi}
\vskip -0.cm
\end{figure}

In Fig.~\ref{fig:FigS6_MISvsi}, we dwell on the details of the ground states in the TQSL regions of two representative clusters, by showing the weights and phases of the most significant states in the superposition. Since these states lie within the TQSL phase, the highest-weight states belong to the MIS subspace, with all the MIS configurations included in the superposition. Furthermore, the perfect TQSL state should have equal weights and equal phases for all of the MIS configurations. This is indeed very close to the actual ground state realized in the typical cluster of Fig.~\ref{fig:FigS6_MISvsi}(a) where all the MIS states participate in the ground state with equal phases and with weights within $2\times10^{-4}$ of the relative spread. The magnitude of the weights themselves is $\sim 0.86$ of the perfect TQSL value of $1/\sqrt{486}$, with the remaining weights being distributed mostly among states in the $|\mathrm{MIS}|-1$ subspace with an order-of-magnitude smaller individual configuration weights (see Fig.~\ref{fig:Fig3_fidelity}(c) in the main text). This is the main cause of the decrease in fidelity relative to the perfect TQSL state and shows the significance of trimer-monomer fluctuations that, in turn, determine the effective physics within the MIS subspace. We emphasize that this remarkable similarity to the perfect TQSL state is shown here for the cluster with the largest MIS degeneracy, and it corresponds to the typical case of clusters with high TQSL fidelities from Fig.~\ref{fig:FigS4_shapefidelity}. The near worst-case scenario exemplified by the cluster of  Fig.~\ref{fig:FigS6_MISvsi}(b) still shows a similar structure with MIS admixtures being dominant with equal phases. The difference arises due to higher weight inequality, with half of the weights being close to the perfect TQSL value and the other half at $\sim 0.75$ thereof. In contrast, deep in the VBS region of the $10 \times 6$ shape 1 cluster [Fig.~\ref{fig:FigS6_MISvsi}(c)], one observes a ground state that is predominantly a superposition of only half of the MIS configurations, with the contribution from the rest of the states in the MIS subspace strongly suppressed such that the next highest weights states arise from the $|\mathrm{MIS}|-1$ subspace. The $|\mathrm{MIS}|-1$ configurations are also accompanied by a relative phase of $\pi$ with respect to the MIS configurations in the wavefunction.

\subsection{Effect of open boundaries \label{sec:OBC_ED}}

\begin{figure}[tb]
\vspace*{-0.0cm}
\centering
\includegraphics[width=1.0\columnwidth]{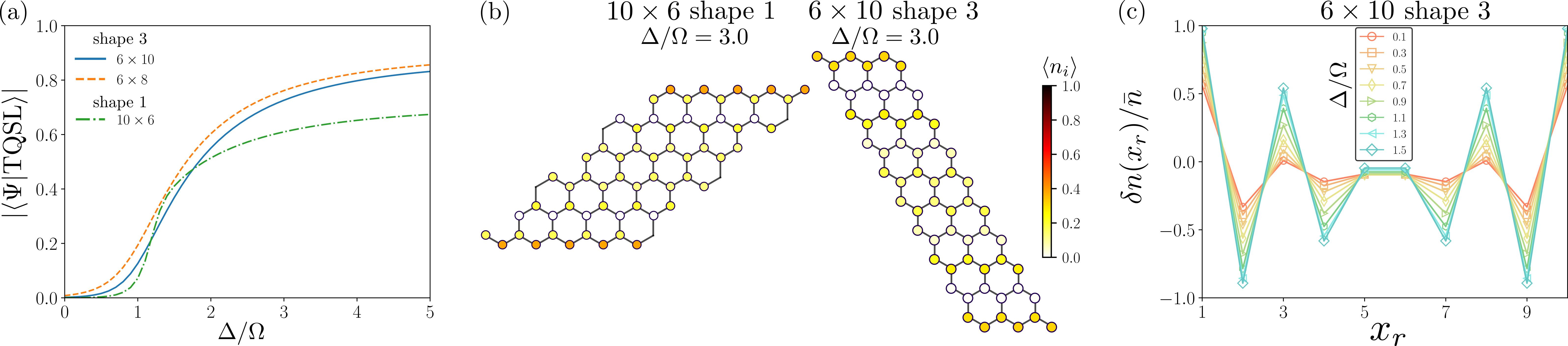}
\vskip -0.cm
\caption{\textbf{Effect of open boundaries.} (a) The TQSL overlap for several representative clusters with cylindrical boundary conditions drops somewhat compared to the system on a torus, but the TQSL nature of the state is still preserved. (b) The density profile of the clusters in the TQSL state shows the primary effect of the open boundaries, namely, inducing the density fluctuations. (c) Columnar density fluctuations are enhanced in the TQSL region compared to the trivial state, leading to similar qualitative behavior as seen in the DMRG calculations [Fig.~\ref{fig:FigS3_dmrgTQSL}(c)] with the equivalent boundary conditions. 
}
\label{fig:FigS7_OBC}
\vskip -0.cm
\end{figure}

In order to probe the robustness of the TQSL state found in ED numerics as well as to compare more directly between the ED and DMRG simulations, we perform exact diagonalization calculations on clusters with cylindrical boundary conditions. For this purpose, we take the cluster shapes and aspect ratios previously considered with periodic boundaries and relax the open boundaries at the shorter edge of the system. The results are presented in Fig.~\ref{fig:FigS7_OBC} for several representative clusters.

Inspecting the TQSL overlap as a function of the detuning in Fig.~\ref{fig:FigS7_OBC}(a) shows that the cylindrical TQSL fidelity remains large for the typical clusters, signifying the preservation of the TQSL phase. The fidelity drops somewhat in the less-favorable cluster ($10 \times 6$), which has a high boundary-to-bulk atom ratio (1/3). Even in this case, however, the phase region at $\Delta/\Omega \gtrsim 1.5$ can be described as a TQSL from the analysis of the ground-state structure. In Fig.~\ref{fig:FigS7_OBC}(b), we show the density profile in the TQSL phase for two clusters. The similarity to the TQSL excitation density profiles found previously with DMRG is apparent, with the overall density in the bulk approaching $1/6$ and the boundary inducing density fluctuations. 
This is further supported by Fig.~\ref{fig:FigS7_OBC}(c), where we show the column-averaged relative density oscillations. These exhibit a strong enhancement across the the trivial--TQSL transition, in qualitative similarity to the ones observed using DMRG [see Sec.~\ref{sec:gap_density} and Fig.~\ref{fig:FigS3_dmrgTQSL}(c)].

\subsection{Effect of interaction tails}

\begin{figure}[htb]
\vspace*{-0.0cm}
\centering
\includegraphics[width=1.0\columnwidth]{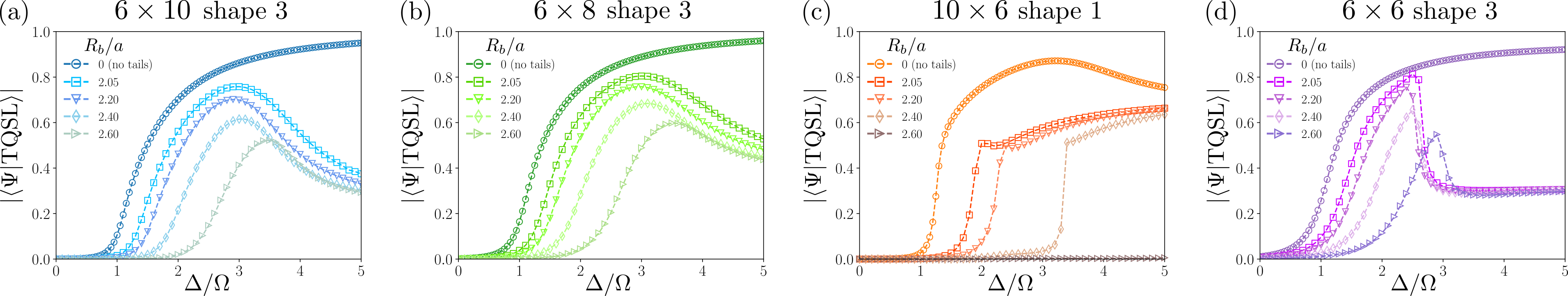}
\vskip -0.cm
\caption{\textbf{Effect of interaction tails.} The TQSL overlap as a function of $\Delta/\Omega$ for several representative clusters with varying interactions, ranging from no tails ($R_b/a=0$) to a tail strength equivalent to $2<R_b/a<\sqrt{7}$ in the $k$\,$=$\,$3$ regime. The clusters that hosted a perfect TQSL state in the case without tails, (a), (b), and (d), show a sizable TQSL regime in the intermediate $\Delta/\Omega$ region for all interaction strengths probed. The relatively less-robust cluster (c) now shows a significantly diminished TQSL region that completely vanishes at larger tail strengths. 
}
\label{fig:FigS8_tailsFid}
\vskip -0.cm
\end{figure}

The second aspect of robustness particularly relevant to experimental Rydberg systems is the effect of the interaction tails initially discarded in the simulation. While we have already summarized the major results in the main text, here, we present additional details with different system sizes and shapes in mind. The results are plotted in Fig.~\ref{fig:FigS8_tailsFid} for several clusters previously studied without tails in Fig.~\ref{fig:FigS4_shapefidelity}. A typical cluster shows the presence of the TQSL region in the intermediate detuning parameter range, with decreasing fidelity as the interaction strength increases. As noted in the main text, at large $\Delta/\Omega$, a VBS phase is always present with the same fidelity plateau reached for different values of the interaction tails' strengths. The worst-case scenario of the $10 \times 6$ shape 1 cluster still shows a TQSL region at intermediate detunings, as exemplified by fidelity peaks and confirmed by considering the wavefunctions' structures, although in a much diminished detuning range and with lower fidelities, vanishing at high $R_b/a$ completely.

\section{Dynamical preparation of TQSL states}

\label{sec:D}

\subsection{Dependence on detuning, system size, and cluster shape}

\begin{figure}[htb]
\vspace*{-0.0cm}
\centering
\includegraphics[width=0.4\columnwidth]{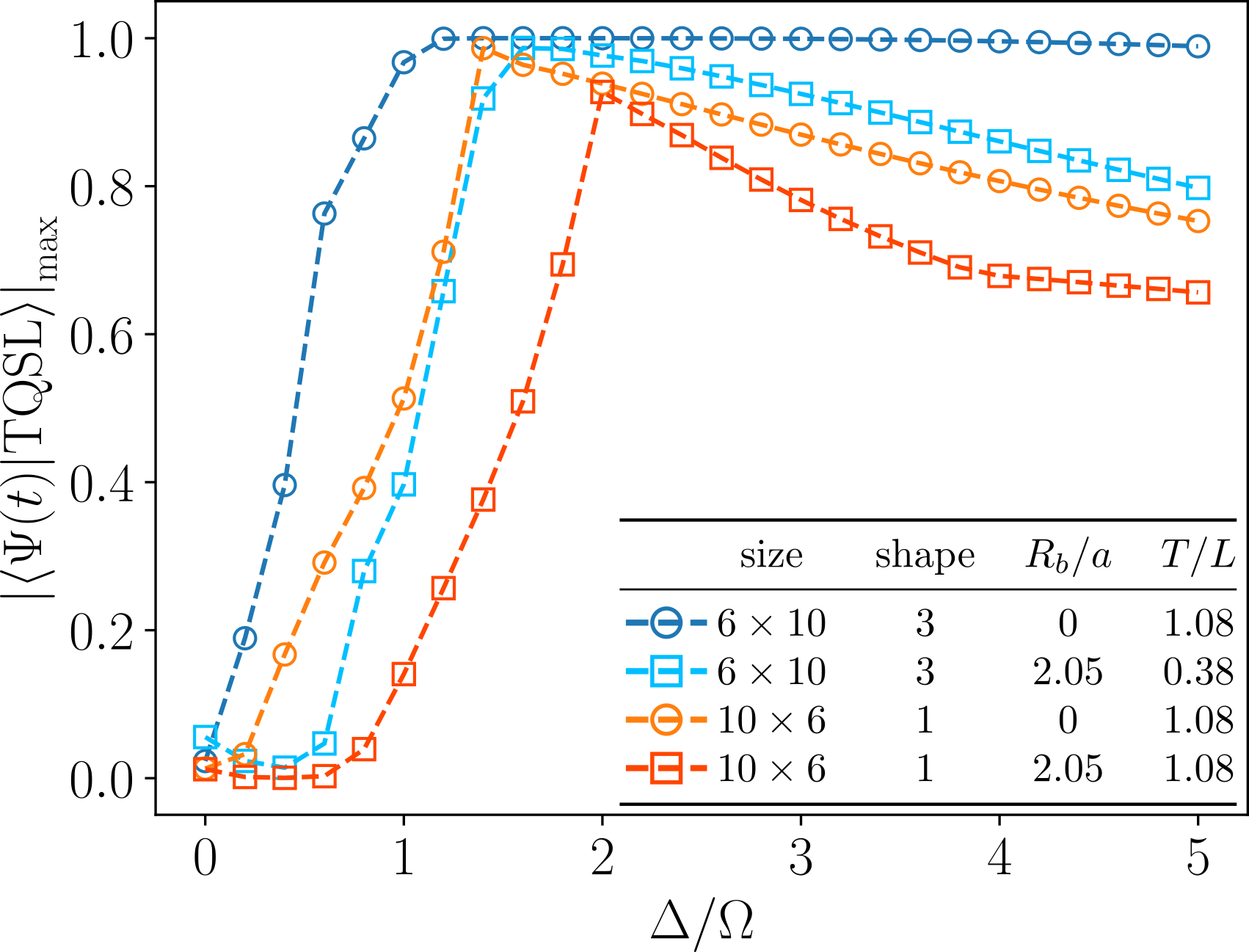}
\vskip -0.cm
\caption{\textbf{Dynamically prepared TQSL fidelity.} The maximum TQSL overlap reached during dynamical state preparation with fixed total time, as a function of the detuning for several system sizes with and without interaction tails. While the dynamical phase diagram obtained from the TQSL fidelity is reminiscent of the ground-state ones, the prepared fidelities are significantly higher, and the presence of a sizable TQSL region is evident even in the $10 \times 6$ shape 1 cluster.
}
\label{fig:FigS8b_prepDelta}
\vskip -0.cm
\end{figure}

In this section, we provide additional data on the dynamical preparation of TQSL states for different clusters. We first show, in Fig.~\ref{fig:FigS8b_prepDelta}, the maximum TQSL overlap reached during dynamical preparation with fixed total time as a function of the detuning for several system sizes with and without interaction tails. This effectively allows us to extract the dynamical phase diagram. The results without tails are qualitatively similar to the ground-state fidelity from Fig.~\ref{fig:FigS8_tailsFid}, with a typical $6 \times 10$ shape 3 cluster showing a large TQSL region and no VBS phase, while the worst-case $10 \times 6$ shape 1 cluster shows a dip in the fidelity at large detunings pointing to the onset of the VBS. The prepared fidelities in the TQSL region are significantly above the ground-state ones. In the system with tails, a sizable region of high TQSL fidelity is observed at intermediate detunings, with the eventual onset of the VBS now visible for all clusters. Perhaps most remarkably, the $10 \times 6$ shape 1 cluster with tails clearly reveals the presence of a sizable TQSL phase with the dynamical preparation scheme, despite the diminished ground-state TQSL regime.

\begin{figure}[htb]
\vspace*{-0.0cm}
\centering
\includegraphics[width=0.75\columnwidth]{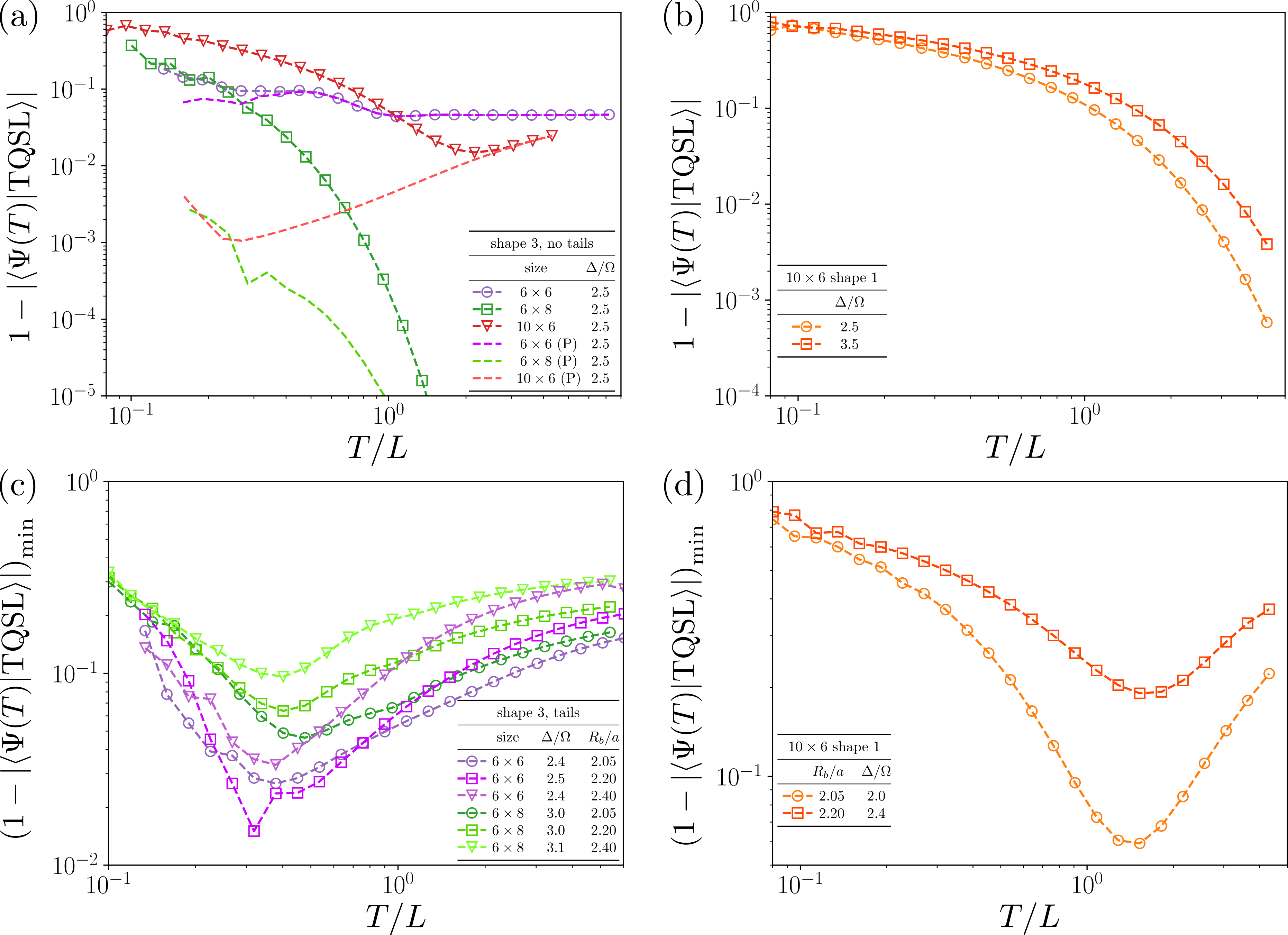}
\vskip -0.cm
\caption{\textbf{Adiabatic preparation for various cluster sizes and shapes.} (a,b) TQSL overlap of the adiabatically prepared state as a function of the total protocol time for different cluster sizes and shapes without interaction tails. The preparation fidelity is higher than the ground-state fidelity in all cases; while the fidelity seems to be able to approach arbitrarily close to 1 for the more robust clusters, it appears to be upper bounded for the less robust ones. The projection fidelity (labeled by P) matches the preparation fidelity in the long time limit, confirming the simple understanding of the off-ramp projection mechanism discussed in the main text. (c,d) Maximum preparation overlap for three clusters with different interaction tail strengths. While the overall fidelity is reduced compared to the case without tails and there now always exists an optimal preparation time, the preparation fidelities with tails are still significantly higher than the corresponding ground-state fidelities.
}
\label{fig:FigS9_prepT}
\vskip -0.cm
\end{figure}

The data for the final prepared TQSL fidelity is shown in Fig.~\ref{fig:FigS9_prepT}(a,b) as a function of the total preparation time for different clusters with no interaction tails. The typical cluster in the TQSL region displays behavior akin to the $6 \times 8$ shape 3 cluster presented here and to the cluster from Fig.~\ref{fig:Fig3b_adiabatic} of the main text. The fidelity appears to be able to approach arbitrarily close to unity with increasing total preparation time. A similar behaviour is observed on the $10 \times 6$ shape 1 cluster shown in Fig.~\ref{fig:FigS9_prepT}(b). In addition, we also examine two clusters with a qualitatively different behavior. The $10 \times 6$ and $6\times 6$ shape 3 clusters exhibit a limited maximum fidelity and correspondingly, the presence of an optimal total preparation time. The optimal prepared overlap is still in excess of $\sim 0.95$ for these cases. The difference is possibly related to the smaller MIS degeneracies for these clusters, which is seen to destabilize the TQSL ground state for the $10 \times 6$ cases (see Sec.~\ref{sec:GS_size}). An upper-bounded fidelity that reaches an optimal value at intermediate preparation times is also the typical behavior observed for the $k$\,$=$\,$3$ PXP model without tails on the ruby lattice \cite{Giudici2022}, as well as for the honeycomb lattice with long-ranged interactions explored here. 

Figure~\ref{fig:FigS9_prepT}(c,d) presents the optimal TQSL overlap reached during the adiabatic preparation sweep as a function of the total preparation time for three additional clusters with different interaction tail strengths. The detuning is selected to correspond to the maxima of the TQSL ground-state fidelities from Fig.~\ref{fig:FigS4_shapefidelity}, which  does not generically correspond to the optimal detunings for the dynamical preparation  seen in Fig.~\ref{fig:FigS8b_prepDelta}. The obtained results are in agreement with the ones described in Fig.~\ref{fig:Fig4_tails} of the main text. The main features observed are: (1) fidelities significantly above the ground-state ones, (2) the existence of an optimal total preparation time, and (3) a decrease in the optimal fidelity with increasing interaction strengths. Moreover, fluctuations in the optimal prepared fidelity are observed for short total preparation times. These stem from the short time-scale effects (see Fig.~\ref{fig:FigS10_preptt}) induced by the preparation protocol that has steeper ramps for smaller $T$ and are more pronounced in smaller clusters.

\subsection{Universal off-ramp projection}

\begin{figure}[htb]
\vspace*{-0.0cm}
\centering
\includegraphics[width=1.0\columnwidth]{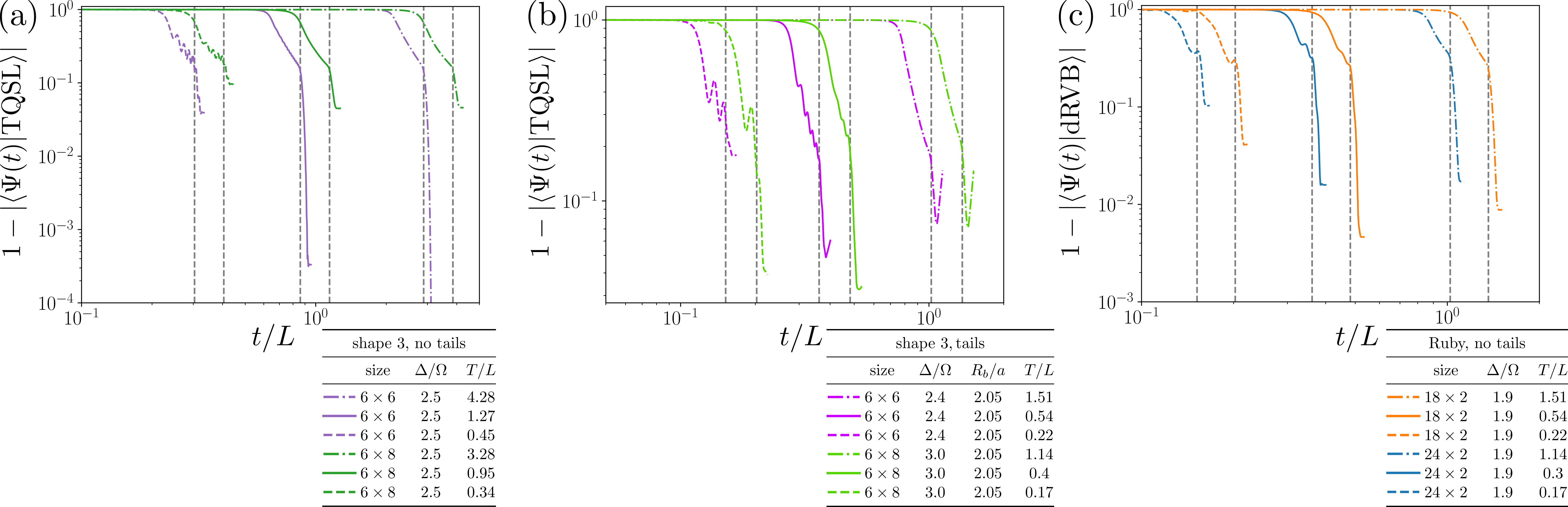}
\vskip -0.cm
\caption{\textbf{Universality of the off-ramp projection mechanism.} Spin liquid overlap during adiabatic preparation for several clusters and total protocol times in the case of (a) the honeycomb lattice without tails (TQSL fidelity), (b) the honeycomb lattice with tails (TQSL fidelity), and (c) the ruby lattice without tails for clusters explored in Ref.~\onlinecite{Giudici2022} (dimer RVB fidelity). In all the cases presented, significant fidelity gain is obtained in the off-ramp part of the protocol, which can be explained by the projection to the MIS subspace during the off ramp. In the case with interaction tails, after $\Omega$ drops below the appropriate value for the next blockade shell, projection to the smaller MIS subspace starts to occur, leading to a final upturn in infidelity.
}
\label{fig:FigS10_preptt}
\vskip -0.cm
\end{figure}

Following our previous discussion, we now show more examples of the off-ramp fidelity-enhancement mechanism in action, including the adiabatic preparation of the dimer RVB (dRVB) state on the ruby lattice, as well as supporting evidence for the projection mechanism of the off ramp. With this in mind, we highlight three distinct cases where the fidelity enhancement mechanism applies in Fig.~\ref{fig:FigS10_preptt}: the PXP model on the honeycomb lattice without tails, the PXP model on the honeycomb lattice with interaction tails, and the no-tail PXP model on the ruby lattice. In all of these cases, the same preparation protocol as in Fig.~\ref{fig:Fig3b_adiabatic} of the main text was used, with $\Delta/\Omega$ in the main part of the ramp corresponding to the TQSL/dRVB region. The clusters employed for the ruby lattice are the same as the 48- and 36-site clusters used in Ref.~\onlinecite{Giudici2022}. In all three scenarios, after the first $0.9T$ of preparation time (denoted by gray dashed lines), the fidelity obtained with this dynamical preparation scheme is similar to that of the ground state (see Figs.~\ref{fig:FigS4_shapefidelity}, \ref{fig:FigS8_tailsFid} and \cite{Giudice2022}). In the last $0.1T$ time segment corresponding to the off ramp, a large gain in the fidelity is observed. In the cases without tails, this can yield several orders of magnitude of fidelity enhancement. We note that the off-ramp part of the pulse is responsible for the high RVB fidelities reported in the recent ruby-lattice simulations \cite{Giudici2022} of the $\mathbb{Z}_2$ spin liquid. More significantly, in the experimental systems with tails, the off ramp still leads to practically relevant fidelity enhancements over the ground state, regardless of the cluster shape and size. An additional upturn appears at the end of the ramp, understanding which in terms of the projection mechanism of the ramp may allow for better protocol optimization, as discussed in the main text.

The conjectured projection mechanism of the ramp was probed quantitatively by comparing the fidelity obtained at the end of the dynamical preparation to the one obtained after projecting the state prepared before the ramp-down to the MIS subspace. The results for the conceptually simpler cases without tails are shown in Fig.~\ref{fig:FigS9_prepT}(a) as a function of the total preparation time and are denoted by (P) in the legend. For the cases explored, the projection fidelity gives an upper bound below 1 to the accessible final preparation fidelity. This bound is indeed approached for very long preparation times where the off ramp is slow enough and long enough to effectively achieve perfect projection. In the system with tails, the limit of long total times is detrimental otherwise, as it eliminates some of the quasi-adiabatic effects that make the interaction tails irrelevant for the prepared state at $0.9T$. Nonetheless, a significant gain is reached during the off-ramp part. In this case, a potentially longer off ramp with a sharp cutoff at $t^*$ (see the main text) in conjunction with a relatively fast main ramp might be the setup that exploits both the quasi-adiabatic and projection effects optimally.

\section{Exact diagonalization of the \texorpdfstring{$k$\,$=$\,$2$}{k=2}  PXP model}

\label{sec:E}

\begin{figure}[htb]
\vspace*{-0.0cm}
\centering
\includegraphics[width=0.6\columnwidth]{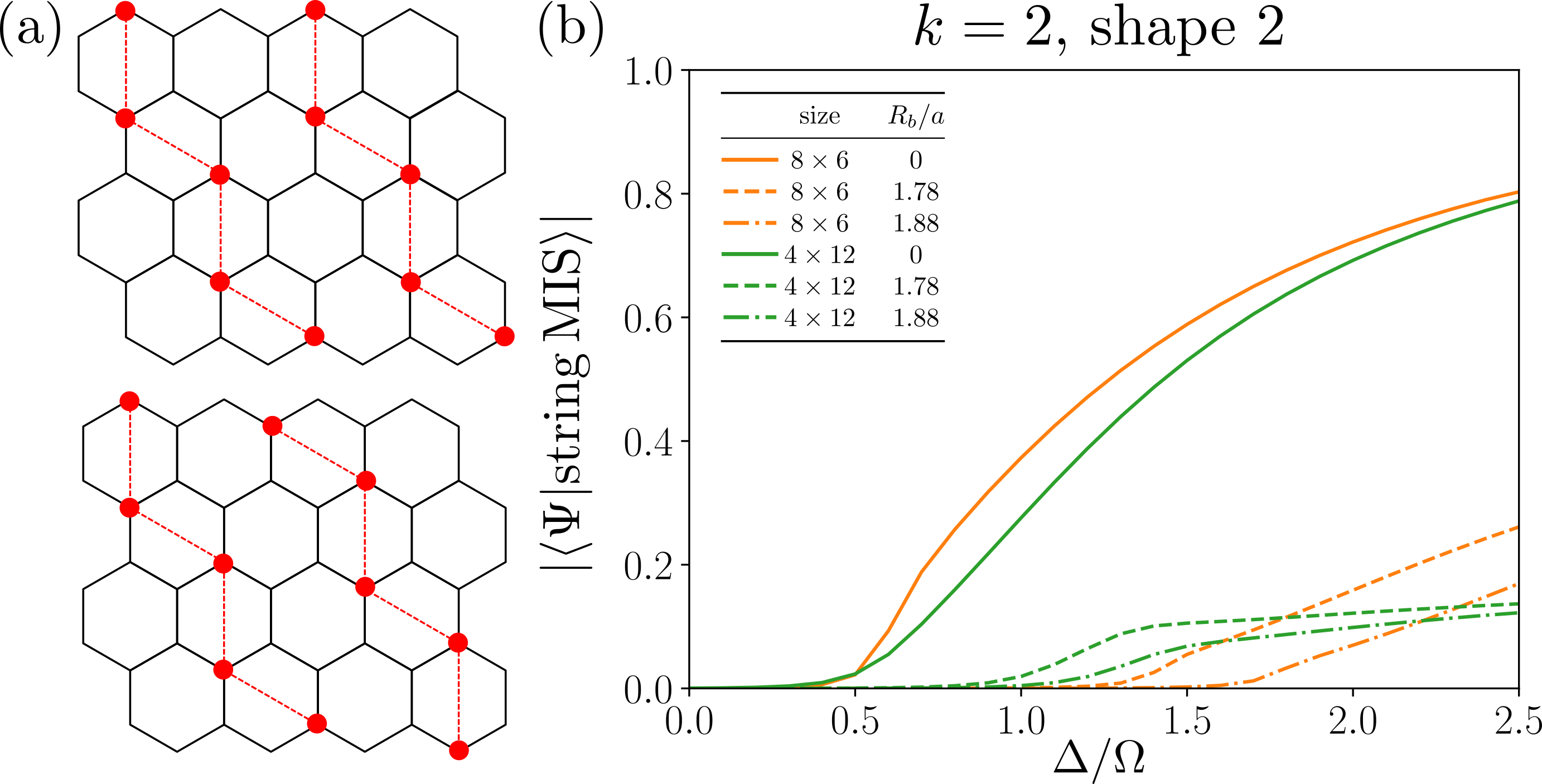}
\vskip -0.cm
\caption{\textbf{String degeneracy and the ground state in the second-neighbor blockade regime.} (a) In the $k$\,$=$\,$2$ regime, there is a classical sliding degeneracy of strings, arising because every two neighboring strings can be in parallel (top) or antiparallel (bottom) configurations. This degeneracy scales exponentially with the linear system size in contrast to the \textit{total} system size scaling of the trimer degeneracy. (b) Overlap of the $k$\,$=$\,$2$ PXP ground state with an equal superposition of string states (``string MIS'') as a function of the detuning. While the string MIS fidelity is quite high for the system without tails, the state is not robust to the interaction tails.
}
\label{fig:FigS11_k2}
\vskip -0.cm
\end{figure}

Lastly, we turn to exploring the $k$\,$=$\,$2$ regime of the honeycomb lattice ($\sqrt{3}<R_b/a<2$), within the PXP model. We are motivated by the  classical version of the $k$\,$=$\,$2$ hard-core boson model at large detuning, where an extensive ``string'' degeneracy appears. This can be seen in Fig.~\ref{fig:FigS11_k2}(a) as for every string composed of excited atoms at a distance $2a$, the neighboring string can be in both parallel or anti-parallel configurations, giving rise to a classical degeneracy that scales exponentially with the linear system size. The only way to flip between distinct string configurations involves flipping a number of sites proportional to the linear system size as well. Thus, although a potentially interesting quantum state, a ``string MIS'' state that is an equal superposition of the classical string configurations is not a resonating state, eliminating the possibility for spin-liquid physics in this regime. We contrast this with a recent study of similar string degeneracy on the kagome lattice \cite{Samajdar2021}, where the classical string configurations could be mapped to a dimer model on the medial triangular lattice, potentially giving rise to a $\mathbb{Z}_2$ spin liquid; however, no such mapping exists for the honeycomb strings. Nonetheless, a disordered quantum string state arising from highly constrained classical dynamics might be an interesting platform for exploring the recent proposal of emergent glassy dynamics \cite{Feldmeier2019} in a nonequilibrium setting.

As discussed in the main text, quantum fluctuations break the string degeneracy and ultimately favor a state with ordered antiparallel strings, the columnar state. The choice of antiparallel strings is not surprising, given that the distance between excited atoms is larger for antiparallel than for parallel strings, leading to a greater energy gain from fluctuating configurations (note that thermal fluctuations make the same  choice of ordered state in the classical model \cite{Thewes2020}). The argument above for maximizing the distance between neighboring excitations also explains why the interaction tails prefer antiparallel strings, leading to the same columnar state.

The columnar order sets in only at $\Delta/\Omega>3.0$, as evidenced by the DMRG phase diagram. Here, we explore whether there is a possibility to access the potentially novel string MIS state in the intermediate detuning regime using the $k$\,$=$\,$2$ PXP model. The effective system size as measured by the MIS degeneracy is significantly smaller than in the $k$\,$=$\,$3$ case, a consequence of the degeneracy scaling with the linear system size, making the ED argument for such a state weaker. Still, as shown in Fig.~\ref{fig:FigS11_k2}(b) for the PXP model without tails, a high string MIS fidelity is observed in the intermediate detuning regime. However, including the interaction tails up to $\mathcal{R}=3a$ leads to a large decrease in fidelity and the loss of the string MIS state, making it unlikely to be realizable in an experiment. The fragility of the string MIS state lies in contrast to the observed remarkable robustness of the TQSL state to interaction tails.

\end{document}